\documentclass{article} 
\usepackage{soul}
\usepackage{iclr2026_conference,times}

\usepackage{amsmath,amsfonts,bm}

\def\eqref#1{equation~\ref{#1}}

\def\1{\bm{1}}

\DeclareMathAlphabet{\mathsfit}{\encodingdefault}{\sfdefault}{m}{sl}
\SetMathAlphabet{\mathsfit}{bold}{\encodingdefault}{\sfdefault}{bx}{n}

\usepackage{amsmath,amssymb,algorithm,algorithmic}
\usepackage{mathtools}
\usepackage{url}
\usepackage{booktabs}
\usepackage{enumitem}
\usepackage{wrapfig}
\usepackage{subcaption}
\usepackage[export]{adjustbox}
\usepackage{multirow}
\usepackage{graphicx}
\usepackage{amsthm}
\usepackage[most]{tcolorbox} 

\usepackage[framemethod=tikz]{mdframed}
\usetikzlibrary{decorations.pathreplacing}
\DeclareRobustCommand{\rectsymbol}[5]{%
  \tikz[baseline=#5]%
    \draw[draw=#1, fill=#2] (0,0) rectangle (#3,#4);%
}
\DeclareRobustCommand{\roundrectsymbol}[5]{%
  \tikz[baseline=#5]%
    \draw[draw=#1, fill=#2, rounded corners=0.15em] (0,0) rectangle (#3,#4);%
}
\definecolor{mydarkyellow}{HTML}{CCCC00}
\definecolor{c1}{HTML}{007FFF}
\definecolor{c2}{HTML}{0066CC}
\definecolor{c3}{HTML}{9999FF}
\definecolor{c4}{HTML}{CC99FF}
\definecolor{c5}{HTML}{B266FF}
\DeclareRobustCommand{\arcsymbol}[6]{%
  \tikz[baseline=#1]{%
    \draw[draw=#5, line width=#6]
      (0,0) arc[start angle=#2, end angle=#3, radius=#4];
  }%
}
\usepackage{xcolor}

\usepackage{colortbl}
\usepackage[normalem]{ulem}
\useunder{\uline}{\ul}{}
\usepackage{hyperref}
\usepackage{xspace}
\newcommand{\xhdr}[1]{\vspace{-1mm}\noindent{{\bf #1.}}}

\newcommand{\algo}{\textsc{GeoBPE}\xspace}
\newcommand{\model}{\textsc{{GeoBPE-Transfer}}\xspace}
\usepackage[font=footnotesize,labelfont=bf]{caption} 

\definecolor{swatchgray}{RGB}{220,220,220}
\definecolor{revision}{RGB}{255,192,203}
\DeclareCaptionLabelFormat{hl}{\revision{#1~#2}}

\DeclareRobustCommand{\revision}[1]{%
  \begingroup
    \setlength{\fboxsep}{0pt}
    \colorbox{revision}{\strut\kern0.3pt #1\kern0.3pt}%
  \endgroup
}
\newtcolorbox{revisionblock}[1][0pt]{%
  enhanced,
  breakable,
  colback=revision,
  colframe=revision,
  boxrule=0pt,
  left=#1, right=#1,
  top=5pt, bottom=0pt,
  boxsep=0pt,
  sharp corners,
}

\DeclareRobustCommand{\grayswatch}{%
  \begingroup\color{swatchgray}\rule{0.8em}{0.6ex}\endgroup}

\newcolumntype{V}[1]{>{\centering\arraybackslash}m{#1}} 

\title{Protein Structure Tokenization via \\ Geometric Byte Pair Encoding}

\author{
Michael Sun$^{1,2}$, Weize Yuan$^{3}$, Gang Liu$^4$, Wojciech Matusik$^1$, Marinka Zitnik$^2$ \\
$^1$MIT CSAIL \quad  $^2$Harvard Medical School \quad $^3$Apple \quad $^4$Notre Dame \\
\texttt{\{msun415,wojciech\}@csail.mit.edu}, \quad
\texttt{marinka@hms.harvard.edu}
}

\iclrfinalcopy 
\newtheoremstyle{mythm}{\topsep}{\topsep}{\itshape}{}{\bfseries}{.}{.5em}{%
  \thmname{#1} \thmnote{(\textbf{#3})}}
\theoremstyle{mythm}
\newmdtheoremenv[
  backgroundcolor=gray!10,
  linecolor=gray!50!black,
  linewidth=0.5pt,
  roundcorner=1mm,
  innertopmargin=6pt, innerbottommargin=6pt,
  innerleftmargin=6pt, innerrightmargin=6pt
]{theorem}{Desideratum}[section]
\begin{document}

\maketitle

\begin{abstract}

%
Protein structure is central to biological function, and enabling multimodal protein models requires joint reasoning over sequence, structure, and function. A key barrier is the lack of principled protein structure tokenizers (PSTs): existing approaches fix token size or rely on continuous vector codebooks, limiting interpretability, multi-scale control, and transfer across architectures. We introduce \algo, a geometry-grounded PST that transforms continuous, noisy, multi-scale backbone conformations into discrete ``sentences'' of geometry while enforcing global constraints. Analogous to byte-pair encoding, \algo generates a hierarchical vocabulary of geometric primitives by iteratively (i) clustering Geo-Pair occurrences with k-medoids to yield a resolution-controllable vocabulary; (ii) quantizing each Geo-Pair to its closest medoid prototype; and (iii) reducing drift through differentiable inverse kinematics that optimizes boundary glue angles under an $\mathrm{SE}(3)$ end-frame loss. \algo offers compression ($>$10× reduction in bits-per-residue at similar distortion rate), data efficiency ($>$10× less training data), and generalization (maintains test/train distortion ratio of $1.0-1.1$). It is architecture-agnostic: (a) its hierarchical vocabulary provides a strong inductive bias for coarsening residue-level embeddings from large PLMs into motif- and protein-level representations, consistently outperforming leading PSTs across $12$ tasks and $24$ test splits; (b) paired with a transformer, \algo supports unconditional backbone generation via language modeling; and (c) tokens align with CATH functional families and support expert-interpretable case studies, offering functional meaning absent in prior PSTs. Code is available at \url{https://github.com/shiningsunnyday/PT-BPE}.

\end{abstract}

\vspace{-7mm}

\section{Introduction}

Protein language models (PLMs) trained on large sequence databases capture evolutionary constraints \citep{rives2021biological} and support de novo sequence design \citep{lin2023evolutionary}, but they do not explicitly model fold geometry and may underperform on tasks where function depends on structural interactions \citep{abramson2024accurate, gelman2025biophysics}. In natural language processing, byte-pair encoding (BPE) constructs a vocabulary by iteratively merging the most frequent symbol pairs, producing a hierarchical representation of text \citep{larsson2002off}. Despite BPE's success on sequential data, there is no geometric analog that can encode and decode protein backbone conformations. The central difficulty is discretizing continuous, noisy structural variability while preserving global consistency. Because protein folds are organized into modular substructures \citep{petsko2004protein}, a protein structure tokenizer should (a) build a hierarchical vocabulary of structural motifs and (b) segment folds into hierarchical decompositions, producing symbolic and interpretable representations of backbone geometry.

Recently, vector-quantized variational autoencoders (VQ-VAEs) have become the most popular class of protein structure tokenizers (PSTs), as adopted by ESM3 \citep{hayes2025simulating} and others. VQ-VAEs learn an autoencoder that compresses and reconstructs a protein structure with $N$ residues to and from $N$ quantized latent codes, which are discrete ``words" drawn from a vocabulary of learnable embeddings \citep{van2017neural}. While powerful, VQ-VAEs lack the efficiency, interpretability and modularity of BPE tokenizers: (1) using a fixed codebook can create performance bottlenecks and imbalance token usage frequency, handicapping downstream performance \citep{yuan2025protein}; (2) using vectors as tokens over real data hinders interpretability, as rows of a 2D matrix do not capture the hierarchical relationships between sub-words like in BPE; (3) lastly, fixing all tokens to have the same size prevents multi-scale resolution, which is key to tasks that identify naturally occurring higher-level functional activity which span variable residue lengths.

\begin{figure}[h!]
    \centering
    \caption{
    \algo tokenizes a protein into discrete motifs linked by boundary glue angles and learns a hierarchical vocabulary of frequent structural primitives via k-medoids and recursively merging Geo-Pairs; at each step glue angles are optimized with differentiable inverse kinematics to preserve the global fold. Tokenization yields a merge tree that provides multi-resolution and interpretable representations of protein structure.
    }
    \fcolorbox{black!20}{white}{\includegraphics[width=\linewidth]{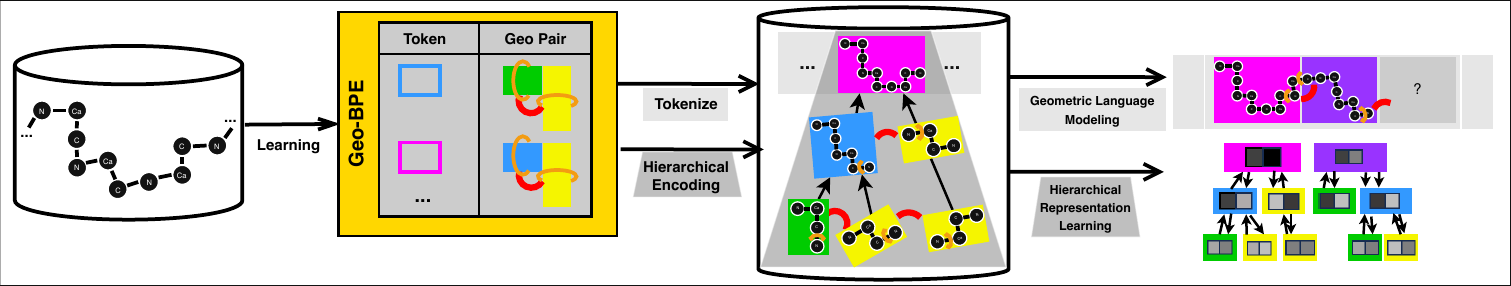}}
    \vspace{-5mm}
    \label{fig:geobpe}
\end{figure}

\xhdr{Present work}
We develop Geometric Byte-Pair Encoding (\algo), a tokenizer that discretizes continuous protein backbones into symbolic ``sentences'' of structural motifs while learning a hierarchical vocabulary. The design is motivated by two requirements: (i) protein folds contain modular substructures that should be captured as reusable tokens, and (ii) discrete approximations must preserve global geometric consistency. To meet these requirements, \algo alternates between local updates and global corrections. At each step, frequent motif pairs are clustered with k-medoids and replaced by representative prototypes, recursively building higher-order motifs. This local quantization inevitably introduces geometric drift, which \algo corrects by optimizing boundary glue angles through differentiable inverse kinematics under an $\mathrm{SE}(3)$ end-frame loss. The output after each iteration is a segmentation of the backbone into quantized motifs and glue parameters; the sequence of iterations yields a hierarchical decomposition of the fold, represented as a merge tree of structural motifs (Fig.~\ref{fig:geobpe}).
%
%
Our contributions are as follows: \textcircled{\small{1}} \algo is the first geometry-grounded BPE analog for protein backbones, which builds a hierarchical vocabulary of motifs and tokenizes structures through an alternating global-local decomposition with glue-aware reconstruction.
\textcircled{\small{2}} On benchmark datasets, \algo traces a smooth Pareto front of compression-distortion tradeoffs, achieving up to 0.27-0.36× the bits-per-residue of ProToken and strong out-of-distribution generalization (test/train RMSD ratio 1.16-1.28 vs. 6.4× for VQ-VAE). It also matches downstream accuracy when trained on as little as 1\% of the pretraining data.
\textcircled{\small{3}} Hierarchical vocabularies from \algo improve representation quality on tasks such as binding site prediction and fold classification, and a transformer trained on its tokens enables unconditional backbone generation.
\textcircled{\small{4}} Tokens align with CATH domain annotations and are supported by expert case studies, providing functional protein insights and multi-resolution interpretability.
%

\vspace{-3mm}

\section{Related Work}

\xhdr{Protein Structural Alphabets}
Structural alphabets approximate protein folds as successions of geometric motifs \citep{branden2012introduction}. \citet{de2000bayesian} introduced 16 five–residue protein blocks from Protein Data Bank (PDB) structures, assigning fragments by RMSD. Later work showed that over 90\% of residues can be covered by such alphabets \citep{de2002extension} and analyzed their quality and specificity \citep{de2005new}. Alphabet strings provide 1D encodings of 3D geometry, enabling the use of sequence alignment for fold analysis and prediction \citep{mahajan2015use, vetrivel2017knowledge}. \citet{camproux1999hidden} proposed 12 building blocks via Hidden Markov Models (HMMs) and extended it to capture whole-protein conformational variability \citep{camproux2004hidden}. HMMs use inter-alpha-carbon distances within four residues as observed variables. Broader tertiary descriptors, such as inter-residue distances or moment invariants \citep{durairaj2020geometricus}, capture non-contiguous context; \citet{mackenzie2016tertiary} found $\sim$600 motifs describe more than half of structural space ($39\cdot 10^6$ conformations), indicating variability collapses into limited modes. Such descriptors extend to protein-level retrieval and classification \citep{durairaj2020geometricus,van2024fast,barrio2023clustering}. \algo builds on these insights by treating structural motifs as extensible primitives and dynamically adjusting alphabet size and token resolution, unlike fixed structural alphabets.

\xhdr{Protein Structure Tokenizers} Modern PSTs, most notably VQ-VAEs, construct structural alphabets by training deep autoencoders with vectorized codebooks that map continuous structure to discrete codes \citep{van2017neural}. Building on this idea, FoldSeek \citep{van2024fast} introduced 3Di alphabets (20 discrete codes learned with VQ-VAE) that compress local structural features for efficient search and homology detection. Subsequent works integrate 3Di alphabets with PLMs: \citet{heinzinger2024bilingual} translate between 3Di and amino acid sequences; \citet{su2023saprot} define ``3Di-residue'' tokens and show pretraining with this vocabulary improves prediction; and \citet{li2024prosst} use disentangled attention to jointly model 3Di and residue tokens with a structure quantization module. End-to-end VQ-VAEs avoid predefined descriptors by training equivariant encoders and decoders to tokenize structure directly, achieving near-perfect reconstruction but at high computational cost. Large-scale efforts such as ESM3 \citep{hayes2025simulating}, trained on 236 million structures, highlight the central role of tokenizers in scaling multimodal PLMs. Recent work benchmarks tokenizer performance itself: AIDO.St and ProTokens show that stronger compression improves retrieval \citep{van2024fast, zhang2024balancing} but reduces reconstruction quality, and both \citet{zhang2024balancing} and \citet{lin2023tokenizing} integrate tokenizers tightly with transformers. \algo differs by using its hierarchical vocabulary as an inductive bias for representation learning and by supporting geometry-grounded language modeling without latent space vector quantization.

\xhdr{Byte-Pair Encoding for Biological Data} BPE underlies modern language models and has been applied to biological sequences with mixed outcomes. On genomes, BPE achieves superior compression and improves over k-mers in language models \citep{dotan2024effect, zhou2023dnabert}, though \citet{nguyen2023hyenadna} find the opposite using Hyena. For functional tasks, BPE often performs best \citep{dotan2024effect}, while on nucleotide-resolution tasks it can underperform \citep{lindsey2025impact}. These results indicate tokenizer utility depends on task scale and architecture, motivating \algo's architecture-agnostic design and multi-scale resolution. Linguistic differences between text and biological sequences further complicate direct transfer: BPE tokens do not align with domain boundaries \citep{suyunu2024linguistic} or regulatory motifs \citep{lindsey2025impact}. Other studies emphasize the importance of vocabulary design, reduced amino acid alphabets impair structure prediction \citep{ieremie2024protein}, while BPE vocabularies of 50–200 tokens are often optimal for sequence tasks \citep{tan2024peta}. Overall, existing tokenizers, including BPE, lack {\em versatility for protein structures.} \algo extends BPE by grounding tokenization in geometry, exposing parameters for quantization, vocabulary, and efficiency, while uniquely providing fine-grained resolution control and a hierarchical motif vocabulary.

\vspace{-3mm}

\section{Methods}

We first establish backbone geometry notations in Sec.~\ref{sec:prelims}. Sec.~\ref{sec:algorithm} presents the \algo algorithm, detailing its components for motif clustering, adaptive quantization, and glue-aware refinement. Finally, Sec.~\ref{sec:pst} formalizes the principles that an ideal protein structure tokenizer should satisfy and evaluates how \algo meets them.

\subsection{Notation \& Preliminaries}\label{sec:prelims}

\xhdr{Global Backbone Formulation}
Let a protein backbone $t^{(\tau)}$ with $N^{(\tau)}$ residues be represented by the Cartesian coordinates
$\{(N_i,\,\mathrm{CA}_i,\,C_i)\in\mathbb{R}^{3\times 3}\}_{i=1}^{N^{(\tau)}}$ of backbone atoms (oxygen and $\mathrm{C}_\beta$ omitted). Define bond lengths, bond angles, and dihedrals:
\[
\begin{aligned}
&
\ell^{N\!-\!CA}_i=\|N_i-\mathrm{CA}_i\|,\ 
\ell^{CA\!-\!C}_i=\|\mathrm{CA}_i-C_i\|\ ;\ 
\ell^{C\!-\!N}_i=\|C_i-N_{i+1}\|.\\
&
\theta^{N\!CA\!C}_i=\angle(N_i,\mathrm{CA}_i,C_i)\ ,\ 
\theta^{CA\!C\!N}_i=\angle(\mathrm{CA}_i,C_i,N_{i+1})\ \ 
\theta^{C\!N\!CA}_i=\angle(C_i,N_{i+1},\mathrm{CA}_{i+1}).\\
&
\psi_i=\angle(N_i,\mathrm{CA}_i,C_i,N_{i+1}),\ 
\omega_i=\angle(\mathrm{CA}_i,C_i,N_{i+1},\mathrm{CA}_{i+1}),\ 
\phi_i=\angle(C_{i},N_{i+1},\mathrm{CA}_{i+1},C_{i+1}).
\end{aligned}
\]
\begin{wrapfigure}[10]{r}{0.45\textwidth}
\vspace{-0.2in}
      {%
      
      \begin{minipage}{0.99\linewidth} 

\setlength{\fboxsep}{0pt}
  \centering
    \fcolorbox{black!20}{white}{\includegraphics[width=0.97\textwidth]{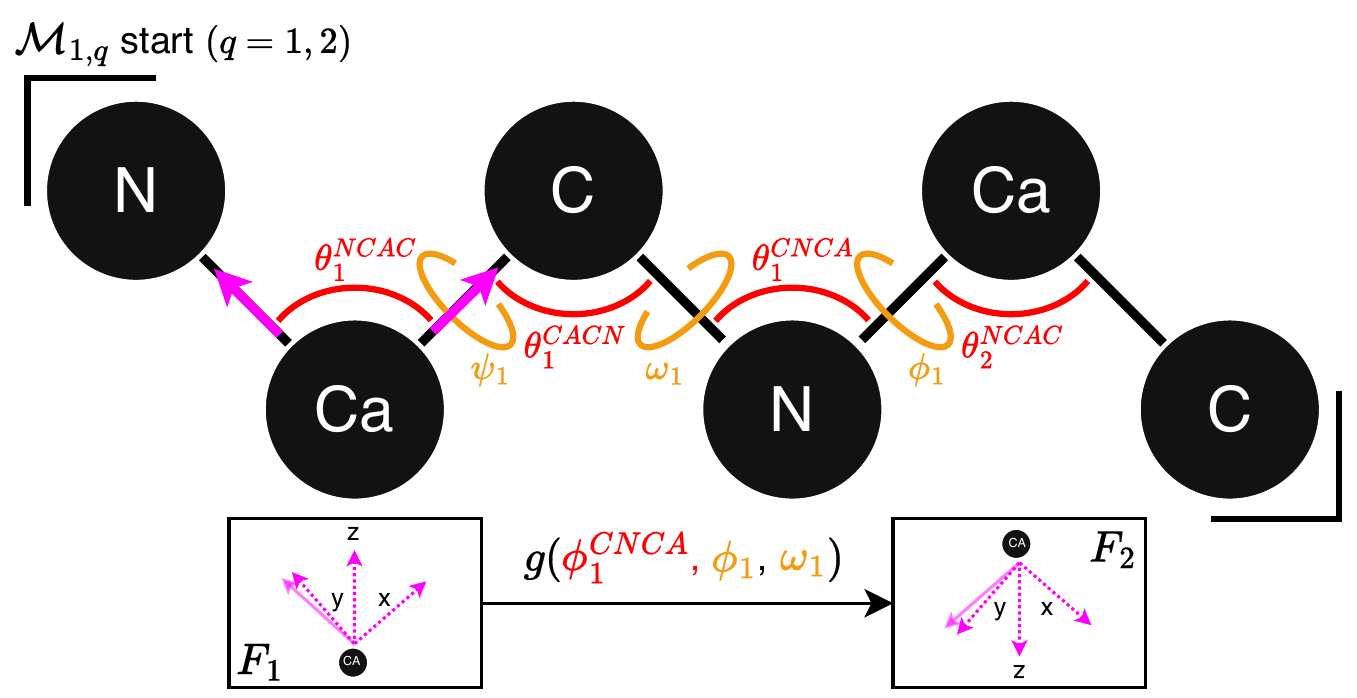}}

  \caption{{Toy backbone, with
internal angles, glues $T_1$, motif $\mathcal{M}_{1:2}$ and per-link transform $G_1$.}}
  \label{fig:notation}
        \end{minipage}%
        
    }

\end{wrapfigure}
{We annotate these definitions in a toy} ${(N^{(\tau)}=2)}$ {example in Fig. \ref{fig:notation} (top).} The full internal representation thus contains $3N^{(\tau)}\!-\!1$ bond lengths, $3N^{(\tau)}\!-\!2$ bond angles, and $3N^{(\tau)}\!-\!3$ dihedrals and is invariant to any $(R,t)\in \mathrm{SE}(3)$.

\xhdr{Local Formulation (Bond--Residue)}
For residue $i$ we define the \emph{bond--residue} as the ordered triple $(N_i-\mathrm{CA}_i), (\mathrm{CA}_i\!-\!C_i), (C_i\!-\!N_{i+1})$ together with its internal angles. For $i<N^{(\tau)}$ this includes the lengths
$\ell^{N\!-\!CA}_i,\ \ell^{CA\!-\!C}_i,\ \ell^{C\!-\!N}_i$,
the bond angles $\theta^{N\!CA\!C}_i,\ \theta^{CA\!C\!N}_i$, and the peptide dihedral $\psi_i$ about {$CA_i\!-\!C_{i}$}. For $i=N^{(\tau)}$, it includes only bond lengths $\ell^{N\!-\!CA}_{N^{(\tau)}}$, $\ell^{CA\!-\!C}_{N^{(\tau)}},$ {and angle $\theta^{N\!CA\!C}_{N^{(\tau)}}$ (the $(C\!-\!N)$ bond, $\theta^{CA\!C\!N}_{N^{(\tau)}}$, $\theta^{C\!NCA}_{N^{(\tau)}}$, and ${\{\psi,\omega,\phi\}}_{{N^{(\tau)}}}$ dihedrals are absent)}.

\xhdr{Glue Parameters Between Neighboring Bond--Residue}
Neighboring bond--residues $i$ and $i{+}1$ are connected by a set of \emph{glue} angles that place the bonds of residue $i{+}1$ relative to residue $i$. These are $\Gamma_i \;=\; \Big\{ \theta^{C\!N\!CA}_i,\ \ {\phi_{i}},\ \ \omega_i \Big\}$, i.e., one bond angle $\theta^{C\!N\!CA}_i$ (to place $N_{i+1}$--$\mathrm{CA}_{i+1}$) and two dihedrals ${\phi_{i}}$ and $\omega_i$ (to orient $\mathrm{CA}_{i+1}\!-\!C_{i+1}$ and the peptide plane). We adopt $(\omega,\phi)$ here to emphasize the two independent dihedral DOFs spanning the peptide and $\mathrm{CA}$ torsions.

\xhdr{Motif Formulation}
A \emph{bond--residue motif} $\mathcal{M}_{p:q}$ is a contiguous block of bond--residues $i=p,\dots,q$ $(1\le p\le q\le N^{(\tau)})$. {Its internal parameter set is the union over the internal bond} {lengths and angles of bond-residues} $p,\ldots,q$
together with the internal glue angles $\{\Gamma_i\}_{i=p}^{q-1}$ that connect consecutive bond--residues inside the motif. {Given $q\le r\le N^{(\tau)}$, we obtain a} {\emph{Geo-Pair occurrence} ($\mathcal{M}_{p:q},\Gamma_q,\mathcal{M}_{q:r}$) from} the internal parameters of $\mathcal{M}_{p:q}$ and $\mathcal{M}_{q+1:r}$, plus the \emph{external} glue angles $\Gamma_q$ connecting the last and first bond-residues of $\mathcal{M}_{p:q}$ and $\mathcal{M}_{q+1:r}$.

\xhdr{Entry/Exit Frames}
For residue $i$, define $F_i=(R_i,t_i)\in\mathrm{SE}(3)$ with origin $t_i=\mathrm{CA}_i$ and axes chosen so that the x-axis points from $\mathrm{CA}_i$ toward the $\mathrm{C}_i$, the $y$-axis is the normalized component of the $\mathrm{CA}_i-\mathrm{N}_i$ direction orthogonal to $x$, and the $z$-axis completes a right-handed triad.


\xhdr{Per-Link Transform}
Define the transform between consecutive residue frames $G_i \ \coloneqq\ F_{i+1}\,F_i^{-1}\ \in\ \mathrm{SE}(3).$
By construction, $G_i$ is a deterministic function of the internal coordinates local to the link $i\!\to\!i{+}1$, namely
$G_i \;=\; g\!\big(\ell^{CA\!-\!C}_i,\ \ell^{C\!-\!N}_i,\ \ell^{N\!-\!CA}_{i+1},\ 
\theta^{N\!CA\!C}_i,\ \theta^{CA\!C\!N}_i,\ \theta^{C\!N\!CA}_i,\ 
\psi_i,\ \omega_i,\ {\phi_{i}}\big)$,
and, in particular, depends on the \emph{glue set} $\Gamma_i=\{\theta^{C\!N\!CA}_i,\ {\phi_{i}},\ \omega_i\}$, {illustrated in Fig. \ref{fig:notation} (bottom).}

\xhdr{Entry/Exit Transforms}
For a motif $\mathcal{M}_{p:q}$, define $F^{\mathrm{entry}}_{p:q} \coloneqq F_p,
F^{\mathrm{exit}}_{p:q} \coloneqq F_q.$ The \emph{internal} entry$\to$exit transform is
$T^{\mathrm{int}}_{p:q} \;=\; F^{\mathrm{exit}}_{p:q}\,(F^{\mathrm{entry}}_{p:q})^{-1}
\;=\; (G_{q-1})\cdots (G_{p}),$ which depends only on the internal coordinates of $\mathcal{M}_{p:q}$. The \emph{external glue} transform between consecutive motifs $\mathcal{M}_{p:q}$ and $\mathcal{M}_{q+1:r}$ is precisely the boundary link $T^{\mathrm{glue}}_{q\to q+1} \;=\; F_{q+1}\,F_q^{-1} \;=\; G_q,$ and is parameterized by the glue set $\Gamma_q$ (and the adjacent three bond lengths).

\begin{figure}[h!]
    \centering
    
    \caption{(Top) GeoBPE tracks a Geo-Pair Encoding, a dictionary mapping Geo-Pair keys to occurrences at all times. Each step pops the most frequent Geo-Pair key, gathers the occurrences {(\roundrectsymbol{blue}{c1}{0.3em}{0.6em}{0.0ex},\roundrectsymbol{blue}{c2}{0.3em}{0.6em}{0.0ex},\roundrectsymbol{blue}{c3}{0.3em}{0.6em}{0.0ex},\roundrectsymbol{blue}{c4}{0.3em}{0.6em}{0.0ex},\roundrectsymbol{blue}{c5}{0.3em}{0.6em}{0.0ex},...)} and fixes $K$ prototypes {(\rectsymbol{blue}{c1}{0.3em}{0.6em}{0.0ex},\rectsymbol{blue}{c3}{0.3em}{0.6em}{0.0ex},\rectsymbol{blue}{c5}{0.3em}{0.6em}{0.0ex})} to add to $\mathcal{V}$. All occurrences are quantized to the closest prototype {(e.g. \roundrectsymbol{blue}{c2}{0.3em}{0.6em}{0.0ex} \textcolor{blue}{$\rightarrow$} \rectsymbol{blue}{c1}{0.3em}{0.6em}{0.0ex})}. Glue angles {(\arcsymbol{-0.5ex}{0}{180}{0.5em}{red}{1.0pt},\arcsymbol{-1.9ex}{90}{270}{0.4em}{orange}{0.6pt})} are optimized to correct for the drift introduced. (Bottom) Toy example with two backbones; we initialize residue-orientation modes using two prototypes {(\rectsymbol{green}{white}{0.3em}{0.6em}{0.0ex},\rectsymbol{mydarkyellow}{white}{0.2em}{0.7em}{0.0ex})}, pop the frequent Geo-Pair {(\rectsymbol{blue}{white}{1.0em}{0.3em}{-0.3ex})}, quantize occurrences {(\roundrectsymbol{blue}{white}{1.0em}{0.3em}{-0.3ex}\textcolor{blue}{$\rightarrow$} \rectsymbol{blue}{blue}{1.0em}{0.3em}{-0.3ex})}, and optimize glue angles.}
    \fcolorbox{black!20}{white}{\includegraphics[width=\linewidth]{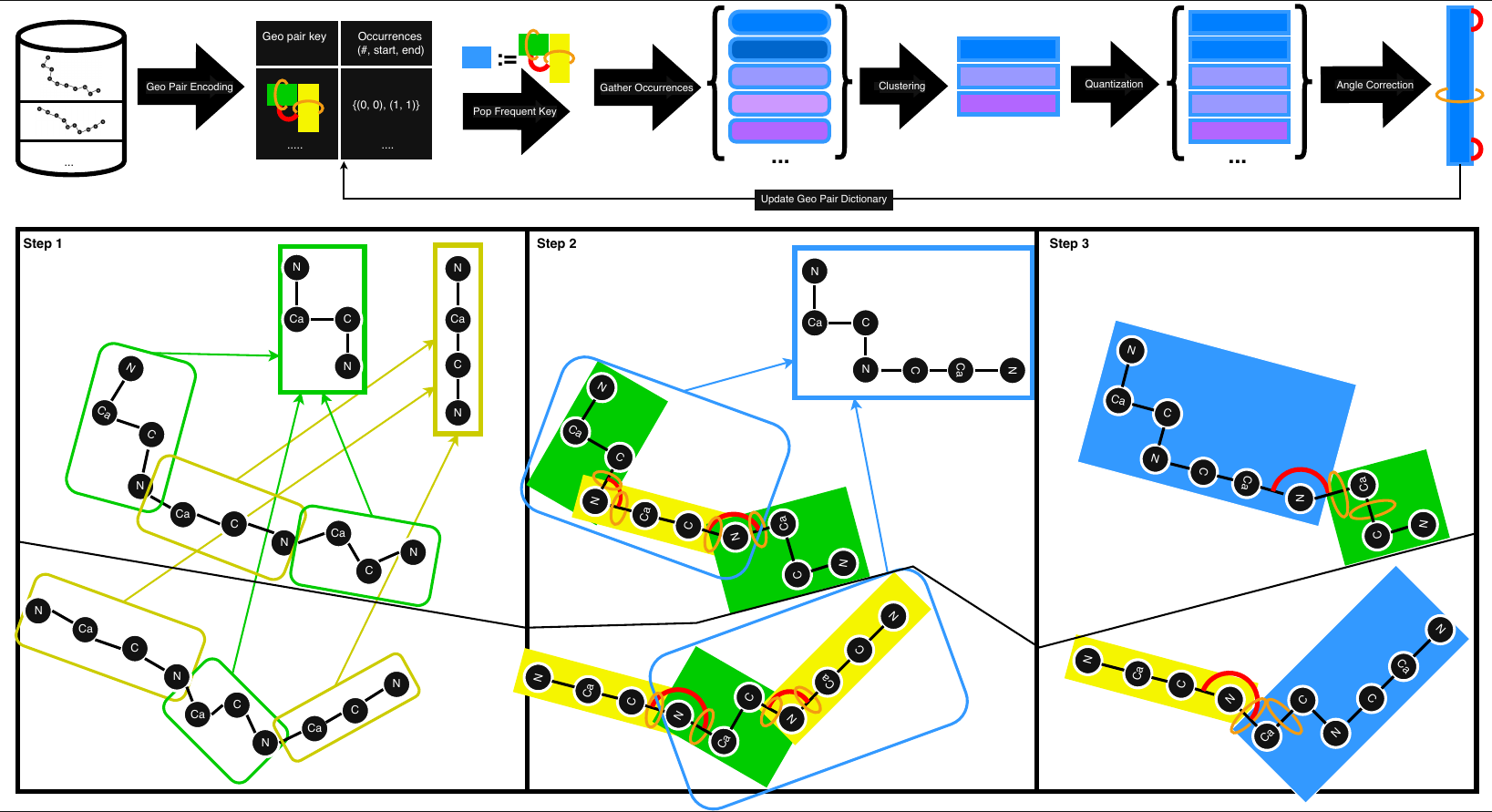}}
    \label{fig:geobpe-step}
    \vspace{-7mm}
\end{figure}
{
\xhdr{Core GeoBPE notation}
We define the main objects used throughout the algorithmic description.
We index training backbones by $\tau=1,\dots,T$, writing the $\tau$-th backbone as
$t^{(\tau)}$ with $N^{(\tau)}$ residues. A \emph{segmentation} of $t^{(\tau)}$ into
bond--residue motifs is the ordered tuple $\mathcal P^{(\tau)}=\big(\mathcal M^{(t_\tau)}_{p_1{:}q_1},\ldots,\mathcal M^{(t_\tau)}_{p_{M_\tau}{:}q_{M_\tau}}\big)$,
with $1=p_1\le q_1 < p_2\le \cdots \le q_{M_\tau}=N^{(\tau)}$. The corresponding \emph{merge hierarchy} $\mathcal F^{(\tau)}$ is a binary forest whose frontier leaves, in order, equal $\mathcal P^{(\tau)}$; each internal node represents a merged motif and stores its span $[p{:}q]$.

\paragraph{Geo-pair keys and occurrences.}
Given two adjacent motifs $(\mathcal M_{p{:}q},\mathcal M_{q+1{:}r})$
and their boundary glue $\Gamma_q$, we define a canonical, hashable
\emph{geo-pair key} $\kappa = \textsc{ComputeGeoKey}(\mathcal M_{p{:}q},\mathcal M_{q+1{:}r})$ (Alg. \ref{alg:compute-geo-key}). For each key $\kappa$ we collect its \emph{occurrence set}
$\mathcal O(\kappa)$ consisting of all such adjacent motif pairs across the dataset.

\paragraph{Prototypes and vocabulary.}
For a geo-pair key $\kappa$, \algo clusters its occurrences and stores a small set
of representative prototypes
$\mathcal A_\kappa = \{\Pi^{(\kappa)}_j\}_{j=1}^{K_{|\kappa|}}$,
where each $\Pi^{(\kappa)}_j$ is the internal-parameter tuple of a medoid occurrence and $K : \mathbb{Z}^+\setminus \{1\}\mapsto \mathbb{Z}^+$ controls how many prototypes by motif (bond) length. The \emph{vocabulary} is the map $\mathcal V:\ \kappa \mapsto \mathcal A_\kappa$, initially containing residue-level codebooks $\mathcal A_3,\mathcal A_2$ and growing as new geo-pair keys are introduced.

\paragraph{Geo-pair dictionary and priorities.}
At any time \algo maintains a \emph{priority-ordered} dictionary $\mathcal D$ that maps
each key $\kappa$ to its occurrence set $\mathcal O(\kappa)$.
Keys are ordered by tuples
\[
\pi(\kappa) = \big(\rho(\kappa), -|\mathcal O(\kappa)|, \kappa\big),
\quad
\rho(\kappa)=\mathbf{1}[\kappa\notin\mathrm{dom}(\mathcal V)],
\]
so that compressible keys with existing prototypes ($\rho=0$) are popped first, followed by a new key with the largest count $|\mathcal O(\kappa)|$ per iteration.
}


\vspace{-3mm}
\subsection{\algo Algorithm}\label{sec:algorithm}
\algo (Algo.~\ref{alg:GeoBPE}, {Fig.~\ref{fig:geobpe-step}}) is organized around four components: (1) clustering motif {(individual} {bond-residues once at the start, Geo-Pairs every step thereafter)} occurrences into representative structural {prototypes}, (2) maintaining an ordered map to track frequent Geo-Pairs, (3) adaptively {hard-}quantizing {noisy Geo-Pairs to their assigned prototypes}, and (4) applying rigid-body refinement to enforce global geometric consistency.

{Components (1)-(4) are designed to answer three new key questions when re-interpreting BPE to work with continuous backbone geometry rather than discrete bytes: (a) how do we \textit{ground} continuous backbone states to discrete keys for Geo-Pair \textit{counting}, (b) how do we \textit{update} the backbone states once we have popped the most frequent Geo-Pair, (c) how do we synchronize the Geo-Pair dictionary with how we updated the backbone states. Component (1) answers to (a), (3) \& (4) to (b), and (2) to (c). The guiding question encompassing (a)-(c) is: \textbf{what is the exact relationship between the Geo-Pair dictionary} (needed for discrete BPE operations) \textbf{and the continuous backbone states} (which should both reflect new keys and preserve original fidelity)? \algo implements a two-way connection through four stages: grounding, in which continuous motif states define discrete prototype keys; quantization, in which internal parameters are overwritten with those of the assigned prototypes; rigid-body refinement, in which backbone internal states self-correct to minimize global distortion; and synchronization, in which the Geo-Pair dictionary is re-synchronized to reflect the high-fidelity backbone states after quantization and refinement.}

\xhdr{(1) Extracting Dominant Modes from a Set of Motif Occurrences}
The core subroutine invoked by \algo is Algo.~\ref{alg:rmsd-partition}, which clusters a set of length $L$ raw backbone fragments into $K$ representative prototypes. This induces a hard quantization of the fragment space, since every possible occurrence is assigned to exactly one prototype. Because RMSD defines a metric over fragments, the clustering yields a Voronoi partition of this space. Importantly, the medoids are themselves observed fragments, so each quantized symbol retains a concrete structural interpretation: it represents the closest empirically observed conformation, providing a denoised approximation of local variability. Each time we quantize, we substitute every non‑medoid occurrence by its assigned medoid, replacing all internal {parameters} with the medoid’s internal {parameters}, making it an exact copy of that medoid (same length and per‑position angles). 

\xhdr{(2) Constructing a Structural Motif Alphabet}
 \algo begins by quantizing all bond-residues and glue angles (Algo.~\ref{alg:res-init-tokens}) and building an \textit{ordered} map {$\mathcal{D}$}
 of discrete geo-pair grounding keys ${\kappa}$ to occurrences ${O(\kappa)}$ (Algo.~\ref{alg:bin}, {see Core GeoBPE notation}). In each iteration (Algo.~\ref{alg:step}), \algo pops the most frequent {Geo-Pair} key, runs Algo.~\ref{alg:rmsd-partition} on mapped occurrences, quantizes the occurrences, runs rigid-body refinement, and updates {$\mathcal{D}$} to account for the new quantized backbone states.

\begin{algorithm}[t]
\caption{\algo: Protein structure tokenizer with geometric byte-pair encoding}
\label{alg:GeoBPE}
\begin{algorithmic}[1]
\REQUIRE Backbones $\{t^{(1)},\dots,t^{(T)}\}$ with lengths $N^{(\tau)}$; {optional backbones $\{t^{(\xi)}\}$ to tokenize}; residue codebook sizes $(K_3,K_2)$; glue-IK weights $(w_R,w_t)$; maximum merge iterations $S_{\max}$.
\ENSURE Final vocabulary $\mathcal V$ (motif prototypes), final segmentations $\{\mathcal P^{(\tau)}\}$, final merge hierarchies $\{\mathcal F^{(\tau)}\}$, and the priority-ordered geo-pair map $\mathcal D$.

\STATE \textbf{Empirical quantizer estimation (once).}
Collect samples over all backbones for the 9 types
$\{\ell^{N\!-\!CA},\ell^{CA\!-\!C},\ell^{C\!-\!N}\}$,
$\{\theta^{N\!CA\!C},\theta^{CA\!C\!N},\theta^{C\!N\!CA}\}$,
$\{\phi,\psi,\omega\}$.
Wrap angles to $[0,2\pi)$ and build circular histograms with edges $0=\beta_0<\cdots<\beta_{B}=2\pi$ that tile the circle; define $Q$ by snapping to bin centers. For lengths, build linear histograms and snap to centers.

\STATE \textbf{Per-residue initialization} (Algo.~\ref{alg:res-init-tokens}).
Cluster interior and terminal bond–residues via \textsc{RMSD\_Partition} to obtain codebooks
$\mathcal A_3,\mathcal A_2$; overwrite each residue’s internals by its assigned prototype.
Set the initial segmentation for each backbone:
\[
\mathcal P^{(\tau)}=(\mathcal M^{(t_\tau)}_{1{:}1},\ldots,\mathcal M^{(t_\tau)}_{N^{(\tau)}{:}N^{(\tau)}}).
\]
\textbf{Initialize hierarchies:} for each $\tau$, create a binary forest $\mathcal F^{(\tau)}$ whose leaves are the bond–residue motifs $\mathcal M^{(t_\tau)}_{i{:}i}$, in order; its frontier equals $\mathcal P^{(\tau)}$.
Initialize the vocabulary with base prototypes:
\[
\mathcal V \leftarrow \{\text{residue-level keys}\mapsto \mathcal A_3,\mathcal A_2\}.
\]

\STATE \textbf{Global glue refinement} (Algo.~\ref{alg:glue-opt-all}).
Optimize all boundary glues $\Gamma_i=\{\theta^{C\!N\!CA}_i,\omega_i,\phi_{i+1}\}$ via differentiable FK with $(w_R,w_t)$; snap each to the nearest bin center using $Q_{\theta^{C\!N\!CA}},Q_{\omega},Q_{\phi}$.

\STATE \textbf{Build the priority-ordered geo-pair map} (Algo.~\ref{alg:bin}).
Using the frontier leaves of each $\mathcal F^{(\tau)}$ (equivalently, $\mathcal P^{(\tau)}$), construct the occurrence sets $\mathcal O(\kappa)$ and insert:
\[
\mathcal D\big[(\rho(\kappa),-|\mathcal O(\kappa)|,\kappa)\big]\leftarrow \mathcal O(\kappa),
\quad \rho(\kappa)=\mathbf{1}[\kappa\notin \mathrm{dom}(\mathcal V)].
\]

\STATE \textbf{BPE loop -- calls (Algo.~\ref{alg:step}) each step.}
\FOR{$s=1$ \TO $S_{\max}$}
  \STATE $(\{\mathcal P^{(\tau)}\},\ \{\mathcal F^{(\tau)}\},\ \mathcal D,\ \mathcal V)\leftarrow
  \textsc{Step}\big(\{\mathcal P^{(\tau)}\},\ \{\mathcal F^{(\tau)}\},\ \mathcal D,\ \mathcal V,\ \{Q_{\theta^{C\!N\!CA}},Q_{\omega},Q_{\phi}\},\ (w_R,w_t)\big)$
\ENDFOR
\STATE {\textbf{Tokenize new/unseen backbones (Algo.~\ref{alg:tokenize})} for each $\xi$,} \[{(\mathcal P^{(\xi)}, \mathcal F^{(\xi)}) \leftarrow \textsc{Tokenize}\big(t^{(\xi)},\mathcal A_3,\mathcal A_2,\mathcal{V},\{Q_{\theta^{C\!N\!CA}},Q_{\omega},Q_{\phi}\}\big)}\]
\STATE \textbf{return} $\mathcal V$, $\{\mathcal P^{(\tau)}\}$, $\{\mathcal F^{(\tau)}\}$, $\mathcal D$ and {(if given) $\{\mathcal P^{(\xi)}\}$, $\{\mathcal F^{(\xi)}\}$}
\end{algorithmic}
\end{algorithm}

\xhdr{(3) Multi-Resolution \& Adaptive (Re-)Quantization}
One-time quantization is a lossy procedure and is only needed to index Geo-Pairs occurrences in the current step. Thus, each \algo iteration can re-quantize occurrences by referencing the original, even if prior iterations have quantized the same regions already. This allows resolution to adapt based on the size of the motif (e.g., coarse-grained for smaller motifs, fine-grained for larger ones), providing precise control over compression-reconstruction tradeoffs (see App. \ref{app:GeoBPE-ablation}).

\xhdr{(4) Minimizing Distortion via Rigid-Body Refinement} Let $T^{\mathrm{int}}_{i{:}j}$ denote the entry$\to$exit SE(3) map of a motif $\mathcal M_{i{:}j}$ determined by its internal coordinates. For an occurrence $u$ with original motif $\mathcal M^{(t_u)}_{i_u{:}k_u}$, the rounding step replaces it by its assigned medoid segment:
\[
\mathcal M^{(t_u)}_{i_u{:}k_u}
\;\longrightarrow\;
\mathcal M^{(t_{\widehat m_{c(u)}})}_{\,i_{\widehat m_{c(u)}}{:}k_{\widehat m_{c(u)}}},
\]
where $\widehat m_{c(u)}$ is the medoid index returned by \textsc{RMSD\_Partition} (an index into $\mathcal S$).
Let $T^{\mathrm{occ}}_u \;\coloneqq\; T^{\mathrm{int}}_{i_u{:}k_u}$ and $T^{\mathrm{med}}_u \;\coloneqq\; T^{\mathrm{int}}_{\,i_{\widehat m_{c(u)}}{:}k_{\widehat m_{c(u)}}}$.
Rounding thus replaces $T^{\mathrm{occ}}_u$ by $T^{\mathrm{med}}_u$, and the induced discrepancy $\Delta T_u \;\coloneqq\; T^{\mathrm{occ}}_u \big(T^{\mathrm{med}}_u\big)^{-1}$ is the \emph{drift} introduced by quantization. If left uncompensated, products of such $\Delta T_u$ across a chain accumulate and move exit frames off their original targets. Each boundary provides 3 \emph{gluing} degrees of freedom ($\Gamma_i$) that can absorb this drift. To exactly recover the original exit (in the idealized case), the boundary transform at the link $i_u\!-\!1\to i_u$ should satisfy:
\[
\overbrace{G^{\mathrm{new}}_{i_u-1}}^{\text{opt vars}}\; T^{\mathrm{med}}_{u}
\;\approx\;
G^{\mathrm{orig}}_{i_u-1}\; T^{\mathrm{occ}}_{u}
\qquad\Longrightarrow\qquad
G^{\mathrm{new}}_{i_u-1}\;\approx\;G^{\mathrm{orig}}_{i_u-1}\,\Delta T_{u},
\]
where the quantization drift is $\Delta T_{u}\;\coloneqq\;T^{\mathrm{occ}}_{u}\,\big(T^{\mathrm{med}}_{u}\big)^{-1}.
$ Since $G_{i_u-1}$ is controlled by only three gluing DOFs, we solve for $G^{\mathrm{new}}_{i_u-1}$ in least squares via the end-frame fitting objective:
\[
\mathcal{L}_u(\Gamma_{i_u-1}) \;=\;
w_R\,\big\|\log\!\big((\widehat R_{k_u})^\top R^\star_{k_u}\big)\big\|_2^2
\;+\;
w_t\,\big\|\widehat t_{k_u}-t^\star_{k_u}\big\|_2^2,
\]
with forward kinematics
$\widehat F_{k_u}
\;=\;
F^\star_{i_u-1}\,G^{\mathrm{new}}_{i_u-1}\,T^{\mathrm{med}}_{u},
\qquad
F^\star_{k_u}
\;=\;
F^\star_{i_u-1}\,G^{\mathrm{orig}}_{i_u-1}\,T^{\mathrm{occ}}_{u}.$ When quantizing many motifs on the same backbone, performing this optimization each time can become computationally prohibitive. Instead, we adopt a global (batch) alternative which treats all gluing DOFs as parameters, with a global end-frame fitting loss. This provides maximum flexibility in drift compensation. The algorithmic details are in Algo.~\ref{alg:glue-opt} and \ref{alg:glue-opt-all}.

\xhdr{{Transferring Hierarchical Inductive Biases}}
\algo adapts the receptive field of a base {residue}-{level} feature extractor $\Theta$ to that of the whole structure, connecting {multiple scales} through recursive aggregation. Algo.~\ref{alg:GeoBPE} emits merge hierarchies $\mathcal {F}$ as a forest: leaf nodes represent residues and parent nodes represent motifs. {The key insight is to use ${F}$ as a recursive computation tree.} Leaf nodes are initialized with pretrained features,
then embeddings propagate up along the parent-child relations of $\mathcal{F}$ until the forest roots {(aligned with $\mathcal{P}$)}; a final step aggregates the forest roots into a protein-level contextual embedding; then they are propagated down until the leaf nodes. {The final leaf nodes output residue-level embeddings induced by the hierarchical $\mathcal{V}$ and informed by} {multi-scale \algo tokenization. These features support supervised learning on fine-grained} {residue-level tasks} (e.g., active site prediction) {and coarse-grained global predictions} (e.g., fold classification). See Algo.~\ref{alg:updown-encoder} for details.

\subsection{Principles of Protein Structure Tokenization}\label{sec:pst}
Let $\mathcal{X}=(\mathbb{R}^{3\times3})^*$ be the space of backbone coordinate tensors and let $\mathcal{V}$ be a finite codebook. A tokenizer is a tuple $\mathsf{T}=(\mathcal{V},\mathrm{Enc},\mathrm{Dec})$: $\mathrm{Enc}:\mathcal{X}\to\mathcal{V}^\ast$ mapping a structure $\mathbf{x}$ to a finite token sequence $\mathbf{q}=\mathrm{Enc}(\mathbf{x})$, $\mathrm{Dec}:\mathrm{Im}(\mathrm{Enc})\to\mathcal{X}$ mapping $\tilde{\mathbf{x}}=\mathrm{Dec}(\mathrm{Enc}(\mathbf{x}))$.
For dataset $\mathcal{D}\subset\mathcal{X}$ and distortion $d:\mathcal{X}\times\mathcal{X}\to[0,\infty)$ (e.g., Kabsch-aligned RMSD per residue), define:
\[
\begin{aligned}
\Delta(\mathsf{T};\mathcal{D})
&=\frac{1}{|\mathcal{D}|}\sum_{\mathbf{x}\in\mathcal{D}} d\!\left(\mathbf{x},\,\mathrm{Dec}(\mathrm{Enc}(\mathbf{x}))\right),\qquad
\operatorname{BPR}(\mathsf{T};\mathcal{D})
=\frac{\mathrm{L}(\mathsf{T})+\sum_{\mathbf{x}\in\mathcal{D}}\mathrm{L}(\mathrm{Enc}(\mathbf{x}))}{\sum_{\mathbf{x}\in\mathcal{D}} N(\mathbf{x})}\ \text{bits/res}
\end{aligned}
\]
where $\mathrm{L}(\mathsf{T})\!\ge\!0$ is the description length of $(\mathcal{V},\mathrm{Enc},\mathrm{Dec})$ and $N(\mathbf{x})$ is the residue count; under a uniform per-token code, $\mathrm{L}(\mathrm{Enc}(\mathbf{x}))=|\mathrm{Enc}(\mathbf{x})|\log_2|\mathcal{V}|$. We setup the following principles for an ideal tokenizer $\widehat{\mathsf{T}}$ and empirically explore the degree \algo satisfies them.

\xhdr{Principle 1: Pareto-optimal on $\mathcal{D}$} $\widehat{\mathsf{T}}$ is Pareto-optimal on $\mathcal{D}$ iff no $\mathsf{T}'$ satisfies $\operatorname{BPR}(\mathsf{T}';\mathcal{D})\le \operatorname{BPR}(\widehat{\mathsf{T}};\mathcal{D})$ and $\Delta(\mathsf{T}';\mathcal{D})\le \Delta(\widehat{\mathsf{T}};\mathcal{D})$, with at least one strict. We empirically explore this principle by evaluating Pareto-efficiency among leading PSTs and codebook configurations in Fig. \ref{fig:pareto}.

\xhdr{Priciple 2: Out-of-distribution (OOD) generalization} $\widehat{\mathsf{T}}$ generalizes OOD if, on unseen test set $\mathcal{D}_{\mathrm{test}}\subset\mathcal{X}, \Delta(\mathsf{T};\mathcal{D}_{\mathrm{test}}) \approx \Delta(\mathsf{T};\mathcal{D})$. We depict generalization gaps of leading PSTs in Fig. \ref{fig:pareto}.


\xhdr{Principle 3: Downstream transfer via codebook/vocabulary} Let $\mathcal{V}$ be the vocabulary of $\mathsf{T}$ and let $N(\mathbf{x})$ be the residue count. Let $\Theta$ parameterize a pretrained feature extractor. The codebook/vocabulary $\mathcal{V}$ \textit{induces} per-residue features $r_{\mathcal{V}}(\mathbf{x})=\Psi_{\mathcal{V}}(F_{\Theta}(\mathbf{x}))\in(\mathbb{R}^d)^{N(\mathbf{x})}$. An ideal tokenizer of protein \textit{structures} should go beyond pure compression; it should learn useful signals related to function. We loosely define the ability of a PST to transfer useful signals by test performance on a battery of downstream tasks when parameterizing samples $\mathbf{x}$ by the vocabulary $\mathcal{V}$ together with a feature extractor $\Theta$. We benchmark downstream transfer of \algo against others in Table \ref{tab:transfer} (\model).
\vspace{-3mm}

\section{Experiments}\label{sec:experiments}
We answer ten research questions (Q1-Q10) to benchmark the performance, efficiency, and application integration potential of \algo against other popular tokenizers.
\begin{itemize}[leftmargin=*,topsep=0pt,noitemsep]
    \item \textbf{Tokenization Performance}: (Q1) How many bits are needed to store the tokenizer and tokenized inputs? (Q2) How faithful is the reconstruction? (Q3) How does performance generalize to unseen data? (Q4) How many samples are needed to train the tokenizer?
    \item \textbf{Token Efficiency}: (Q5) How frequent and balanced is vocabulary utilization? (Q6) Does small-scale {language modeling generate better structures with \algo or VQ-VAE tokens?}
    \item \textbf{Downstream Transfer}: (Q7) How much transferrable signal does the tokenizer capture about the data? (Q8) How much does the vocabulary help on representation learning tasks?
    \item \textbf{Interpretability}: (Q9) How well do \algo tokens agree with ``ground-truth" domain annotations? (Q10) Can experts \textit{understand} \algo through real-world case studies?
\end{itemize}
\xhdr{Datasets} We follow the same dataset splits as in \citet{yuan2025protein}. Pretraining uses structures from the Protein Data Bank following OpenFold2’s protocol and retained a non-redundant subset of $\approx$48K protein chains, which were split into training/validation sets, with CAMEO and CASP14 reserved as held-out test sets for evaluating OOD generalization and token efficiency. For downstream evaluation, we use 8 datasets, spanning residue-level classification (ligand binding, catalytic, conserved, repeat, and epitope), residue-level regression (structural flexibility prediction), and protein-level classification. Together these datasets probe functional relevance, structural variability, token distinctiveness, and efficiency across a wide range of proteins. For citations and details, see App. \ref{app:dataset}.

\xhdr{Baselines} We compare with VQ-VAEs, the leading family of discrete PSTs \citep{hayes2025simulating, van2024fast, lin2023tokenizing, yuan2025protein}. They consist of
(1) a structure encoder maps structure $\bm{x}$ into a continuous representation 
$z \in \mathbb{R}^{N \times D}$;  (2) a vector quantization layer discretizes each $z_i$ by selecting $k_i = \arg\min_j d(z_i, q_j)$ from a learnable codebook $Q \in \mathbb{R}^{K \times D}$; and (3) a structure decoder reconstructs $\tilde{\bm{x}}\approx \textbf{x} $ from the discrete codes $\bm{q}_k = \{\bm{q}_{k_j}\}_{j=1}^L$. We also compare with Inverse Folding (IF) \textit{continuous} PSTs, which skips the quantization step $\textbf{z}\rightarrow \bm{q}_{k}$ and trained to recover the amino acid sequence from $\bm{z}$ \citep{dauparas2022robust, yang2023masked}.


\xhdr{Downstream Transfer} For VQ-VAEs, $\Theta$ and $\mathcal{V}$ are \textit{jointly} learned, so we set $r^{\text{VQ-VAE}}_\mathcal{V}(\textbf{x}):= r^{\text{VQ-VAE}}(\textbf{x}) \leftarrow \text{Enc}(\textbf{x})$. For \model, we use $\Theta \leftarrow \text{ESM3}$ to demonstrate how $\mathcal{V}^{\text{\algo}}$ can transfer useful signals from $\mathcal{F}_{\Theta}(\textbf{x})$ to $r_{\mathcal{V}}(\mathcal{F}_{\Theta}(\textbf{x}))$. 

\xhdr{Performance Metrics} \textit{Compression} measures \uline{Bits-Per-Residue (BPR)}, as defined in Sec.~\ref{sec:pst}. \textit{Distortion} ($\Delta$) use standard \uline{RMSD} and \uline{LDDT}. \textit{Token Efficiency} uses \uline{Codebook Utility Rate (UR)}, \uline{Perplexity} (details in App.~\ref{app:token-efficiency}) and \textit{Small Structure Language Model Evaluation (SSLM-Eval)} (details in App.~\ref{app:lm-eval}). SSLM-Eval {compares} {tokenizers (\algo vs VQ-VAEs) via integration} with a small $\sim$7.3M Transformer architecture after {respectively tokenizing the pretraining data splits} (Algo.~\ref{alg:geom-lm}). {Under the same data, model, training and compute resources}, the {respective} models generate new sentences, detokenizes them into structures, and we {compare \textit{relative} generation metrics} (Algo.~\ref{alg:length-start-priors}, \ref{alg:uncond-generate}, \ref{alg:dequantize-assemble}). \textit{Downstream Transfer} covers 12 tasks (24 test splits) using \uline{AUROC} (\%) for functional site prediction, \uline{Spearman's} $\rho$ (\%) for flexibility prediction, and \uline{Macro F1} (\%) for fold classification. \textit{Expert Agreement} measures Domain \& Segment \uline{Recall/Precision/F1/IOU} (details in App.~\ref{app:expert-agreement-eval}).

\xhdr{Computational Complexity / Implementation Details} We analyze the theoretical complexity of \algo in App.~\ref{app:computational-complexity} and justify the steps we took towards efficient implementation and use.
\vspace{-3mm}

\section{Results}
\begin{figure}
\setlength{\fboxsep}{0pt}
    {%
      \begin{minipage}{0.99\linewidth} 
    \includegraphics[width=\linewidth]{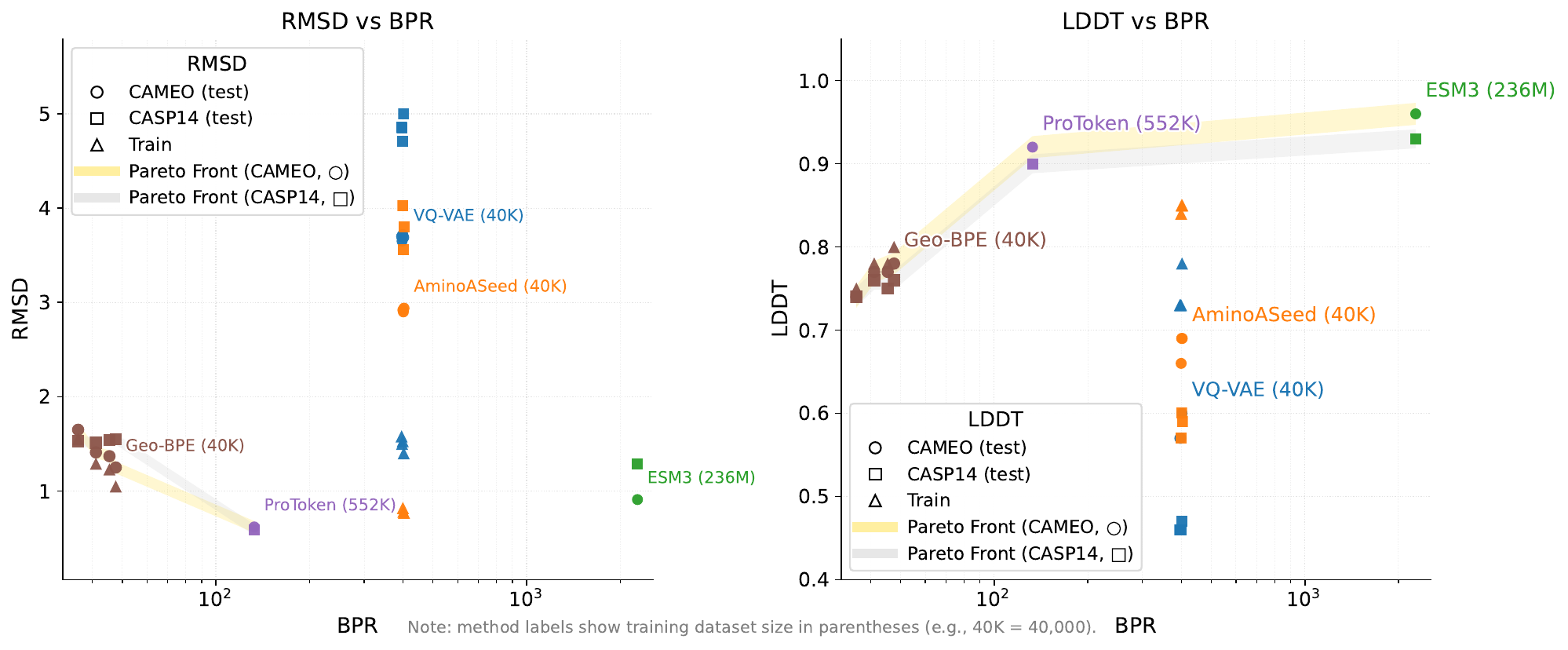}
          \end{minipage}%
    }
    \caption{Plots of $(\operatorname{BPR}(\mathsf{T};\mathcal{D}),\ \Delta(\mathsf{T};\mathcal{D}_{\mathrm{test}}))$ across tokenizers for $\Delta\in \{\text{RMSD},\text{LDDT}\}$. We vary $|\mathcal{V}|\in \{128,256,512,1024\}$ for VQ-VAE/AminoASeed and $|\mathcal{V}|\in \{600,2500,6000,21000\}$ for \algo to sample multiple points; we observe \algo sweeps a smooth tradeoff curve. {Hyperparameters in App. \ref{app:hyperparameters}.}}
    \label{fig:pareto}
    \vspace{-5mm}
\end{figure}
\xhdr{Tokenizer Performance} We find \algo and ProToken form the Pareto front under both $\Delta \in \{\text{RMSD}, \text{LDDT}\}$. \algo achieves $0.271-0.358$x and $0.016-0.021$x the BPR of ProToken and ESM3, dropping LDDT by only $18-22\%$ and $22-25\%$, which are impressive feats considering \algo's training data was only $\approx 7\%$ and $0.02\%$ the size. We also observe \algo's strong OOD generalization, with test/train RMSD peaking at $1.16$ (CAMEO) and $1.28$ (CASP), showing negligible degradation reconstructing unseen data; VQ-VAE/AminoASeed, using identical data splits, show degradation as high as $6.4$x test RMSD. Crucially, as the \algo codebook grows, the variants trace a near-linear path along the Pareto front toward ProToken, elastically trading off BPR for lower distortion, a feature other tokenizers do not have (as codebook dimensions are fixed).

\xhdr{Token Efficiency} We report UR \& Perplexity averaged over held-out test sets to gauge codebook/vocabulary usage on unseen data, the setting where the tokenizer is deployed. In Table \ref{tab:token-efficiency}, we see all methods except VQ-VAE and ESM3 achieve an average UR of $>40\%$; all except VQ-VAE achieve $0.2$ average Perplexity. An ideal tokenizer avoids codebook collapse, but exactly uniform token usage may not be desirable. We introduce SSLM-Eval to stress test whether codebook efficiency actually leads to generative efficiency. SSLM-Eval is a holistic way to compare tokenizers using both encoder token efficiency \textit{and} decoder's generative efficiency. In Table \ref{tab:lm-eval}, we find \model is capable of generating $99\%$ unique and designable backbones, achieving up to $49\%$ higher scTM and maintaining higher diversity than both VQ-VAE methods using the same data splits. We visualize some realistic, novel backbones \model generated in App. \ref{sec:gen-quality}. 
Interestingly, the ``less-efficient" VQ-VAE generated $58\%$ more diverse backbones, demonstrating uniform token usage can be counterproductive to language modeling.

\xhdr{Downstream Task Transfer}
In Table \ref{tab:transfer}, we see \algo-induced features rank first, on average, across both function and structure property prediction tasks. The relative performance gaps $15.44\%$ and $43.28\%$ quantify the add-on benefits of \algo-induced features. \algo-induced features reverse the trend that discrete PSTs produce less informative representations for downstream tasks (due to quantization-related issues \citep{yuan2025protein}), highlighting that \textit{hierarchical structure} from discrete vocabularies raises the ceiling on downstream transfer. 

{\xhdr{Further Ablations}
We include a comprehensive series of ablation studies in App.~\ref{app:GeoBPE-ablation} demonstrating \algo's \textit{data-efficiency}, \model's \textit{task-agnosticism}, \algo tokenizer's \textit{scalability}, \algo tokens' \textit{adaptive resolution} over iterations, and \textit{performance vs runtime tradeoffs} in components (1), (3) \& (4). Key findings include: (i) \algo shows \textit{better} OOD generalization when fitted on 1\% training data; (ii) \model predictions are no worse when \algo was fitted with (a) 1\% of the pretraining PDBs, (b) the downstream task-specific PDBs; (iii) \algo performance gains diminish beyond $M_{\max}=5000$ randomly sampled motif occurrences used to extract prototypes, taming a complexity term that depends on $M_{\max}$.}
\begin{table}[h!]
\caption{Downstream transfer performance benchmark.  We \underline{underline} and \textbf{bold} the best continuous and discrete PSTs, respectively; 
\grayswatch\ indicates the best method across both. The relative performance v.s. ESM3 for \model is included. Omitted rows in Table \ref{tab:transfer-additional}; {\algo hyperparameters are in App. \ref{app:hyperparameters}.}}
\label{tab:transfer}
\resizebox{\textwidth}{!}{
\begin{tabular}{@{}llcccccccc@{}}
\toprule
\multicolumn{1}{l|}{\textbf{Task}}               & \multicolumn{1}{l|}{\textbf{Split}} & \multicolumn{2}{c|}{\textbf{Continuous PST}}                                                   & \multicolumn{6}{c}{\textbf{Discrete PST}}                                                                                                                                                                               \\ \midrule
\multicolumn{1}{l|}{}                            & \multicolumn{1}{l|}{}               & ProteinMPNN                         & \multicolumn{1}{c|}{MIF}                                 & FoldSeek & ProTokens & ESM3                                   & VanillaVQ      & \multicolumn{1}{c|}{AminoAseed}                             & \model (v.s. ESM3)                                                      \\ \midrule
\multicolumn{10}{c}{\textbf{Functional Site Prediction (AUROC\%)}}                                                                                                                                                                                                                                                                                                                                                \\ \midrule
\multicolumn{1}{l|}{BindInt}                     & \multicolumn{1}{l|}{Fold}           & {\ul 51.83}                         & \multicolumn{1}{c|}{50.38}                               & 53.18    & 44.66     & 44.30                                  & 47.25          & \multicolumn{1}{c|}{47.11}                                  & \cellcolor[HTML]{C0C0C0}\textbf{59.19 (+33.61\%)}                        \\
\multicolumn{1}{l|}{}                            & \multicolumn{1}{l|}{SupFam}         & 94.00                               & \multicolumn{1}{c|}{\cellcolor[HTML]{C0C0C0}{\ul 94.56}} & 46.26    & 86.05     & 90.77                                  & 86.71          & \multicolumn{1}{c|}{90.53}                                  & \textbf{91.31 (+0.59\%)}                                                 \\
\multicolumn{1}{l|}{BindBio}                     & \multicolumn{1}{l|}{Fold}           & 78.42                               & \multicolumn{1}{c|}{{\ul 85.79}}                         & 32.37    & 58.47     & 62.84                                  & 62.02          & \multicolumn{1}{c|}{65.73}                                  & \cellcolor[HTML]{C0C0C0}\textbf{94.94 (+51.08\%)}                        \\
\multicolumn{1}{l|}{}                            & \multicolumn{1}{l|}{SupFam}         & 81.00                               & \multicolumn{1}{c|}{{\ul 87.27}}                         & 52.44    & 60.47     & 65.22                                  & 62.92          & \multicolumn{1}{c|}{68.30}                                  & \cellcolor[HTML]{C0C0C0}\textbf{95.94 (+47.10\%)}                        \\
\multicolumn{1}{l|}{BindShake}                   & \multicolumn{1}{l|}{Org}            & 75.52                               & \multicolumn{1}{c|}{{\ul 79.90}}                         & 53.43    & 59.10     & 66.10                                  & 67.04          & \multicolumn{1}{c|}{69.61}                                  & \cellcolor[HTML]{C0C0C0}\textbf{87.73 (+32.72\%)}                        \\
\multicolumn{1}{l|}{CatInt}                      & \multicolumn{1}{l|}{Fold}           & {\ul 61.05}                         & \multicolumn{1}{c|}{59.62}                               & 53.43    & 58.16     & 61.09                                  & 58.89          & \multicolumn{1}{c|}{62.19}                                  & \cellcolor[HTML]{C0C0C0}\textbf{66.21 (+8.38\%)}                         \\
\multicolumn{1}{l|}{}                            & \multicolumn{1}{l|}{SupFam}         & 93.40                               & \multicolumn{1}{c|}{\cellcolor[HTML]{C0C0C0}{\ul 96.49}} & 51.41    & 83.85     & 89.82                                  & 85.00          & \multicolumn{1}{c|}{\textbf{91.91}}                         & 88.65 (-1.30\%)                                                          \\
\multicolumn{1}{l|}{CatBio}                      & \multicolumn{1}{l|}{Fold}           & 82.49                               & \multicolumn{1}{c|}{{\ul 85.85}}                         & 56.33    & 67.68     & 65.33                                  & 67.58          & \multicolumn{1}{c|}{65.95}                                  & \cellcolor[HTML]{C0C0C0}\textbf{95.01 (+45.43\%)}                        \\
\multicolumn{1}{l|}{}                            & \multicolumn{1}{l|}{SupFam}         & 93.19                               & \multicolumn{1}{c|}{\cellcolor[HTML]{C0C0C0}{\ul 96.97}} & 53.78    & 64.05     & 74.65                                  & 70.92          & \multicolumn{1}{c|}{87.59}                                  & \textbf{95.90 +28.47\%}                                                  \\
\multicolumn{1}{l|}{Con}                         & \multicolumn{1}{l|}{Fold}           & 57.18                               & \multicolumn{1}{c|}{{\ul 58.43}}                         & 49.20    & 57.20     & 55.22                                  & 56.98          & \multicolumn{1}{c|}{57.23}                                  & \cellcolor[HTML]{C0C0C0}\textbf{71.96 (+30.32\%)}                        \\
\multicolumn{1}{l|}{}                            & \multicolumn{1}{l|}{SupFam}         & 84.68                               & \multicolumn{1}{c|}{\cellcolor[HTML]{C0C0C0}{\ul 92.66}} & 51.31    & 70.64     & 80.53                                  & 74.60          & \multicolumn{1}{c|}{\textbf{86.60}}                         & 84.84 (+5.35\%)                                                          \\
\multicolumn{10}{c}{\small\textit{...2 tasks omitted (Rep, Ept)...}\ }\\\midrule
\multicolumn{1}{l|}{\textbf{Average AUROC\%}}    & \multicolumn{1}{l|}{}               & 75.92                               & \multicolumn{1}{c|}{{\ul 79.82}}                         & 51.90    & 65.37     & 69.24                                  & 68.30          & \multicolumn{1}{c|}{72.43}                                  & \cellcolor[HTML]{C0C0C0}{\color[HTML]{333333} \textbf{{80.20 (+18.13\%)}}} \\ \midrule
\multicolumn{10}{c}{\textbf{Physicochemical Property Prediction (Spearman’s $\rho$\%)}}                                                                                                                                                                                                                                                                                                                           \\ \midrule
\multicolumn{1}{l|}{FlexRMSF}                    & \multicolumn{1}{l|}{Fold}           & \cellcolor[HTML]{C0C0C0}{\ul 62.37} & \multicolumn{1}{c|}{59.60}                               & 15.35    & 13.81     & 44.53                                  & 44.22          & \multicolumn{1}{c|}{\textbf{44.63}}                         & 40.89 (-8.17\%)                                                          \\
\multicolumn{1}{l|}{}                            & \multicolumn{1}{l|}{SupFam}         & \cellcolor[HTML]{C0C0C0}{\ul 59.24} & \multicolumn{1}{c|}{56.80}                               & 11.99    & 7.62      & 39.08                                  & 38.98          & \multicolumn{1}{c|}{40.99}                                  & \textbf{47.17 (20.70\%)}                                                 \\
\multicolumn{10}{c}{\small\textit{...2 tasks omitted (FlexBFactor, FlexNEQ)...}\ }\\\midrule
\multicolumn{1}{l|}{\textbf{Average $\rho$\%}}   & \multicolumn{1}{l|}{}               & \cellcolor[HTML]{C0C0C0}{\ul 54.41} & \multicolumn{1}{c|}{52.73}                               & 7.80     & 9.84      & 37.35                                  & 33.49          & \multicolumn{1}{c|}{38.08}                                  & \textbf{45.26 (+21.18\%)}                                                \\ \midrule
\multicolumn{10}{c}{\textbf{Structure Property Prediction (Macro F1\%)}}                                                                                                                                                                                                                                                                                                                                          \\ \midrule
\multicolumn{1}{l|}{Homo}                        & \multicolumn{1}{l|}{Fold}           & {\ul 25.66}                         & \multicolumn{1}{c|}{22.56}                               & 11.57    & 5.84      & \cellcolor[HTML]{C0C0C0}\textbf{30.02} & 18.17          & \multicolumn{1}{c|}{29.87}                                  & 23.60 (-21.39\%)                                                         \\
\multicolumn{1}{l|}{}                            & \multicolumn{1}{l|}{SupFam}         & 30.83                               & \multicolumn{1}{c|}{{\ul 33.86}}                         & 4.67     & 6.17      & 24.89                                  & 22.10          & \multicolumn{1}{c|}{38.38}                                  & \cellcolor[HTML]{C0C0C0}\textbf{47.28 (+89.96\%)}                        \\
\multicolumn{1}{l|}{}                            & \multicolumn{1}{l|}{Fam}            & 63.33                               & \multicolumn{1}{c|}{{\ul 74.22}}                         & 15.34    & 18.33     & 54.42                                  & 47.18          & \multicolumn{1}{c|}{69.78}                                  & \cellcolor[HTML]{C0C0C0}\textbf{85.75 (+57.47\%)}                        \\ \midrule
\multicolumn{1}{l|}{\textbf{Average Macro F1\%}} & \multicolumn{1}{l|}{}               & 39.94                               & \multicolumn{1}{c|}{{\ul 43.55}}                         & 10.51    & 10.11     & 36.44                                  & 29.15          & \multicolumn{1}{c|}{46.01}                                  & \cellcolor[HTML]{C0C0C0}\textbf{52.21 (+43.28\%)}                        \\ \bottomrule
\end{tabular}
}
\end{table}
\vspace{-3mm}

\section{Discussion}

\xhdr{Case Study: Agreement with PFAM Annotations}
\begin{table}[h!]
\caption{We annotate 100 PDBs from each dataset and report \% of 1,000 random equal-length segmentations that \algo matches or outscores. Omitted columns are in Table \ref{tab:agreement-additional}. {Secondary structure analysis in Table \ref{tab:sec-agreement-avg}}.}
\label{tab:agreement}
\resizebox{\textwidth}{!}{
\begin{tabular}{@{}llcccccc V{12mm}|c@{}}
\toprule
& & \multicolumn{1}{c}{BindInt} & \multicolumn{1}{c}{BindBio} & \multicolumn{1}{c}{BindShake} &
      \multicolumn{1}{c}{CatInt} & \multicolumn{1}{c}{CatBio} & \multicolumn{1}{c}{Con} &
      & \multicolumn{1}{c}{Average} \\
\midrule
\multicolumn{1}{l|}{Domain} & Mean Recall
  & 99.95 (98.35) & 100 (100.0) & 100 (100.0) & 100 (100.0) & 99.99 (93.55) & 99.95 (98.0)
  & \multirow{7}{*}{\centering\raisebox{0pt}[0pt][0pt]{%
  ...
      \rotatebox[origin=c]{90}{\small\textit{4 columns omitted}}%
    }
    ...
    }
  & 99.97 (97.97) \\
\multicolumn{1}{l|}{} & Mean Precision
  & 98.9 (53.87) & 99.62 (71.92) & 99.76 (68.24) & 99.28 (50.49) & 99.33 (42.78) & 99.19 (63.89)
  & & 99.25 (54.59) \\
\multicolumn{1}{l|}{} & Mean F1
  & 99.42 (86.48) & 99.81 (83.63) & 99.88 (76.49) & 99.64 (62.56) & 99.66 (61.04) & 99.57 (87.03)
  & & 99.61 (76.82) \\
\multicolumn{1}{l|}{} & Mean IOU
  & 98.86 (86.32) & 99.62 (83.63) & 99.76 (76.54) & 99.28 (62.44) & 99.32 (60.94) & 99.14 (86.97)
  & & 99.22 (76.75) \\
\multicolumn{1}{l|}{Segment} & Mean Recall
  & 100 (100.0) & 100 (100.0) & 100 (100.0) & 100 (100.0) & 100 (100.0) & 100 (100.0)
  & & 100.00 (100.00) \\
\multicolumn{1}{l|}{} & Mean Precision
  & 97.16 (72.04) & 81.84 (61.82) & 97.64 (68.33) & 90.4 (63.09) & 98.87 (74.76) & 98.92 (92.0)
  & & 95.11 (65.62) \\
\multicolumn{1}{l|}{} & Mean F1
  & 98.54 (72.04) & 89.05 (61.82) & 98.8 (68.33) & 94.04 (63.09) & 99.43 (74.76) & 99.45 (92.0)
  & & 97.23 (65.62) \\
\bottomrule
\end{tabular}
}
\vspace{-3mm}
\end{table}
We ran CATH Functional Families (FunFams) \citep{das2015cath} to obtain domain boundaries and compared them against \algo-derived motifs. Because sequence conservation is linked to structural preservation, we expect overlap between predicted motifs and functional domains. In Table \ref{tab:agreement}, \algo achieves 99.97\% domain recall with mean F1 = 0.996 and IOU = 0.992, showing near-perfect agreement across 10 datasets. {\em The agreement is not only geometric but also functional:} \algo tokens frequently coincide with boundaries of ligand-binding grooves, transmembrane cavities, and scaffolding helices, capturing motifs that underlie molecular recognition and catalysis. This suggests \algo does more than segment folds consistently: it surfaces interpretable structural primitives that map onto biochemical roles, offering a functional vocabulary absent in prior PSTs. Details are in App.~\ref{app:expert-agreement-eval}

\xhdr{Case Study: Human Expert Analysis of Interpretability} 
We conducted three expert evaluations of \algo-derived hierarchies (App.~\ref{app:expert-case-study}). Across proteins, the discovered motifs align with functionally meaningful substructures, including regions mediating ligand binding, molecular recognition, and structural gating. In the SLC25A20 transporter (Fig.~\ref{fig:case-study-4xk4}), \algo isolates a transmembrane binding cavity formed by helices and polar residues. In the 14-3-3:Tau complex (Fig.~\ref{fig:case-study-6FI4}), it identifies a canonical phospho-binding groove stabilized by charged side chains. Recurrent local motifs (aromatic cages, polar bridges, helix-helix clamps) are combined into higher-order scaffolds that mirror established biochemical organization. {\em These hierarchies capture geometric regularities and also modular design principles conserved across folds and families.} Even in compact domains, such as nucleotide-recognition modules, \algo motifs reveal the coupling between geometric curvature and chemical specificity, meaning that \algo surfaces reusable motifs that are both interpretable and evolutionarily grounded.

{\xhdr{Limitations} \algo currently does not incorporate sequence or side chains, but \textit{can} via direct extensions, e.g. taking the Cartesian product of the current vocabulary with amino acid types and augmenting the backbone formulation with type-dependent $\chi$-angle spans. The present integration with small-scale Transformers is set up to compare tokenizers' compability with language modeling on structure tokens; only \textit{relative} backbone design metrics are relevant. Generative performance depends on model capacity and data scale, which are \textit{orthogonal} to the tokenizer. In the separate SSLM scalability study, we see \textit{steep gains in generative performance} when both LM parameter count and pretraining data increase ten-fold, conforming with scaling law expectations. The improved numbers are preliminary evidence of \algo's promise as a tokenizer for lage-scale PLMs, but are not competitive with state-of-the-art backbone design models.}
\vspace{-3mm}

\section{Conclusion}

We present \algo, a principled geometry-grounded analog of BPE for protein folds. \algo (a) captures natural conformational variability in protein backbones, (b) constructs a hierarchical vocabulary of structural motifs, and (c) produces hierarchical views of folds for downstream representation learning. Its hierarchies reveal conserved modular design principles that connect structure to function. Empirically, \algo advances the state of the art in tokenizer performance, out-of-distribution generalization, token and generative efficiency, downstream transfer, and interpretability. These results establish \algo as a foundation for structure-native protein language models. 

\section*{Reproducibility Statement}
Code for \algo and steps to reproduce all experiments in the paper are available at \url{https://github.com/shiningsunnyday/PT-BPE/}. We include detailed descriptions for understanding our method in the main text, with mathematical descriptions and pseudocodes in the Appendix. In App.~\ref{app:hyperparameters}, we list the key hyperparameters, their effects on algorithm behavior, the default values used in our experiments, and any deviations from the default values used to obtain the results reported in the main text. In App. \ref{app:computational-complexity}, we analyze the computational complexity of our method and describe practical implementation choices used to make the method efficient in practice.

\subsubsection*{Acknowledgments}

{\footnotesize M.S. and M.Z. gratefully acknowledge partial support by NSF CAREER Award 2339524, ARPA-H Biomedical Data Fabric (BDF) Toolbox Program, Amazon Faculty Research, Google Research Scholar Program, AstraZeneca Research, GlaxoSmithKline Award, Roche Alliance with Distinguished Scientists (ROADS) Program, Sanofi iDEA-iTECH Award, Boehringer Ingelheim Award, Merck Award, Optum AI Research Collaboration Award, Pfizer Research, Gates Foundation (INV-079038), Chan Zuckerberg Initiative, Collaborative Center for XDP at Massachusetts General Hospital, John and Virginia Kaneb Fellowship at Harvard Medical School, Biswas Computational Biology Initiative in partnership with the Milken Institute, Harvard Medical School Dean’s Innovation Fund for the Use of Artificial Intelligence, and the Kempner Institute for the Study of Natural and Artificial Intelligence at Harvard University. Any findings, conclusions or recommendations expressed in this material are those of the authors and do not necessarily reflect the views of the funders.}


\bibliography{iclr2026_conference}
\bibliographystyle{iclr2026_conference}
\clearpage
\appendix
\section{Ablation Studies}\label{app:GeoBPE-ablation}
\xhdr{\algo is task-agnostic, and using task-specific data does not increase {downstream perfor-} {mance of \model}}
For each task $i$, let $\mathsf{T}^{\text{task}}_i$ be a tokenizer fitted using only $\mathcal{D}^{\text{train}}_i$ (with its own vocabulary $\mathcal{V}_i$ but the same feature extractor $F_{\Theta}$), and define $r_{\mathcal{V}_i}(\mathbf{x})=\Psi_{\mathcal{V}_i}(F_{\Theta}(\mathbf{x}))$. We follow the same downstream transfer evaluation. We find an interesting result in Table \ref{tab:GeoBPE-ablation-data-efficiency}, where directly training on the task-specific dataset does not meaningful change downstream prediction results. A closer look reveals the underlying reason is because the individual tokens do not differ significantly; motifs added to $\mathcal{V}$, in order, are similar across both \model and \model (task-specific). We can interpret this both positively and negatively. \algo is insensitive to task-specific data and learns the ``language" of protein folds consistently. This may be desirable for reusability of a tokenizer, as one does not need to retrain it for different data distributions, as all protein folds obey the same universal principles \citep{petsko2004protein}. At the same time, this upper bound tests whether the tokenizer can tailor its vocabulary to individual datasets for potentially higher scores, indicating \algo by itself may lack the parameter capacity to overfit to individual tasks.

\begin{table}[h!]
\small
\centering
\caption{{\model (1\%) runs Algo. \ref{alg:GeoBPE} with 1\% of the pretrain training set, then uses the output} vocabulary {to induce features}; \model (task-specific) {does not use pretraining data; instead it} {runs Algo. \ref{alg:GeoBPE} with downstream data to learn a vocabulary}. All use default value parameters in App. \ref{app:hyperparameters}.}
\label{tab:GeoBPE-ablation-data-efficiency}

\centering
\resizebox{\textwidth}{!}{%
\begin{tabular}{@{}l *{16}{c} c@{}}
\toprule
& \multicolumn{16}{c}{\textbf{Functional Site Prediction (AUROC\%)}}  \\
\cmidrule(lr){2-17}\cmidrule(l){18-18}
\textbf{Model}
& BindInt (Fold) & BindInt (SupFam)
& BindBio (Fold) & BindBio (SupFam)
& BindShake (Org)
& CatInt (Fold) & CatInt (SupFam)
& CatBio (Fold) & CatBio (SupFam)
& Con (Fold) & Con (SupFam)
& Rep (Fold) & Rep (SupFam)
& Ept (Fold) & Ept (SupFam)
& Avg \\
\midrule
\model (1\%)
& 59.98 & 90.17
& 95.00 & 95.89
& 87.73
& 66.28 & 88.87
& 94.95 & 95.95
& 71.75 & 84.56
& 56.37 & 72.87
& 63.83 & 77.55
& 80.12  \\
\model
& 59.19 & 91.31
& 94.94 & 95.94
& 87.73
& 66.21 & 88.65
& 95.01 & 95.90
& 71.96 & 84.84
& 56.44 & 72.98
& 64.78 & 77.06
& {80.20} \\
\model (task-specific)
& 60.16 & 89.93
& 95.05 
& 95.92 
& 87.73 
& 66.28 & 88.82
& 94.98 & 95.90
&  71.85 
& 85.92 
& 56.33 & 72.72
& 64.78 & 77.04
& 80.23 
\\
\bottomrule
\end{tabular}
}

\resizebox{\textwidth}{!}{%
\begin{tabular}{@{}l *{6}{c} c *{3}{c} c@{}}
\toprule
& \multicolumn{7}{c}{\textbf{Physicochemical Property Prediction (Spearman’s $\rho$\%)}} & \multicolumn{4}{c}{\textbf{Structure Property Prediction (Macro F1\%)}} \\
\cmidrule(lr){2-8}\cmidrule(l){9-12}
\textbf{Model}
& FlexRMSF (Fold) & FlexRMSF (SupFam)
& FlexBFactor (Fold) & FlexBFactor (SupFam)
& FlexNEQ (Fold) & FlexNEQ (SupFam)
& Avg
& Homo (Fold) & Homo (SupFam) & Homo (Fam)
& Avg \\
\midrule
\model (1\%)
& 40.42 & 47.55
& 34.74 & 32.21
& 56.78 & 55.32
& 44.50
& 21.65 & 50.25 & 84.87
& 52.26 \\
\model
& 40.89 & 47.17
& 37.28 & 35.61
& 56.65 & 53.98
& 45.26
& 23.60 & 47.28 & 85.75
& 52.21 \\
\model (task-specific)
& 39.39 & 44.00
& 37.94 & 38.36
& 56.22 & 54.22
& 45.02
& 24.22 & 46.58 & 84.57
& 51.79 \\
\bottomrule
\end{tabular}
}
\end{table}
\xhdr{\model maintains comparable downstream transfer performance {even when} {{\algo was fitted on 1\% of pretraining data}}} In Table \ref{tab:GeoBPE-ablation-data-efficiency}, we see \algo fitted on just $1\%$ of the pretraining data is enough to transfer, on average, the same amount of performance downstream as \algo {fitted} on the full dataset. There are no meaningful differences between \model and \model (1\%), with \model doing 1.7\% better on physicochemical property prediction and \model (1\%) doing better 0.2\% better on functionals ite prediction. These findings can be interpreted both positively and negatively for \algo {learned vocabularies}: (1) they are \textit{extremely} {informative}, learning useful signals to transfer downstream with as few as 300 PDB structures; (2) they \textit{underfit} the data, with no noticeable improvements {from using more data to learn the tokenizer}. Taken together, these findings imply \algo is a lightweight add-on on top of any pretrained features $\Theta$, but feeding more data to \algo yields diminishing {downstream} returns quickly.

\begin{table}[h]
\small
\centering

\caption{{We use \model from Table \ref{tab:transfer} (reported as 100\%), then vary only the percent of downst-} {ream task training data available, fixing the same valid and test sets.}}
\label{tab:GeoBPE-transfer-data-efficiency}

{
\begin{tabular}{@{}llllllc@{}}
\toprule
\textbf{Task}                                    & \textbf{Split}              & \multicolumn{5}{l}{\model}      \\ \midrule
\multicolumn{1}{l|}{}                            & \multicolumn{1}{l|}{}       & 20\%  & 40\%  & 60\%  & 80\%  & 100\%          \\ \midrule
\multicolumn{7}{c}{\textbf{Functional Site Prediction (AUROC\%)}}                                                               \\ \midrule
\multicolumn{1}{l|}{BindInt}                     & \multicolumn{1}{l|}{Fold}   & 58.78 & 61.18 & 59.79 & 59.45 & 59.19          \\
\multicolumn{1}{l|}{}                            & \multicolumn{1}{l|}{SupFam} & 89.55 & 89.88 & 90.42 & 90.71 & 91.31          \\
\multicolumn{1}{l|}{BindBio}                     & \multicolumn{1}{l|}{Fold}   & 94.86 & 94.82 & 94.96 & 94.97 & 94.94          \\
\multicolumn{1}{l|}{}                            & \multicolumn{1}{l|}{SupFam} & 95.69 & 95.79 & 95.88 & 95.97 & 95.94          \\
\multicolumn{1}{l|}{BindShake}                   & \multicolumn{1}{l|}{Org}    & 87.40 & 87.59 & 87.65 & 87.64 & 87.73          \\
\multicolumn{1}{l|}{CatInt}                      & \multicolumn{1}{l|}{Fold}   & 64.66 & 65.55 & 66.94 & 66.29 & 66.21          \\
\multicolumn{1}{l|}{}                            & \multicolumn{1}{l|}{SupFam} & 87.96 & 88.27 & 88.78 & 88.73 & 88.65          \\
\multicolumn{1}{l|}{CatBio}                      & \multicolumn{1}{l|}{Fold}   & 94.95 & 94.92 & 94.94 & 94.96 & 95.01          \\
\multicolumn{1}{l|}{}                            & \multicolumn{1}{l|}{SupFam} & 95.67 & 95.81 & 95.96 & 95.97 & 95.90          \\
\multicolumn{1}{l|}{Con}                         & \multicolumn{1}{l|}{Fold}   & 71.42 & 71.37 & 71.72 & 71.74 & 71.96          \\
\multicolumn{1}{l|}{}                            & \multicolumn{1}{l|}{SupFam} & 83.77 & 84.02 & 84.76 & 84.76 & 84.84          \\
\multicolumn{1}{l|}{Rep}                         & \multicolumn{1}{l|}{Fold}   & 55.04 & 54.25 & 56.02 & 56.41 & 56.44          \\
\multicolumn{1}{l|}{}                            & \multicolumn{1}{l|}{SupFam} & 73.24 & 75.18 & 75.89 & 71.58 & 72.98          \\
\multicolumn{1}{l|}{Ept}                         & \multicolumn{1}{l|}{Fold}   & 63.63 & 53.28 & 61.72 & 61.66 & 64.78          \\
\multicolumn{1}{l|}{}                            & \multicolumn{1}{l|}{SupFam} & 71.21 & 49.39 & 73.59 & 76.01 & 77.06          \\ \midrule
\multicolumn{1}{l|}{\textbf{Average AUROC\%}}    & \multicolumn{1}{l|}{}       & 79.19 & 77.42 & 79.93 & 79.79 & \textbf{80.20} \\ \midrule
\multicolumn{7}{c}{\textbf{Physicochemical Property Prediction (Spearman’s $\rho$\%)}}                                          \\ \midrule
\multicolumn{1}{l|}{FlexRMSF}                    & \multicolumn{1}{l|}{Fold}   & 41.49 & 43.33 & 38.48 & 39.02 & 40.89          \\
\multicolumn{1}{l|}{}                            & \multicolumn{1}{l|}{SupFam} & 34.74 & 45.41 & 44.61 & 47.06 & 47.17          \\
\multicolumn{1}{l|}{FlexBFactor}                 & \multicolumn{1}{l|}{Fold}   & 23.90 & 27.86 & 33.83 & 34.82 & 37.28          \\
\multicolumn{1}{l|}{}                            & \multicolumn{1}{l|}{SupFam} & 23.80 & 25.46 & 37.70 & 36.49 & 35.61          \\
\multicolumn{1}{l|}{FlexNEQ}                     & \multicolumn{1}{l|}{Fold}   & 54.40 & 56.07 & 55.98 & 57.53 & 56.65          \\
\multicolumn{1}{l|}{}                            & \multicolumn{1}{l|}{SupFam} & 51.12 & 53.77 & 52.52 & 54.96 & 53.98          \\ \midrule
\multicolumn{1}{l|}{\textbf{Average $\rho$\%}}   & \multicolumn{1}{l|}{}       & 38.24 & 41.98 & 43.85 & 44.98 & \textbf{45.26} \\ \midrule
\multicolumn{7}{c}{\textbf{Structure Property Prediction (Macro F1\%)}}                                                         \\ \midrule
\multicolumn{1}{l|}{Homo}                        & \multicolumn{1}{l|}{Fold}   & 14.35 & 20.55 & 23.74 & 24.25 & 23.60          \\
\multicolumn{1}{l|}{}                            & \multicolumn{1}{l|}{SupFam} & 27.63 & 35.11 & 43.09 & 43.98 & 47.28          \\
\multicolumn{1}{l|}{}                            & \multicolumn{1}{l|}{Fam}    & 65.79 & 73.66 & 80.77 & 82.04 & 85.75          \\ \midrule
\multicolumn{1}{l|}{\textbf{Average Macro F1\%}} & \multicolumn{1}{l|}{}       & 35.92 & 43.11 & 49.20 & 50.09 & \textbf{52.21} \\ \bottomrule
\end{tabular}
}
\end{table}
\xhdr{{\model does not underfit the data for structure-related downstream tasks}} {In Table \ref{tab:GeoBPE-transfer-data-efficiency}, we see for physicochemical (residue-level regression) and fold-level tasks, the model indicates a clear propensity for more training data, with step-wise gains for every 20\% of training data. $20\% \rightarrow 100\%$ training data sees performance lift significantly ($+18.89\%$ average $\rho\%$ and $+45.35\%$, respectively). For residue-level classification tasks, the lift is only marginal ($+1.28\%$). We hypothesize the cause is not limited capacity, but rather that the tasks are localized label predictions; a residue-level receptive field is sufficient when features are informative.}
{
This shows GeoBPE’s data-efficiency at learning a vocabulary does not limit its capacity on downstream tasks that require multi-scale resolution (e.g. structural flexibility or fold-level classification), which existing fixed-size tokenizers cannot at both residue and structure-level. Thus, the data-efficiency strengths of GeoBPE \textit{training} is orthogonal to downstream modeling. The structure-related tasks in Table \ref{tab:GeoBPE-transfer-data-efficiency} see large gains in performance with more training data, implying \model scales to complex, hierarchical structural patterns that can only be learned from more data.}

\begin{table}[h!]
\small
\centering

{\caption{We rerun the \algo experiments used to trace out the Pareto Front in Fig. \ref{fig:pareto} by using $1\%$ of pretraining data. We include raw numbers of Fig. \ref{fig:pareto} (bottom rows) for comparison. All other hyperparameter settings are kept the same (App. \ref{app:hyperparameters}).}
\label{tab:GeoBPE-data-efficiency}}
{
\resizebox{\textwidth}{!}{
\begin{tabular}{@{}lllllllll@{}}
\toprule
                         & Train                 &                       & Valid                 &                       & CAMEO                 &                       & CASP14                &                       \\ \midrule
\multirow{2}{*}{}        & \multirow{2}{*}{RMSD} & \multirow{2}{*}{LDDT} & \multirow{2}{*}{RMSD} & \multirow{2}{*}{LDDT} & \multirow{2}{*}{RMSD} & \multirow{2}{*}{LDDT} & \multirow{2}{*}{RMSD} & \multirow{2}{*}{LDDT} \\
                         &                       &                       &                       &                       &                       &                       &                       &                       \\ \midrule
GeoBPE (1\%)  ($|V|=600$)   & 1.72                  & 0.74                  & 1.63                  & 0.73                  & 1.66                  & 0.73                  & 1.53                  & 0.72                  \\
GeoBPE ~~~~~~~~~($|V|=600$)         & 1.66                  & 0.73                  & 1.71                  & 0.72                  & 1.77                  & 0.72                  & 1.53                  & 0.72                  \\ \midrule
GeoBPE (1\%)  ($|V|=2278$)  & 1.57                  & 0.75                  & 1.51                  & 0.71                  & 1.51                  & 0.74                  & 1.43                  & 0.73                  \\
GeoBPE ~~~~~~~~~($|V|=2500$)        & 1.41                  & 0.75                  & 1.50                  & 0.74                  & 1.57                  & 0.74                  & 1.51                  & 0.73                  \\ \midrule
GeoBPE (1\%)  ($|V|=5278$)  & 1.36                  & 0.77                  & 1.34                  & 0.76                  & 1.35                  & 0.75                  & 1.30                  & 0.74                  \\
GeoBPE ~~~~~~~~~($|V|=6000$)        & 1.37                  & 0.76                  & 1.46                  & 0.75                  & 1.52                  & 0.74                  & 1.54                  & 0.72                  \\ \midrule
GeoBPE (1\%)  ($|V|=20278$) & 1.29                  & 0.77                  & 1.28                  & 0.76                  & 1.28                  & 0.76                  & 1.37                  & 0.73                  \\
GeoBPE ~~~~~~~~~($|V|=21000$)       & 1.21                  & 0.77                  & 1.28                  & 0.76                  & 1.40                  & 0.75                  & 1.55                  & 0.72                  \\ \bottomrule
\end{tabular}
}
}
\end{table}
\xhdr{{\algo is data-efficient OOD, but more training data can lower training distortion}}
{
In Table \ref{tab:GeoBPE-data-efficiency}, we see \algo (1\%) consistently achieves lower distortion ($\downarrow 7.7\%$ RMSD averaged, $\uparrow 0.06$ LDDT summed across all four runs and both test splits) than \algo. This suggests a small, well-chosen set of structures is enough for \algo to achieve superior reconstruction on \textit{OOD} structures, and more training data can introduce noise and hinder generalization. However, \algo (1\%) obtains $\uparrow 5.1\%$ averaged, $\downarrow 0.02$ LDDT summed across all four runs on the respective training splits. 
A lower RMSD suggests \algo better captures global structural fidelity; by constructing the vocabulary from the full pretraining dataset, it can choose more representative prototypes; hence, its vocabulary better preserves global fold. Meanwhile, a slightly lower LDDT indicates \algo (1\%) can capture a few local details in the 1\% subset of structures better than \algo. This suggests \algo (1\%) is more sensitive to the individual local interactions of the small set of structures it fitted with; \algo considers vastly more structures.
In summary, \algo is preferred for fold-preserving \textit{compression} of whole datasets, but \algo (1\%) can be feasible if not superior when \algo is primarily used to tokenize unseen data.
}

\begin{figure}[h!]
    \centering
    \includegraphics[width=\linewidth]{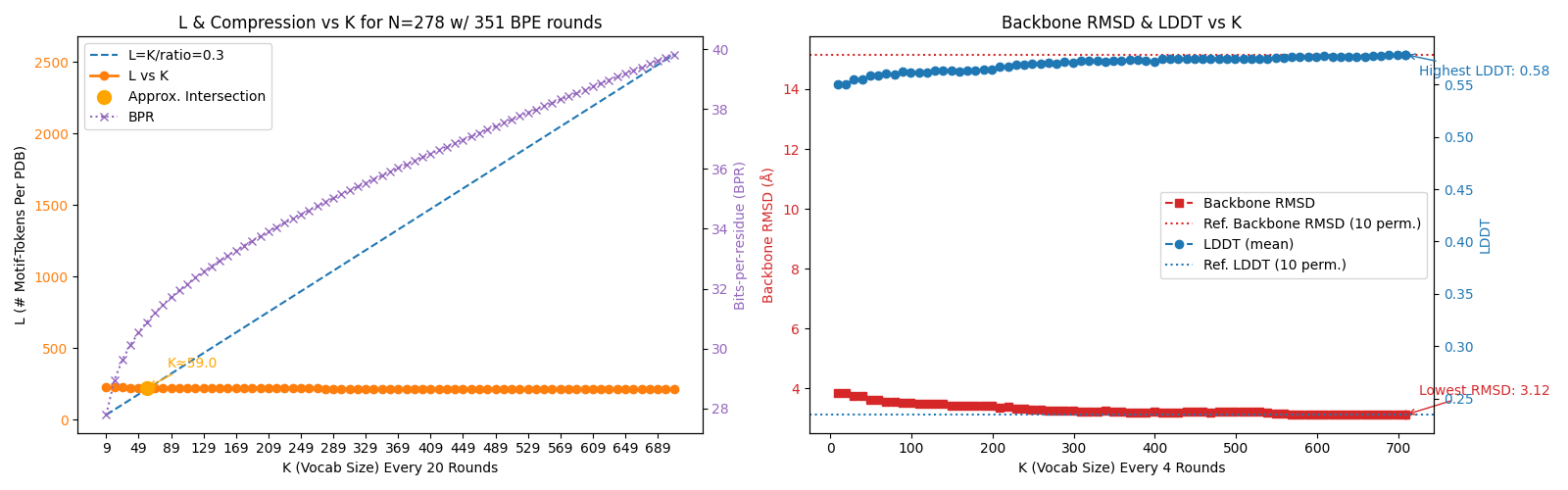}
    \caption{We plot the BPR (\textcolor{purple}{purple}), length (\textcolor{orange}{orange}), backbone distortion (\textcolor{red}{RMSD}, \textcolor{blue}{LDDT}) as $|\mathcal{V}|$ across BPE steps. Ref. backbone RMSD/LDDT (dotted lines) uses random angle values for all internal angles, sampled from the empirical angle distribution.}
    \label{fig:super-res}
\end{figure}

\xhdr{\algo is multi-resolution, revealing finer details as more tokens are introduced} In Fig. \ref{fig:super-res}, we run a coarse-grained version of \algo (small initial $|\mathcal{V}|$) to observe an interesting feature of \algo's design. As newly introduced tokens \textit{re}-quantize the occurrences from the original data (span gathering step in Alg. \ref{alg:step}, tokenization can adaptively \textit{increase} the resolution if the new prototypes better capture the modes of variability for those occurrences than their previous quantization. We expose this via hyperparameters \texttt{bins} \& \texttt{num\_p} (see App. \ref{app:hyperparameters}), which tradeoff the super-resolution effect against coarse-graining effect at different token sizes, offering fine-grained control.

\begin{figure}[h!]
  \centering  
  \begin{subfigure}[t]{0.31\linewidth}
    \centering
    \includegraphics[width=\linewidth]{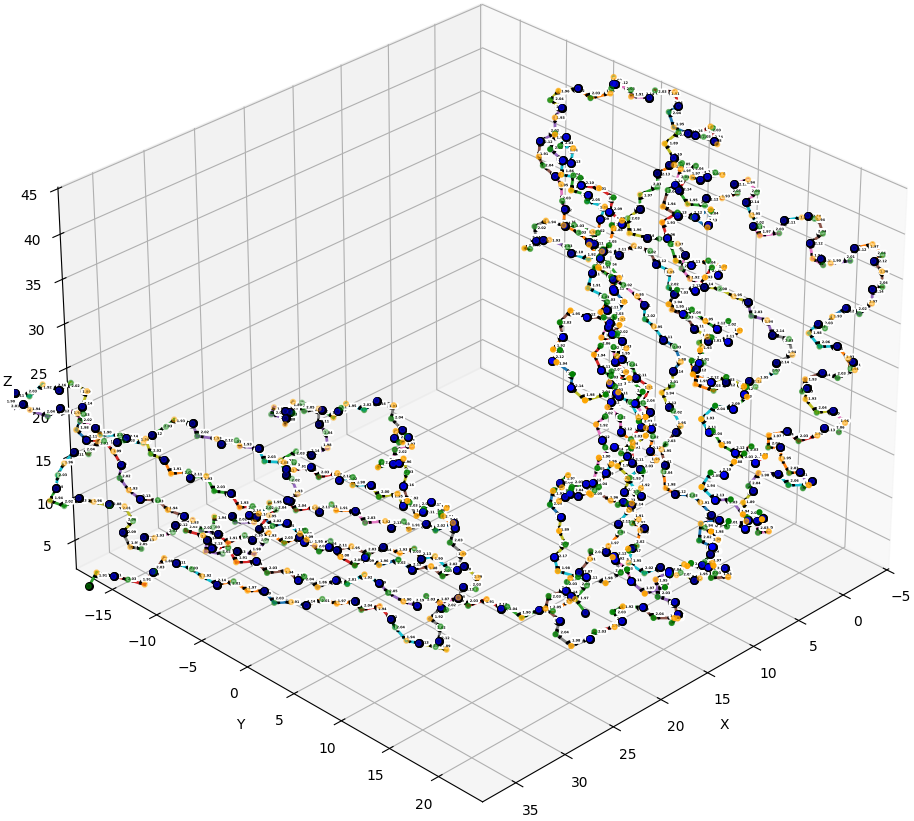}
    \caption{Quantized backbone (\algo)}
    \label{fig:ex-w-glue-opt}
  \end{subfigure}
  \begin{subfigure}[t]{0.31\linewidth}
    \centering
    \includegraphics[width=\linewidth]{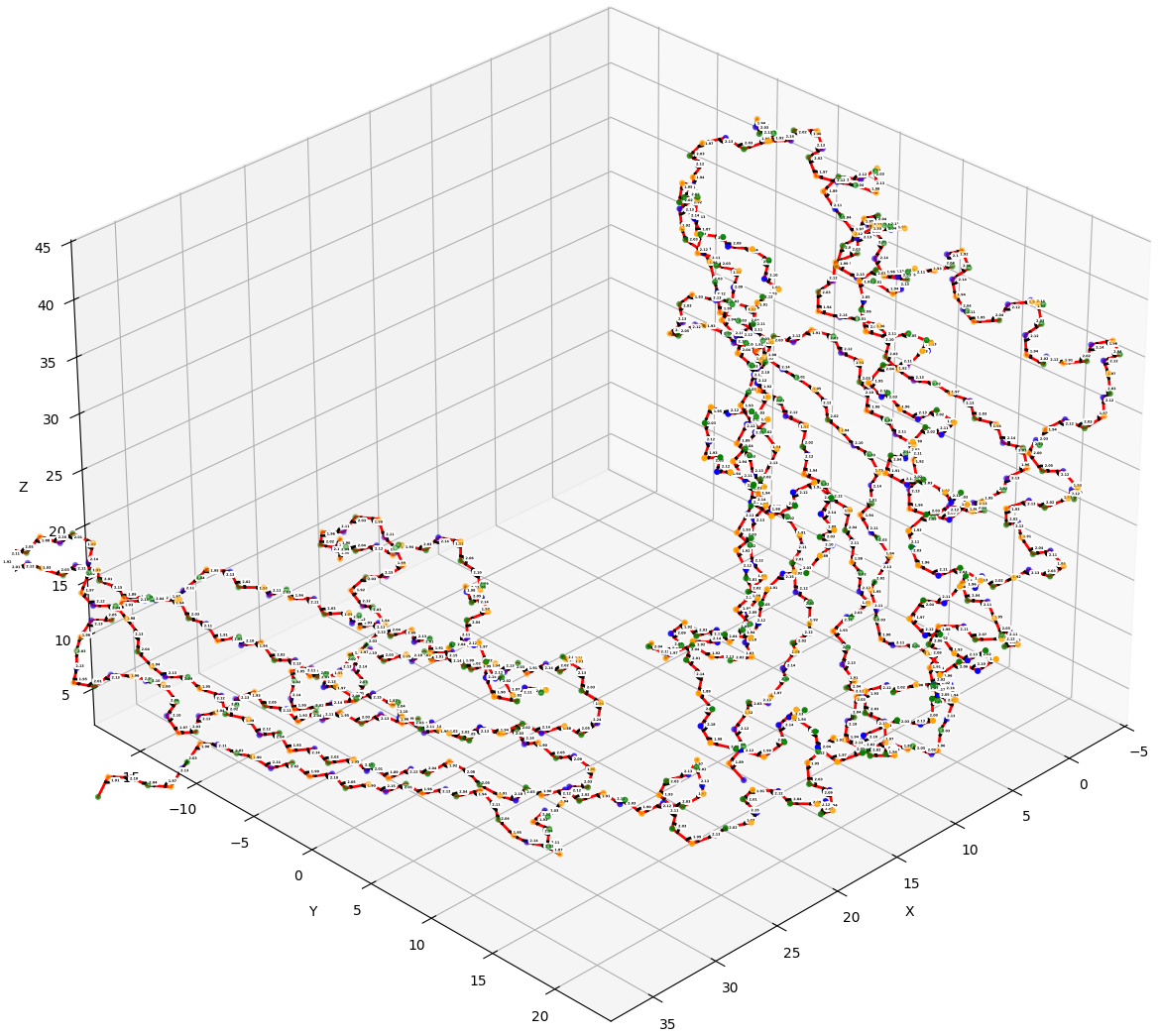}
    \caption{Original backbone}
    \label{fig:ex-orig}
  \end{subfigure}  
  \begin{subfigure}[t]{0.31\linewidth}
    \centering
    \includegraphics[width=\linewidth]{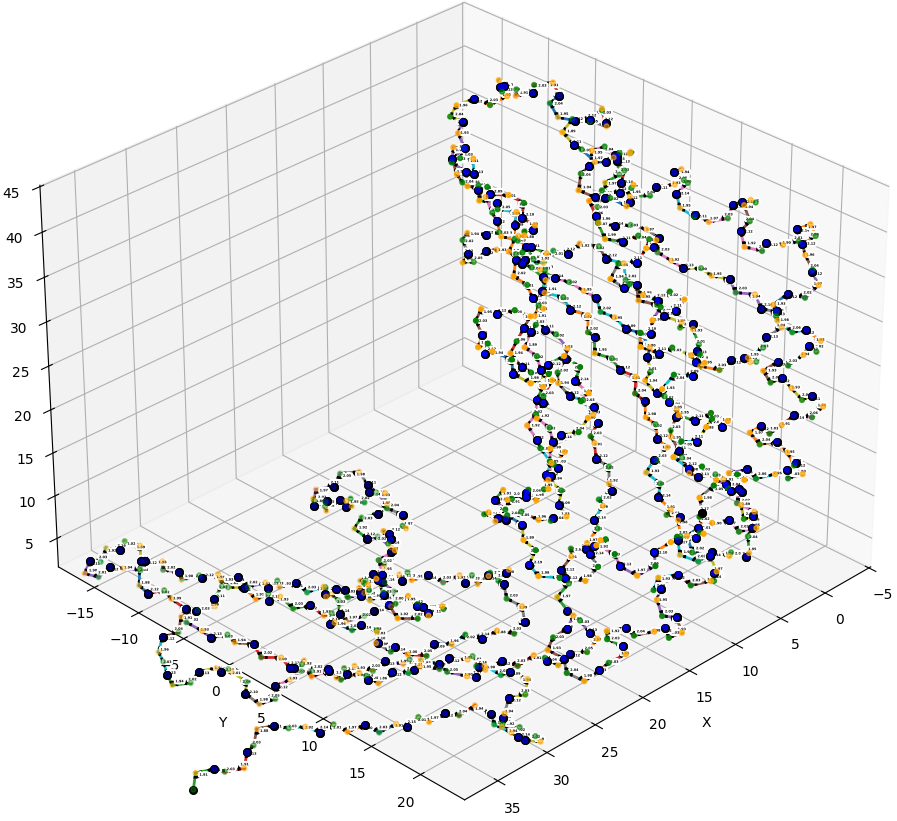}
    \caption{Quantized backbone (\algo - Glue Opt)}
    \label{fig:ex-wo-glue-opt}
  \end{subfigure}  
  \caption{We ran an ablation for \algo version with $|\mathcal{V}|=600$, keeping all parameters the same but toggling whether glue opt is skipped in Alg. \ref{alg:res-init-tokens}. We visualize the original (center), \algo (left) and \algo without glue opt (right) backbone states.}
  \label{fig:ex}
\end{figure}

\xhdr{Rigid body refinment as an essential step for preserving fold integrity} If we omit the glue optimization from Algs \ref{alg:res-init-tokens} and \ref{alg:step} altogether, we see the effects in Fig. \ref{fig:ex}. For that experiment, we find avg. RMSD increase $1.66\rightarrow 4.39$, and avg. LDDT drop $0.73\rightarrow 0.69$ when glue opt is turned off. Rigid body refinement preserves the overall fold and modular architecture; turning it off causes individual domains to distort -- the parallel strands drift apart -- as well as the overall configuration to lose its integrity. Over the course of many time steps, global drift accumulate as local rounding occurs. Rigid body refinement is an indispensable subroutine for ensuring the overall quantization faithfully reproduces the fold integrity.

\xhdr{Increasing $M_{\text{max}}$ beyond a certain threshold does not yield additional distortion benefits} We did a study comparing \algo ($\mathcal{V}=6000$, $M_{\text{max}}\leftarrow 5000$, full settings in App. \ref{app:hyperparameters}) with ``higher-resolution" settings $\texttt{bins} \leftarrow \{1:5000\}$, $M_{\text{max}}\leftarrow 20000$. Interestingly, we found overall RMSD/LDDT did \textit{not} improve ($1.40$ vs $1.39$, $0.76$ vs $0.75$, both in favor of the incumbent) despite increased computational expenditure spent on Alg. \ref{alg:k-medoids}. The most likely explanation is there is no marginal utility increasing $M_{\text{max}}$ beyond $5000$, and differences in distortion rates are likely due to the numerical stability of Alg. \ref{alg:glue-opt-all} more so than the hyperparameters.

\section{Dataset Details}\label{app:dataset}
\xhdr{Training} For training \algo, we started with the pretraining data splits released by \citet{yuan2025protein}, which follows the same criteria used to train the OpenFold2 model \cite{ahdritz2024openfold}. For VanillaVQ and AminoASeed baselines, we use the same splits as \citet{yuan2025protein} directly. For \algo, we further filtered the data down to only ones with complete backbone information (e.g. backbone dihedrals are not NaN, each residue contains N, CA and C), resulting in $34818$ structures. We further excluded structures shorter than $40$ or longer than $512$ residues, resulting in $33992$ structures for training GeoBPE and $3810$ for validation (only used for \ref{app:lm-eval}).

\xhdr{Held-out testing} We use CAMEO and CASP14 test sets for evaluating the generalization of tokenizers \citep{robin2021continuous, kryshtafovych2021critical}. For CASP14, we follow \citet{yuan2025protein} and select only proteins released after the pretraining data cutoff date.

\xhdr{Downstream Tasks} Our 8 downstream tasks cover a breadth of structure and function-related predictions. They are divided into 3 categories and are assembled from 6 sources: InterPro (BindInt, Con, Rep) \citep{blum2025interpro}, BioLIP2 (BindBio, CatBio) \citep{zhang2024biolip2}, ProteinShake (BindShake) \citep{kucera2023proteinshake}, ProteinGLUE (Ept) \citep{capel2022proteinglue}, TAPE (Homo) \citep{rao2019evaluating} and ATLAS (FlexRMSF, FlexBFactor, FlexNEQ) \citep{vander2024atlas}.
\begin{enumerate}[leftmargin=*,topsep=0pt]
    \item \textbf{Functional site prediction}: Binding site prediction (BindInt), catalytic site (CatInt), conserved site prediction (Con), repeat motif prediction (Rep), epitope region prediction (Ept)
    \item \textbf{Physicochemical property prediction}: Structural flexibility prediction, measured using metric RMSF (FlexRMSF), B-factor (FlexBFactor) and Neq (FlexNEQ)
    \item \textbf{Structure classification} (protein-level): Remote homology detection (Homo)
\end{enumerate}

\textbf{Functional site prediction} tasks predict whether each residue is in a site of functional importance (binding, catalytic activity or antibody recognition) or part of an evolutionary motif (conserved site or part of a repeated motif). PSTs which learn semantically meaningful signals like motif boundaries are expected to perform well on these tasks.

\textbf{Physicochemical property prediction} tasks predict the flexibility of each residue as a continuous value. Higher flexibility can be a clue that the residue may be more amenable to functional activity. PSTs that capture a fine-grained view of the localized protein dynamics are expected to predict residue-level flexibility well.

\textbf{Remote homology detection} is a multi-class fold classification problem. Proteins which belong to the same fold class can be distantly related or share similar functions on the whole. Therefore, PSTs that capture the overall fold-level geometry are expected to do well on this task.

For more dataset statistics and preparation details, see \citet{yuan2025protein}.

\section{Additional Results}\label{app:additional-results}

\begin{table}[h!]
\caption{Additional downstream transfer performance tasks. Setup follows Table \ref{tab:transfer}.}
\label{tab:transfer-additional}
\resizebox{\textwidth}{!}{
\begin{tabular}{@{}llcccccccc@{}}
\toprule
\multicolumn{1}{l|}{\textbf{Task}} & \multicolumn{1}{l|}{\textbf{Split}} & \multicolumn{2}{c|}{\textbf{Continuous PST}}                                                   & \multicolumn{6}{c}{\textbf{Discrete PST}}                                                                                                                       \\ \midrule
\multicolumn{1}{l|}{}              & \multicolumn{1}{l|}{}               & ProteinMPNN                         & \multicolumn{1}{c|}{MIF}                                 & FoldSeek & ProTokens & ESM3  & VanillaVQ      & \multicolumn{1}{c|}{AminoAseed}                             & \model (v.s. ESM3)                               \\ \midrule
\multicolumn{10}{c}{\textbf{Functional Site Prediction (AUROC\%)}}                                                                                                                                                                                                                                                                          \\ \midrule
\multicolumn{1}{l|}{Rep}           & \multicolumn{1}{l|}{Fold}           & \cellcolor[HTML]{C0C0C0}{\ul 77.63} & \multicolumn{1}{c|}{74.53}                               & 47.71    & 53.20     & 74.70 & \textbf{75.99} & \multicolumn{1}{c|}{74.97}                                  & 56.44 (-24.44\%)                                  \\
\multicolumn{1}{l|}{}              & \multicolumn{1}{l|}{SupFam}         & 80.71                               & \multicolumn{1}{c|}{{\ul 83.11}}                         & 52.54    & 77.25     & 82.36 & 82.09          & \multicolumn{1}{c|}{\cellcolor[HTML]{C0C0C0}\textbf{84.57}} & 72.98 (-11.39\%)                                  \\
\multicolumn{1}{l|}{Ept}           & \multicolumn{1}{l|}{Fold}           & 62.84                               & \multicolumn{1}{c|}{\cellcolor[HTML]{C0C0C0}{\ul 68.78}} & 54.56    & 52.49     & 63.69 & 59.28          & \multicolumn{1}{c|}{62.16}                                  & \textbf{64.78 (+1.71)\%}                            \\
\multicolumn{1}{l|}{}              & \multicolumn{1}{l|}{SupFam}         & 64.84                               & \multicolumn{1}{c|}{\cellcolor[HTML]{C0C0C0}{\ul 82.98}} & 50.53    & 61.92     & 61.97 & 67.24          & \multicolumn{1}{c|}{72.02}                                  & \textbf{77.06 (+24.35)\%}                           \\ \midrule
\multicolumn{10}{c}{\textbf{Physicochemical Property Prediction (Spearman’s $\rho$\%)}}                                                                                                                                                                                                                                                     \\ \midrule
\multicolumn{1}{l|}{FlexBFactor}   & \multicolumn{1}{l|}{Fold}           & 31.88                               & \multicolumn{1}{c|}{{\ul 34.60}}                         & 4.17     & 6.67      & 23.60 & 22.32          & \multicolumn{1}{c|}{21.30}                                  & \cellcolor[HTML]{C0C0C0}\textbf{37.28 (+57.97\%)} \\
\multicolumn{1}{l|}{}              & \multicolumn{1}{l|}{SupFam}         & 34.56                               & \multicolumn{1}{c|}{{\ul 35.23}}                         & 6.99     & 5.47      & 25.80 & 23.73          & \multicolumn{1}{c|}{21.76}                                  & \cellcolor[HTML]{C0C0C0}\textbf{35.61 (+38.02\%)} \\
\multicolumn{1}{l|}{FlexNEQ}       & \multicolumn{1}{l|}{Fold}           & \cellcolor[HTML]{C0C0C0}{\ul 69.69} & \multicolumn{1}{c|}{65.32}                               & 5.71     & 12.98     & 45.05 & 35.95          & \multicolumn{1}{c|}{49.64}                                  & \textbf{56.65 (+25.75\%)}                         \\
\multicolumn{1}{l|}{}              & \multicolumn{1}{l|}{SupFam}         & \cellcolor[HTML]{C0C0C0}{\ul 68.69} & \multicolumn{1}{c|}{64.82}                               & 2.66     & 10.51     & 35.45 & 35.61          & \multicolumn{1}{c|}{50.15}                                  & \textbf{53.98 (+52.27\%)}                         \\ \bottomrule
\end{tabular}
}
\end{table}

\begin{table}[h!]
\caption{Additional expert agreement results. Setup follows Table \ref{tab:agreement}.}
\label{tab:agreement-additional}
\resizebox{\textwidth}{!}{
\begin{tabular}{@{}llllll@{}}
\toprule
                             &                & \multicolumn{1}{c}{Rep} & \multicolumn{1}{c}{Ept} & \multicolumn{1}{c}{Atlas} & \multicolumn{1}{c}{Homo} \\ \midrule
\multicolumn{1}{l|}{Domain}  & Mean Recall    & 99.93 (99.34)              & 100 (100.0)             & 99.93 (90.47)             & 99.98 (100.0)            \\
\multicolumn{1}{l|}{}        & Mean Precision & 99.2 (43.59)               & 99.75 (77.89)           & 98.44 (31.29)             & 99 (41.92)               \\
\multicolumn{1}{l|}{}        & Mean F1        & 99.56 (81.86)              & 99.87 (82.68)           & 99.17 (67.37)             & 99.48 (79.07)            \\
\multicolumn{1}{l|}{}        & Mean IOU       & 99.12 (82.18)              & 99.75 (82.68)           & 98.37 (67.07)             & 98.98 (78.78)            \\
\multicolumn{1}{l|}{Segment} & Mean Recall    & 100 (100.0)                & 100 (100.0)             & 100 (100.0)               & 100 (100.0)              \\
\multicolumn{1}{l|}{}        & Mean Precision & 98.76 (61.93)              & 95.52 (60.8)            & 96.09 (47.29)             & 95.91 (54.13)            \\
\multicolumn{1}{l|}{}        & Mean F1        & 99.38 (61.93)              & 97.68 (60.8)            & 98 (47.29)                & 97.91 (54.13)            \\ \bottomrule
\end{tabular}
}
\end{table}

\xhdr{Main Text Tables} Table \ref{tab:transfer-additional} contains additional tasks Rep, Ept, FlexRMSF, and FlexBFactor. Table \ref{tab:agreement-additional} contains additional task data Repeat, Ept, Atlas, Homo. Task abbrevations are defined in App. \ref{app:dataset}.

\begin{table}[h!]{

\caption{{Secondary structure element (SSE) agreement results. Setup is the same as in Sec. \ref{app:expert-agreement-eval}, but stratified} {over 8 basic secondary structure building blocks. Annotations are obtained from DSSP.}}
\label{tab:sec-agreement}
\resizebox{\textwidth}{!}{
{
\begin{tabular}{@{}lcccccccccc@{}}
\toprule
Metric / SSE                             & BindInt       & BindBio       & BindShake     & CatInt        & CatBio        & Con           & Repeat        & Ept           & Atlas         & Homo          \\ \midrule
\multicolumn{1}{l|}{Mean recall H}       & 29.11 (98.22) & 33.13 (97.42) & 30.76 (96.39) & 33.97 (98.1)  & 32.93 (96.48) & 29.11 (98.22) & 33.13 (97.42) & 30.76 (96.39) & 33.97 (98.1)  & 32.93 (96.48) \\
\multicolumn{1}{l|}{Mean recall G}       & 19.15 (99.4)  & 31.13 (98.61) & 30.87 (98.55) & 34.66 (98.75) & 32.29 (98.71) & 19.15 (99.4)  & 31.13 (98.61) & 30.87 (98.55) & 34.66 (98.75) & 32.29 (98.71) \\
\multicolumn{1}{l|}{Mean recall I}       & 10.42 (100.0) & 30.75 (100.0) & 17.59 (100.0) & 10.85 (99.26) & 30.85 (99.68) & 10.42 (100.0) & 30.75 (100.0) & 17.59 (100.0) & 10.85 (99.26) & 30.85 (99.68) \\
\multicolumn{1}{l|}{Mean recall E}       & 41.24 (99.56) & 34.75 (98.46) & 36.17 (97.49) & 30.82 (97.51) & 34.93 (97.99) & 41.24 (99.56) & 34.75 (98.46) & 36.17 (97.49) & 30.82 (97.51) & 34.93 (97.99) \\
\multicolumn{1}{l|}{Mean recall B}       & 38.13 (100.0) & 32.24 (100.0) & 38.72 (100.0) & 40.22 (100.0) & 31.67 (100.0) & 38.13 (100.0) & 32.24 (100.0) & 38.72 (100.0) & 40.22 (100.0) & 31.67 (100.0) \\
\multicolumn{1}{l|}{Mean recall T}       & 32.78 (98.74) & 33.08 (97.4)  & 32.26 (96.41) & 33.75 (96.67) & 32.91 (97.08) & 32.78 (98.74) & 33.08 (97.4)  & 32.26 (96.41) & 33.75 (96.67) & 32.91 (97.08) \\
\multicolumn{1}{l|}{Mean recall S}       & 33.92 (99.32) & 33.29 (98.0)  & 32.91 (97.72) & 36.01 (97.4)  & 33.12 (97.96) & 33.92 (99.32) & 33.29 (98.0)  & 32.91 (97.72) & 36.01 (97.4)  & 33.12 (97.96) \\
\multicolumn{1}{l|}{Mean recall -}       & 32.7 (97.81)  & 32.76 (97.17) & 33.02 (96.16) & 34.71 (96.86) & 32.65 (96.03) & 32.7 (97.81)  & 32.76 (97.17) & 33.02 (96.16) & 34.71 (96.86) & 32.65 (96.03) \\
\multicolumn{1}{l|}{Mean precision H}    & 31.34 (54.95) & 34.06 (48.88) & 31.33 (51.22) & 35.54 (48.86) & 33.32 (46.04) & 31.34 (54.95) & 34.06 (48.88) & 31.33 (51.22) & 35.54 (48.86) & 33.32 (46.04) \\
\multicolumn{1}{l|}{Mean precision G}    & 17.93 (79.42) & 29.28 (65.93) & 28.95 (62.87) & 32.3 (65.07)  & 30.32 (64.77) & 17.93 (79.42) & 29.28 (65.93) & 28.95 (62.87) & 32.3 (65.07)  & 30.32 (64.77) \\
\multicolumn{1}{l|}{Mean precision I}    & 10.09 (94.67) & 29.93 (87.7)  & 16.55 (87.53) & 10.37 (93.27) & 30.16 (88.4)  & 10.09 (94.67) & 29.93 (87.7)  & 16.55 (87.53) & 10.37 (93.27) & 30.16 (88.4)  \\
\multicolumn{1}{l|}{Mean precision E}    & 39.54 (47.86) & 33.59 (46.21) & 34.84 (45.6)  & 29.58 (53.03) & 33.53 (45.56) & 39.54 (47.86) & 33.59 (46.21) & 34.84 (45.6)  & 29.58 (53.03) & 33.53 (45.56) \\
\multicolumn{1}{l|}{Mean precision B}    & 26.8 (74.03)  & 21.86 (71.01) & 26.66 (68.64) & 26.21 (61.46) & 21.2 (69.63)  & 26.8 (74.03)  & 21.86 (71.01) & 26.66 (68.64) & 26.21 (61.46) & 21.2 (69.63)  \\
\multicolumn{1}{l|}{Mean precision T}    & 28.59 (53.66) & 28.55 (50.1)  & 27.98 (50.94) & 29.18 (52.52) & 28.13 (48.36) & 28.59 (53.66) & 28.55 (50.1)  & 27.98 (50.94) & 29.18 (52.52) & 28.13 (48.36) \\
\multicolumn{1}{l|}{Mean precision S}    & 26.78 (56.59) & 25.28 (50.82) & 24.66 (49.22) & 26.52 (45.95) & 25.0 (50.96)  & 26.78 (56.59) & 25.28 (50.82) & 24.66 (49.22) & 26.52 (45.95) & 25.0 (50.96)  \\
\multicolumn{1}{l|}{Mean precision -}    & 27.07 (54.17) & 26.5 (48.75)  & 26.32 (48.44) & 28.07 (51.65) & 25.99 (48.15) & 27.07 (54.17) & 26.5 (48.75)  & 26.32 (48.44) & 28.07 (51.65) & 25.99 (48.15) \\
\multicolumn{1}{l|}{Mean f1 H}           & 29.42 (61.9)  & 33.12 (60.16) & 30.7 (62.58)  & 34.0 (59.45)  & 32.79 (58.18) & 29.42 (61.9)  & 33.12 (60.16) & 30.7 (62.58)  & 34.0 (59.45)  & 32.79 (58.18) \\
\multicolumn{1}{l|}{Mean f1 G}           & 18.44 (83.1)  & 30.03 (72.52) & 29.75 (70.88) & 33.34 (72.73) & 31.15 (72.15) & 18.44 (83.1)  & 30.03 (72.52) & 29.75 (70.88) & 33.34 (72.73) & 31.15 (72.15) \\
\multicolumn{1}{l|}{Mean f1 I}           & 10.25 (95.59) & 30.24 (89.98) & 17.04 (89.31) & 10.59 (94.02) & 30.38 (90.53) & 10.25 (95.59) & 30.24 (89.98) & 17.04 (89.31) & 10.59 (94.02) & 30.38 (90.53) \\
\multicolumn{1}{l|}{Mean f1 E}           & 40.21 (59.65) & 33.88 (60.28) & 35.25 (59.17) & 29.98 (63.68) & 33.99 (60.01) & 40.21 (59.65) & 33.88 (60.28) & 35.25 (59.17) & 29.98 (63.68) & 33.99 (60.01) \\
\multicolumn{1}{l|}{Mean f1 B}           & 30.57 (75.28) & 25.3 (73.05)  & 30.65 (71.04) & 30.86 (64.57) & 24.67 (71.8)  & 30.57 (75.28) & 25.3 (73.05)  & 30.65 (71.04) & 30.86 (64.57) & 24.67 (71.8)  \\
\multicolumn{1}{l|}{Mean f1 T}           & 30.22 (62.79) & 30.27 (61.7)  & 29.61 (63.41) & 30.93 (64.04) & 29.95 (61.29) & 30.22 (62.79) & 30.27 (61.7)  & 29.61 (63.41) & 30.93 (64.04) & 29.95 (61.29) \\
\multicolumn{1}{l|}{Mean f1 S}           & 29.24 (61.77) & 28.04 (58.79) & 27.46 (57.52) & 29.76 (54.36) & 27.79 (59.56) & 29.24 (61.77) & 28.04 (58.79) & 27.46 (57.52) & 29.76 (54.36) & 27.79 (59.56) \\
\multicolumn{1}{l|}{Mean f1 -}           & 28.82 (61.51) & 28.59 (60.08) & 28.57 (59.89) & 30.35 (63.16) & 28.25 (60.75) & 28.82 (61.51) & 28.59 (60.08) & 28.57 (59.89) & 30.35 (63.16) & 28.25 (60.75) \\
\multicolumn{1}{l|}{Mean iou H}          & 28.4 (61.54)  & 32.22 (59.93) & 29.94 (62.38) & 32.95 (59.09) & 31.96 (57.96) & 28.4 (61.54)  & 32.22 (59.93) & 29.94 (62.38) & 32.95 (59.09) & 31.96 (57.96) \\
\multicolumn{1}{l|}{Mean iou G}          & 17.75 (83.07) & 28.87 (72.45) & 28.59 (70.78) & 31.99 (72.63) & 29.98 (72.11) & 17.75 (83.07) & 28.87 (72.45) & 28.59 (70.78) & 31.99 (72.63) & 29.98 (72.11) \\
\multicolumn{1}{l|}{Mean iou I}          & 10.09 (95.59) & 29.58 (89.98) & 16.52 (89.31) & 10.31 (94.02) & 29.67 (90.51) & 10.09 (95.59) & 29.58 (89.98) & 16.52 (89.31) & 10.31 (94.02) & 29.67 (90.51) \\
\multicolumn{1}{l|}{Mean iou E}          & 39.1 (59.46)  & 32.76 (59.84) & 34.14 (58.81) & 29.02 (63.37) & 32.9 (59.53)  & 39.1 (59.46)  & 32.76 (59.84) & 34.14 (58.81) & 29.02 (63.37) & 32.9 (59.53)  \\
\multicolumn{1}{l|}{Mean iou B}          & 26.79 (74.03) & 21.86 (71.14) & 26.66 (68.8)  & 26.21 (61.53) & 21.2 (69.74)  & 26.79 (74.03) & 21.86 (71.14) & 26.66 (68.8)  & 26.21 (61.53) & 21.2 (69.74)  \\
\multicolumn{1}{l|}{Mean iou T}          & 28.36 (61.64) & 28.26 (59.8)  & 27.69 (61.71) & 28.9 (62.33)  & 27.92 (59.16) & 28.36 (61.64) & 28.26 (59.8)  & 27.69 (61.71) & 28.9 (62.33)  & 27.92 (59.16) \\
\multicolumn{1}{l|}{Mean iou S}          & 26.63 (59.27) & 25.15 (53.97) & 24.47 (52.46) & 26.41 (48.85) & 24.9 (54.31)  & 26.63 (59.27) & 25.15 (53.97) & 24.47 (52.46) & 26.41 (48.85) & 24.9 (54.31)  \\
\multicolumn{1}{l|}{Mean iou -}          & 26.17 (57.07) & 25.93 (54.13) & 25.83 (53.73) & 27.63 (57.52) & 25.56 (53.91) & 26.17 (57.07) & 25.93 (54.13) & 25.83 (53.73) & 27.63 (57.52) & 25.56 (53.91) \\
\multicolumn{1}{l|}{Segment recall H}    & 29.35 (99.64) & 33.14 (99.88) & 30.94 (99.93) & 33.77 (99.65) & 33.07 (99.91) & 29.35 (99.64) & 33.14 (99.88) & 30.94 (99.93) & 33.77 (99.65) & 33.07 (99.91) \\
\multicolumn{1}{l|}{Segment recall G}    & 19.18 (100.0) & 31.2 (100.0)  & 30.93 (100.0) & 34.89 (100.0) & 32.37 (99.85) & 19.18 (100.0) & 31.2 (100.0)  & 30.93 (100.0) & 34.89 (100.0) & 32.37 (99.85) \\
\multicolumn{1}{l|}{Segment recall I}    & 10.42 (100.0) & 30.79 (100.0) & 17.63 (100.0) & 10.91 (100.0) & 30.94 (100.0) & 10.42 (100.0) & 30.79 (100.0) & 17.63 (100.0) & 10.91 (100.0) & 30.94 (100.0) \\
\multicolumn{1}{l|}{Segment recall E}    & 41.28 (100.0) & 34.8 (99.36)  & 36.21 (99.8)  & 30.84 (100.0) & 35.0 (99.68)  & 41.28 (100.0) & 34.8 (99.36)  & 36.21 (99.8)  & 30.84 (100.0) & 35.0 (99.68)  \\
\multicolumn{1}{l|}{Segment recall B}    & 38.01 (99.69) & 31.7 (98.92)  & 38.05 (98.75) & 39.75 (99.02) & 31.2 (98.87)  & 38.01 (99.69) & 31.7 (98.92)  & 38.05 (98.75) & 39.75 (99.02) & 31.2 (98.87)  \\
\multicolumn{1}{l|}{Segment recall T}    & 32.81 (99.4)  & 33.09 (99.07) & 32.25 (98.3)  & 33.74 (98.46) & 32.86 (98.61) & 32.81 (99.4)  & 33.09 (99.07) & 32.25 (98.3)  & 33.74 (98.46) & 32.86 (98.61) \\
\multicolumn{1}{l|}{Segment recall S}    & 33.73 (99.03) & 33.02 (98.7)  & 32.5 (98.17)  & 35.39 (97.75) & 32.77 (98.16) & 33.73 (99.03) & 33.02 (98.7)  & 32.5 (98.17)  & 35.39 (97.75) & 32.77 (98.16) \\
\multicolumn{1}{l|}{Segment recall -}    & 33.29 (98.57) & 32.94 (98.35) & 33.07 (98.3)  & 34.87 (98.83) & 32.72 (98.52) & 33.29 (98.57) & 32.94 (98.35) & 33.07 (98.3)  & 34.87 (98.83) & 32.72 (98.52) \\
\multicolumn{1}{l|}{Segment precision H} & 39.24 (63.48) & 34.23 (61.03) & 34.32 (61.08) & 29.43 (59.98) & 33.02 (59.47) & 39.24 (63.48) & 34.23 (61.03) & 34.32 (61.08) & 29.43 (59.98) & 33.02 (59.47) \\
\multicolumn{1}{l|}{Segment precision G} & 2.64 (82.35)  & 4.22 (72.63)  & 4.4 (71.91)   & 4.56 (75.06)  & 4.16 (70.14)  & 2.64 (82.35)  & 4.22 (72.63)  & 4.4 (71.91)   & 4.56 (75.06)  & 4.16 (70.14)  \\
\multicolumn{1}{l|}{Segment precision I} & 1.96 (93.94)  & 2.25 (83.44)  & 1.2 (90.89)   & 0.68 (93.18)  & 1.68 (82.77)  & 1.96 (93.94)  & 2.25 (83.44)  & 1.2 (90.89)   & 0.68 (93.18)  & 1.68 (82.77)  \\
\multicolumn{1}{l|}{Segment precision E} & 34.7 (62.4)   & 23.29 (64.43) & 23.79 (64.45) & 17.3 (65.59)  & 20.66 (63.88) & 34.7 (62.4)   & 23.29 (64.43) & 23.79 (64.45) & 17.3 (65.59)  & 20.66 (63.88) \\
\multicolumn{1}{l|}{Segment precision B} & 1.28 (99.67)  & 1.08 (98.39)  & 1.18 (98.47)  & 1.2 (98.86)   & 0.96 (98.46)  & 1.28 (99.67)  & 1.08 (98.39)  & 1.18 (98.47)  & 1.2 (98.86)   & 0.96 (98.46)  \\
\multicolumn{1}{l|}{Segment precision T} & 9.58 (70.95)  & 9.36 (72.49)  & 9.05 (71.02)  & 9.59 (71.11)  & 9.21 (72.8)   & 9.58 (70.95)  & 9.36 (72.49)  & 9.05 (71.02)  & 9.59 (71.11)  & 9.21 (72.8)   \\
\multicolumn{1}{l|}{Segment precision S} & 8.12 (78.71)  & 6.83 (83.39)  & 6.37 (82.98)  & 7.36 (83.94)  & 6.53 (83.21)  & 8.12 (78.71)  & 6.83 (83.39)  & 6.37 (82.98)  & 7.36 (83.94)  & 6.53 (83.21)  \\
\multicolumn{1}{l|}{Segment precision -} & 15.47 (81.35) & 16.31 (81.4)  & 15.15 (80.22) & 16.62 (79.37) & 15.71 (83.16) & 15.47 (81.35) & 16.31 (81.4)  & 15.15 (80.22) & 16.62 (79.37) & 15.71 (83.16) \\
\multicolumn{1}{l|}{Segment f1 H}        & 30.95 (63.48) & 32.18 (61.08) & 31.04 (61.09) & 30.71 (60.02) & 32.09 (59.54) & 30.95 (63.48) & 32.18 (61.08) & 31.04 (61.09) & 30.71 (60.02) & 32.09 (59.54) \\
\multicolumn{1}{l|}{Segment f1 G}        & 4.54 (82.35)  & 7.27 (72.63)  & 7.56 (71.91)  & 7.96 (75.06)  & 7.24 (70.1)   & 4.54 (82.35)  & 7.27 (72.63)  & 7.56 (71.91)  & 7.96 (75.06)  & 7.24 (70.1)   \\
\multicolumn{1}{l|}{Segment f1 I}        & 3.17 (93.94)  & 4.06 (83.44)  & 2.2 (90.89)   & 1.28 (93.18)  & 3.15 (82.77)  & 3.17 (93.94)  & 4.06 (83.44)  & 2.2 (90.89)   & 1.28 (93.18)  & 3.15 (82.77)  \\
\multicolumn{1}{l|}{Segment f1 E}        & 35.43 (62.51) & 26.32 (64.46) & 27.09 (64.47) & 21.22 (65.66) & 24.68 (63.9)  & 35.43 (62.51) & 26.32 (64.46) & 27.09 (64.47) & 21.22 (65.66) & 24.68 (63.9)  \\
\multicolumn{1}{l|}{Segment f1 B}        & 2.46 (99.67)  & 2.06 (98.38)  & 2.25 (98.44)  & 2.31 (98.86)  & 1.85 (98.46)  & 2.46 (99.67)  & 2.06 (98.38)  & 2.25 (98.44)  & 2.31 (98.86)  & 1.85 (98.46)  \\
\multicolumn{1}{l|}{Segment f1 T}        & 14.37 (71.22) & 14.33 (73.02) & 13.94 (71.35) & 14.72 (71.68) & 14.18 (73.37) & 14.37 (71.22) & 14.33 (73.02) & 13.94 (71.35) & 14.72 (71.68) & 14.18 (73.37) \\
\multicolumn{1}{l|}{Segment f1 S}        & 12.6 (79.62)  & 10.96 (83.99) & 10.36 (83.6)  & 11.91 (84.44) & 10.68 (83.93) & 12.6 (79.62)  & 10.96 (83.99) & 10.36 (83.6)  & 11.91 (84.44) & 10.68 (83.93) \\
Segment f1 -                             & 20.45 (84.94) & 21.18 (86.73) & 20.3 (86.08)  & 22.14 (85.95) & 20.82 (88.68) & 20.45 (84.94) & 21.18 (86.73) & 20.3 (86.08)  & 22.14 (85.95) & 20.82 (88.68) \\ \bottomrule
\end{tabular}

}
}
}
\end{table}

\begin{table}[h!]{

\caption{{Table \ref{tab:sec-agreement} with averages over 10 datasets. We also report global averages over all 8 SSEs.}}
\label{tab:sec-agreement-avg}
\resizebox{\textwidth}{!}{
{
\begin{tabular}{@{}lccccccccc@{}}
\toprule
Metric            & H             & G             & I              & E             & B              & T             & S             & -             & Avg (HGIEBTS-) \\ \midrule
Mean recall       & 31.98 (97.32) & 29.62 (98.80) & 20.09 (99.79)  & 35.58 (98.20) & 36.20 (100.00) & 32.96 (97.26) & 33.85 (98.08) & 33.17 (96.81) & 31.68 (98.28)  \\
Mean precision    & 33.12 (49.99) & 27.76 (67.61) & 19.42 (90.31)  & 34.22 (47.65) & 24.55 (68.95)  & 28.49 (51.12) & 25.65 (50.71) & 26.79 (50.23) & 27.50 (59.57)  \\
Mean f1           & 32.01 (60.45) & 28.54 (74.28) & 19.70 (91.89)  & 34.66 (60.56) & 28.41 (71.15)  & 30.20 (62.65) & 28.46 (58.40) & 28.92 (61.08) & 28.86 (67.56)  \\
Mean iou          & 31.09 (60.18) & 27.44 (74.21) & 19.23 (91.88)  & 33.58 (60.20) & 24.54 (69.05)  & 28.23 (60.93) & 25.51 (53.77) & 26.22 (55.27) & 26.98 (65.69)  \\
Segment recall    & 32.05 (99.80) & 29.71 (99.97) & 20.14 (100.00) & 35.63 (99.77) & 35.74 (99.05)  & 32.95 (98.77) & 33.48 (98.36) & 33.38 (98.51) & 31.64 (99.28)  \\
Segment precision & 34.05 (61.01) & 4.00 (74.42)  & 1.55 (88.84)   & 23.95 (64.15) & 1.14 (98.77)   & 9.36 (71.67)  & 7.04 (82.45)  & 15.85 (81.10) & 12.12 (77.80)  \\
Segment f1        & 31.39 (61.04) & 6.91 (74.41)  & 2.77 (88.84)   & 26.95 (64.20) & 2.19 (98.76)   & 14.31 (72.13) & 11.30 (83.12) & 20.98 (86.48) & 14.60 (78.62)  \\ \bottomrule
\end{tabular}

}
}
}
\end{table}

{
\xhdr{{Secondary Structure Element (SSE) Agreement Results}}
We ran a new expert agreement evaluation against the 8 basic SSEs (from DSSP). The summary is in \ref{tab:sec-agreement-avg}. We do see above-random enrichment of SSEs in our tokens across all-metrics. The recall of existing SSEs is exceptional: $98.28\%$ block-level, $99.28\%$ segment-level, averaged over datasets and SSEs. This indicates the ability to recapitulate the 8 known elements, while the milder precision ($59.57$ block, $77.80$ segment) hints that GeoBPE goes beyond SSEs. As prior agreement results and case studies show, GeoBPE can find biologically meaningful regions (e.g. conserved homology, functional sites) and is not constrained to SSE boundaries, with the data dictating the exact high-level clusters.}

\section{Token Efficiency Metrics}\label{app:token-efficiency}
Let $v=\{1,\dots,K\}$ denote the codebook (size $K$). Given a corpus tokenized into a flat list of code indices, let $c_j$ be the count of code $j$ and $N=\sum_{j=1}^K c_j$ the total token count. We define the empirical unigram distribution
\[
p(j)=\frac{c_j}{N}\quad\text{for }j\in v.
\]

\xhdr{Utilization rate (UR)} UR measures how many distinct codes are actually used:
\[
\mathrm{UR} \;=\; \frac{1}{K}\,\bigl|\{\,j\in v:\,c_j>0\,\}\bigr| \;\in [0,1].
\]
We report UR in percent. UR is important for diagnosing codebook collapse, a well-known phenomenon in VQ-VAEs where only a small number of codes are actively used \cite{zhang2024codebook}. This creates a quantization bottleneck, handicapping the tokenizer's performance and efficiency \cite{yuan2025protein}.

\xhdr{Unigram entropy and perplexity} Using the Shannon entropy (natural logarithm),
\[
H \;=\; -\sum_{j\in v} p(j)\,\log p(j),
\qquad
\mathrm{PPL} \;=\; \exp\!\bigl(H\bigr).
\]
This \emph{codebook perplexity} reflects how uniformly codes are used (model-free, ignores sequence context).

\xhdr{Max-normalized perplexity} Because the maximum entropy at uniform is $\log K$ (hence $\mathrm{PPL}_{\max}=K$), we also report the scale-free ratio
\[
\widetilde{\mathrm{PPL}} \;=\; \frac{\mathrm{PPL}}{K}
\;=\; \exp\!\Bigl(\frac{H}{\log K}\cdot\log K\Bigr)\,\frac{1}{K}
\;=\; \exp\!\Bigl(H-\log K\Bigr)
\;\in (0,1].
\]

\begin{table}[h!]
\small
\caption{We evaluate token efficiency of \algo across varying $|\mathcal{V}|\in \{600,2500,6000\}$, as reported in Figure \ref{fig:pareto}, {with Hyperparameter Settings \ref{hparams:1}.}}

\label{tab:token-efficiency} 
\centering 

\begin{tabular}{@{}l|c|cc|cc|@{}}
\toprule
                        & \multicolumn{1}{r|}{}              & \multicolumn{2}{c|}{UR (\%)}                            & \multicolumn{2}{c|}{Perplexity}                         \\ \midrule
Method                  & \multicolumn{1}{r|}{Codebook Size} & \multicolumn{1}{r}{CAMEO} & \multicolumn{1}{r|}{CASP14} & \multicolumn{1}{r}{CAMEO} & \multicolumn{1}{r|}{CASP14} \\ \midrule
VQ-VAE                  & 512                                & 5.55                      & 5.60                        & 0.034                     & 0.0337                      \\
AminoASeed              & 512                                & 64.45                     & 68.87                       & 0.495                     & 0.5119                      \\
ESM3                    & 4096                               & 27.60                     & 32.10                       & 0.249                     & 0.2841                      \\
FoldSeek                & 20                                 & 99.00                     & 100.00                      & 0.755                     & 0.7435                      \\
ProToken                & 512                                & 69.88                     & 75.56                       & 0.537                     & 0.5697                      \\ \midrule
\multirow{3}{*}{\algo} & 600                                & 59.81                     & 39.48                       & 0.397                     & 0.403                       \\
                        & 2500                               & 58.24                     & 38.22                       & 0.274                     & 0.264                       \\
                        & 6000                               & 53.73                     & 31.30                       & 0.242                     & 0.222                       \\ \bottomrule
\end{tabular}

\end{table}

\begin{table}[h!]
\caption{We adopt the Small Structure Language Model evaluation protocol described in App. \ref{app:lm-eval}. We sample 100 PDB structures. {Best scTM and Designability scores are \textbf{bolded}. \textit{Underlined} methods follow the evaluat} {-ion protocol in App. \ref{app:scalability}.}}
\label{tab:lm-eval}
\resizebox{\textwidth}{!}{
\begin{tabular}{@{}lccccc@{}}
\toprule
\multicolumn{6}{c}{Small Structure Language Model Evaluation}                                                                                                                                                                         \\ \midrule
\multicolumn{1}{l|}{Method}                                      & \multicolumn{1}{r|}{Codebook Size} & scTM  & Designability (scTM $> 0.5$) & Diversity (1 - mean TM) & Uniqueness (TM=0.5) \\ \midrule
\multicolumn{1}{l|}{VQ-VAE}                                      & \multicolumn{1}{c|}{512}           & 0.205 & 1\%                                                                   & 0.752                   & 98\%                \\
\multicolumn{1}{l|}{AminoASeed}                                  & \multicolumn{1}{c|}{512}           & 0.186 & 1\%                                                                   & 0.476                   & 16\%                \\ \midrule
\multicolumn{1}{l|}{\multirow{3}{*}{\algo}}       & \multicolumn{1}{c|}{600}           & 0.268 & 3\%                                                                   & 0.768                   & 99\%                \\
\multicolumn{1}{l|}{}                                            & \multicolumn{1}{c|}{2500}          & 0.267 & 3\%                                                                   & 0.766                   & 99\%                \\
\multicolumn{1}{l|}{}                                            & \multicolumn{1}{c|}{6000}          & 0.277 & 4\%                                                                   & 0.763                   & 98\%                \\ \midrule
\multicolumn{1}{l|}{{\algo (x10 data)}}             & \multicolumn{1}{c|}{{600}}           & {0.376} & {12\%}                                                                  & {0.743}                   & {83\%}                \\ \midrule
\multicolumn{1}{l|}{{\algo (x10 data, x10 params)}} & \multicolumn{1}{c|}{{600}}           & {\textbf{0.405}} & {\textbf{21\%}}                                                                  & {0.731}                   & {76\%}                \\ \bottomrule
\end{tabular}
}
\end{table}


\section{Small Structure Language Model Evaluation}\label{app:lm-eval}

{\xhdr{Goals \& Aims} This section specifies a protocol used to compare the language modeling efficiency of \algo vs VQ-VAE tokens. The goal of SSLM was to create a small, isolated environment for language model integration. Modeling is \textbf{intentionally minimalistic}--we train a small decoder-only Transformer (7.3M) parameters over the same PDB splits used in Sec. \ref{sec:experiments} ($48$k structures). Fixing the same data splits, training, model architecture and hyperparameters (App. \ref{app:vqvae-sampling}), the aim of this experiment is to compare tokenizer options for a structure token language model. We emphasize the generation quality of the resulting models should be compared \textit{relatively}; all models trained would be insufficient for real-world backbone design tasks. To bridge the gap with large-scale models, we provide an orthogonal study on \textit{scalability} in App. \ref{app:scalability}, where we reran SSLM with $10$x more data and model parameters.}

\xhdr{{Setup}} {For \algo, we use the joint geometric vocabulary learned by \algo; for VQ-VAE,} {we use their codebook.} \algo+SSLM incorporates the mask constraints used at generation time (Alg.~\ref{alg:uncond-generate}) during training to ensure consistency between training and sampling. The same procedure is used for evaluating VQ-VAEs. For training and sampling, the only difference is dropping the mask constraints. For inference, the samples are passed through the VQ-VAE decoder to construct backbone coordinates instead of assembling the backbone directly via Alg. \ref{alg:dequantize-assemble}. This required considerably more resources, and we discuss how we implemented this in App. \ref{app:vqvae-sampling}.

\begin{algorithm}[h!]
\small
\caption{\textsc{GeoLM-Pretrain} — decoder-only next-token prediction on geometric tokens}
\label{alg:geom-lm}
\begin{algorithmic}[1]
\REQUIRE Corpus of proteins $\{t_\tau\}_{\tau=1}^T$ with final segmentations $\{\mathcal P^{(\tau)}\}$ and assigned medoids; joint vocabulary $\Sigma$ and tokenizers from Alg.~\ref{alg:build-joint-vocab}, \ref{alg:backbone-to-seq}; a decoder-only Transformer $\mathsf{Tr}_\theta:\Sigma^\ast\!\to\!\Delta^{|\Sigma|}$ with causal mask; special BOS/EOS (optional); training steps $S$, optimizer $\mathcal O$.
\ENSURE Trained parameters $\theta$.

\STATE \textbf{Dataset construction.} For each $\tau$, build $x^{(\tau)}=\textsc{BackboneToSequence}(t_\tau)$ (Alg.~\ref{alg:backbone-to-seq}). Let $L_\tau=|x^{(\tau)}|$.
\STATE \textbf{Objective.} For any sequence $x=(x_1,\ldots,x_L)$, define
\[
\mathcal{L}_{\mathrm{NTP}}(\theta;\ x)\;=\;-\sum_{t=1}^{L-1}\log p_\theta\big(x_{t+1}\ \big|\ x_{\le t}\big),\quad
p_\theta(\cdot|x_{\le t})=\mathrm{softmax}\big(\mathsf{Tr}_\theta(x_{\le t})\big).
\]
\STATE \textbf{Training loop.} 
\FOR{$s=1$ \TO $S$}
  \STATE Sample a minibatch $\mathcal B\subset\{1,\ldots,T\}$.
  \STATE $\displaystyle \mathcal{L}\leftarrow \frac{1}{|\mathcal B|}\sum_{\tau\in\mathcal B}\mathcal{L}_{\mathrm{NTP}}(\theta;\ x^{(\tau)})$.
  \STATE Update $\theta \leftarrow \mathcal O\big(\theta,\ \nabla_\theta \mathcal{L}\big)$.
\ENDFOR
\STATE \textbf{return} $\theta$.
\end{algorithmic}
\end{algorithm}
\begin{algorithm}[h!]
\small
\caption{\textsc{BuildEmpiricalPriors} — length prior and first-token prior}
\label{alg:length-start-priors}
\begin{algorithmic}[1]
\REQUIRE Training corpus of tokenized backbones $\{x^{(\tau)}=(x^{(\tau)}_1,\ldots,x^{(\tau)}_{L_\tau})\}_{\tau=1}^T$ constructed by Alg.~\ref{alg:backbone-to-seq} (motif, then $\theta,\omega,\phi$, repeating); valid sequence lengths satisfy $L_\tau\equiv 1\ (\mathrm{mod}\ 4)$ and end in a \emph{terminating} motif token.
\ENSURE Discrete priors $\Pi_L$ on lengths $K$ and $\Pi_{\mathrm{start}}$ on the first token.

\STATE \textbf{Length prior:} for every $K$ with $K\equiv 1\ (\mathrm{mod}\ 4)$, set
\[
\Pi_L(K)\ \propto\ \Big|\big\{\tau:\ L_\tau=K\big\}\Big|\ \ \text{and normalize } \sum_K \Pi_L(K)=1.
\]
\STATE \textbf{First-token prior:} over motif tokens only, set
\[
\Pi_{\mathrm{start}}(i)\ \propto\ \Big|\big\{\tau:\ x^{(\tau)}_1=i\big\}\Big|,\ \ \ i\in\Sigma_{\mathrm{med}};\ \ \ \sum_{i\in\Sigma_{\mathrm{med}}}\Pi_{\mathrm{start}}(i)=1.
\]
\STATE \textbf{return} $\Pi_L,\ \Pi_{\mathrm{start}}$.
\end{algorithmic}
\end{algorithm}
\begin{algorithm}[h!]
\small
\caption{\textsc{UnconditionalGeoLMGenerate} — motif/glue token generation}
\label{alg:uncond-generate}
\begin{algorithmic}[1]
\REQUIRE Trained decoder-only Transformer $\mathsf{Tr}_\theta$ with vocabulary $\Sigma$ from Alg.~\ref{alg:build-joint-vocab}; id blocks
\begin{align}
&\Sigma_{\mathrm{med}}=\{1,\ldots,M\}\\
&\Sigma_{\theta}=\{M{+}1,\ldots,M{+}B_\theta\}\\
&\Sigma_{\omega}=\{M{+}B_\theta{+}1,\ldots,M{+}B_\theta{+}B_\omega\}\\
&\Sigma_{\phi}=\{M{+}B_\theta{+}B_\omega{+}1,\ldots,M{+}B_\theta{+}B_\omega{+}B_\phi\};    
\end{align}
terminating-motif set $\Sigma_{\mathrm{term}}\subseteq\Sigma_{\mathrm{med}}$ (motifs in the length-2 bond-residue class); priors $\Pi_L,\Pi_{\mathrm{start}}$ (Alg.~\ref{alg:length-start-priors}); temperature $\tau>0$; maximum length $K_{\max}$; number of samples $S$.
\ENSURE $S$ unconstrained token sequences $\{x^{(s)}\}$ alternating motif and glue tokens and ending in a terminating motif.

\STATE Define the \textbf{type mask by position} ($t$ starts at $1$):
\[
t\equiv 1\! \!\!\!\pmod 4 \Rightarrow \text{motif }(\Sigma_{\mathrm{med}}),\quad
t\equiv 2 \Rightarrow \theta\ (\Sigma_{\theta}),\quad
t\equiv 3 \Rightarrow \omega\ (\Sigma_{\omega}),\quad
t\equiv 0 \Rightarrow \phi\ (\Sigma_{\phi}).
\]
\FOR{$s=1$ \TO $S$} \label{line:sample-loop}
  \STATE Sample a target cap $K^{\mathrm{cap}}\sim \Pi_L$ and set $K^{\mathrm{cap}}\leftarrow \min(K^{\mathrm{cap}},K_{\max})$.
  \STATE Sample the first token $x^{(s)}_1 \sim \Pi_{\mathrm{start}}$ (so $x^{(s)}_1\in\Sigma_{\mathrm{med}}$).
  \FOR{$t=2,3,\ldots,K^{\mathrm{cap}}$}
     \STATE Compute last-position logits $z_t=\mathsf{Tr}_\theta\big(x^{(s)}_{1:t-1}\big)$ with causal masking; let $v=|\Sigma|$.
     \STATE Build a \textbf{hard mask} $m\in\mathbb{R}^v$ initialized to $-\infty$ and set:
     \[
     \begin{cases}
       m_i\leftarrow 0 & \text{if } t\equiv 1\ (4)\ \text{and } i\in\Sigma_{\mathrm{med}},\\
       m_i\leftarrow 0 & \text{if } t\equiv 2\ (4)\ \text{and } i\in\Sigma_{\theta},\\
       m_i\leftarrow 0 & \text{if } t\equiv 3\ (4)\ \text{and } i\in\Sigma_{\omega},\\
       m_i\leftarrow 0 & \text{if } t\equiv 0\ (4)\ \text{and } i\in\Sigma_{\phi}.
     \end{cases}
     \]
     \STATE \textbf{Termination constraint at motif positions:}
       \begin{itemize}[leftmargin=*,topsep=0pt,noitemsep]
       \item If $t\equiv 1\ (4)$ and $t<K^{\mathrm{cap}}$, then \emph{disallow} early stop: set $m_i\leftarrow -\infty$ for all $i\in\Sigma_{\mathrm{term}}$.
       \item If $t\equiv 1\ (4)$ and $t=K^{\mathrm{cap}}$, then \emph{force} stop: set $m_i\leftarrow -\infty$ for all $i\in\Sigma_{\mathrm{med}}\setminus \Sigma_{\mathrm{term}}$.
       \end{itemize}
     \STATE Form masked logits $\tilde z_t = z_t + m$ and sample 
     \[
       x^{(s)}_t \sim \mathrm{Categorical}\!\left(\mathrm{softmax}\!\left(\tilde z_t/\tau\right)\right).
     \]
     \STATE \textbf{(Optional early stop)} If $t\equiv 1\ (4)$ and $x^{(s)}_t\in\Sigma_{\mathrm{term}}$, then \textbf{break}.
  \ENDFOR
\ENDFOR
\STATE \textbf{return} $\{x^{(s)}\}_{s=1}^S$.
\end{algorithmic}
\end{algorithm}
\begin{algorithm}[h!]
\small
\caption{\textsc{DequantizeAndAssemble} — from tokens to a full backbone}
\label{alg:dequantize-assemble}
\begin{algorithmic}[1]
\REQUIRE One generated sequence $x=(x_1,\ldots,x_L)$ from Alg.~\ref{alg:uncond-generate}; medoid dictionary 
$\big\{\mathrm{id}_{\mathrm{med}}(\kappa,j)\mapsto \Pi^{(\kappa)}_j\big\}$ where each prototype $\Pi^{(\kappa)}_j$ is a tuple of internal coordinates for a motif $\mathcal M$; glue bin edges $\{\beta^\theta_b\}_{b=0}^{B_\theta}$, $\{\beta^\omega_b\}_{b=0}^{B_\omega}$, $\{\beta^\phi_b\}_{b=0}^{B_\phi}$ (circular edges for angles, linear for lengths if used); canonical seed triad $(N_\star,\mathrm{CA}_\star,C_\star)$ and \textsc{SeedTriad}.
\ENSURE A complete backbone $\big\{(N_i,\mathrm{CA}_i,C_i)\in\mathbb{R}^3\big\}_{i=1}^{\widehat N}$ assembled from the decoded motifs and glues.

\STATE \textbf{Parse tokens into motifs and glues (fixed 4-cycle).}
Let the motif indices be $t\in\{1,5,9,\ldots\}$; write $x_t=\mathrm{id}_{\mathrm{med}}(\kappa^{(m)},j^{(m)})$ for $m=1,\ldots,M$ where $M=\frac{L+3}{4}$.
For each boundary $m=1,\ldots,M-1$, decode the three bins:
\[
b_\theta = x_{4m-2}-M,\quad
b_\omega = x_{4m-1}-(M{+}B_\theta),\quad
b_\phi   = x_{4m}-\big(M{+}B_\theta{+}B_\omega\big),
\]
and \textbf{dequantize} to the bin midpoints
\[
\bar\theta_m=\tfrac12(\beta^\theta_{b_\theta-1}{+}\beta^\theta_{b_\theta}),\quad
\bar\omega_m=\tfrac12(\beta^\omega_{b_\omega-1}{+}\beta^\omega_{b_\omega}),\quad
\bar\phi_m  =\tfrac12(\beta^\phi_{b_\phi-1}{+}\beta^\phi_{b_\phi}).
\]

\STATE \textbf{Recover internal coordinates.}
For each motif $m$, let $\Pi^{(\kappa^{(m)})}_{j^{(m)}}$ provide the internal bond lengths $\ell$, bond angles $\theta$, and dihedrals $(\psi,\omega,\phi)$ \emph{across its span} $\mathcal M^{(m)}$. 
Construct its internal entry$\to$exit transform $T^{\mathrm{int}}_{(m)}$ (product of link transforms $G_i$ inside the motif; see Preliminaries).

\STATE \textbf{Forward kinematics assembly.}
\STATE Initialize the entry frame by seeding the very first residue: $(N_1,\mathrm{CA}_1,C_1)\leftarrow \textsc{SeedTriad}(1)$ and form $F_1=(R_1,t_1)$ as in the Entry/Exit frame definition.
\STATE \textbf{Motif 1:} Traverse the links inside $\mathcal M^{(1)}$ using its internal coordinates to compute frames $F_2,\ldots,F_{q_1}$ (and atom positions) by repeated $G_i$ multiplications; set the current exit frame $F^{\mathrm{exit}}_{(1)}=F_{q_1}$.
\FOR{$m=1$ \TO $M-1$}
   \STATE \textbf{Boundary glue:} form the boundary transform
   \[
     T^{\mathrm{glue}}_{(m)}\ =\ G_{q_m}\big(\theta^{C\!N\!CA}=\bar\theta_m,\ \omega=\bar\omega_m,\ \phi=\bar\phi_m\big),
   \]
   i.e., the SE(3) map from the exit frame of $\mathcal M^{(m)}$ to the entry frame of $\mathcal M^{(m+1)}$ determined by the three dequantized glue angles (and adjacent bond lengths).
   \STATE Set the entry frame of $\mathcal M^{(m+1)}$ to 
   \[
     F^{\mathrm{entry}}_{(m+1)} \leftarrow T^{\mathrm{glue}}_{(m)}\; F^{\mathrm{exit}}_{(m)}.
   \]
   \STATE \textbf{Motif $(m{+}1)$:} traverse its internal links to produce all residue frames and atom positions; update $F^{\mathrm{exit}}_{(m+1)}$.
\ENDFOR

\STATE \textbf{Concatenate atoms.}
Collect the atoms from all traversals in order, yielding the backbone
$\big\{(N_i,\mathrm{CA}_i,C_i)\big\}_{i=1}^{\widehat N}$, where $\widehat N$ is the total number of residues implied by the concatenated motif spans (the final motif is guaranteed terminating).

\STATE \textbf{return} the complete backbone coordinates.
\end{algorithmic}
\end{algorithm}

\subsection{Data Preparation and Splits}
\xhdr{Tokenization} We construct the joint vocabulary $\Sigma$ (Alg.~\ref{alg:build-joint-vocab}) and convert each protein $t_\tau$ into a token sequence $x^{(\tau)}=(x^{(\tau)}_1,\ldots,x^{(\tau)}_{L_\tau})$ via \textsc{BackboneToSequence} (Alg.~\ref{alg:backbone-to-seq}). 

{We tokenize the validation/test sets (unseen during \algo training) via Algo. \ref{alg:tokenize}, a procedure} {analagous to BPE encoding.} Sequences alternate strictly $\text{motif} \;\rightarrow\; \theta \;\rightarrow\; \omega \;\rightarrow\; \phi \;\rightarrow\; \text{motif} \;\rightarrow\; \cdots$ and end with a \emph{terminating} motif token (length-2 bond–residue class), hence $L_\tau \equiv 1 \pmod 4$.

\xhdr{Splits} We partition proteins at the \emph{protein level} into train/validation/test (e.g., 80/10/10) to prevent leakage across chains.

\subsection{Training Objective with Structural Masks}
We train a causal Transformer $\mathsf{Tr}_\theta$ with teacher forcing. To enforce legality at each position $t$, we apply the same \emph{type mask by position modulo 4} used in generation (Alg.~\ref{alg:uncond-generate}):
\[
t \equiv 1 \!\!\!\!\!\pmod 4 \Rightarrow \Sigma_{\mathrm{med}},\quad
t \equiv 2 \Rightarrow \Sigma_\theta,\quad
t \equiv 3 \Rightarrow \Sigma_\omega,\quad
t \equiv 0 \Rightarrow \Sigma_\phi,
\]
setting logits for all other token types to $-\infty$ before the softmax.

\xhdr{Termination constraint at motif slots} At motif positions ($t \equiv 1 \pmod 4$), we impose the same termination rule as in Alg.~\ref{alg:uncond-generate}:
(i) if $t<L_\tau$, mask out terminating motifs $\Sigma_{\mathrm{term}}$;
(ii) if $t=L_\tau$, mask out non-terminating motifs.

\xhdr{Loss} With masks applied, the negative log-likelihood is
\[
\mathcal{L}_{\mathrm{NTP}}(\theta; x^{(\tau)})
\;=\;
-\sum_{t=1}^{L_\tau-1} \log p_\theta\!\left(x^{(\tau)}_{t+1}\mid x^{(\tau)}_{\le t}\right),
\qquad
p_\theta(\cdot\mid x_{\le t})=\mathrm{softmax}\!\big(\tilde z_t\big),
\]
where $\tilde z_t$ are masked logits. We optimize $\theta$ by minimizing the average NLL over the training set.

\xhdr{Early stopping} We select checkpoints by validation loss with a patience of 5 epochs.


\subsection{Unconditional Sampling for Qualitative Evaluation}
\xhdr{Empirical priors} We form the \emph{length prior} $\Pi_L$ and \emph{first-token prior} $\Pi_{\mathrm{start}}$ from the training corpus using \textsc{BuildEmpiricalPriors} (Alg.~\ref{alg:length-start-priors}). $\Pi_L$ is supported on legal lengths $K \equiv 1 \pmod 4$; $\Pi_{\mathrm{start}}$ is over $\Sigma_{\mathrm{med}}$.

\xhdr{Constrained generation} We sample with \textsc{UnconditionalGeoLMGenerate} (Alg.~\ref{alg:uncond-generate}): draw $K^{\mathrm{cap}}\!\sim\!\Pi_L$ (clipped by a maximum), sample the first motif $x_1\!\sim\!\Pi_{\mathrm{start}}$, then autoregress under the same positional type mask and termination constraint as training. Temperature and nucleus sampling are optional ablations.

\xhdr{\algo Dequantization and assembly} Generated token sequences are mapped to full backbones via \textsc{DequantizeAndAssemble} (Alg.~\ref{alg:dequantize-assemble}): medoid tokens decode to internal coordinates over their motif spans; glue-bin tokens decode to bin-midpoint angles; forward kinematics with the seeded entry frame yields atom coordinates $\{(N_i,\mathrm{CA}_i,C_i)\}_{i=1}^{\widehat N}$.

\subsection{Generative Quality Assessment}
\label{sec:gen-quality}

We evaluate unconditional samples produced by \textsc{UnconditionalGeoLMGenerate} (Alg.~\ref{alg:uncond-generate}) and assembled by \textsc{DequantizeAndAssemble} (Alg.~\ref{alg:dequantize-assemble}) using four structure-centric metrics based on TM‐score.%
\footnote{TM‐score is obtained with a standard implementation (e.g., TM-align); higher is better.}

\xhdr{Setup}
From each model we draw a fixed number of backbones $\{\widehat{\mathcal{B}}_n\}_{n=1}^N$ (legal lengths, terminal motif constraint). Unless noted, metrics are computed on these \emph{backbone geometries} without further post-processing.

\xhdr{(1) \textbf{scTM (self-consistency TM-score)}}
For each generated backbone $\widehat{\mathcal{B}}$, we (i) design a sequence $\widehat{s}$ with a standard inverse-folding model, (ii) predict a structure $\widetilde{\mathcal{B}}$ from $\widehat{s}$ using a single-structure predictor (e.g., ESMFold), and (iii) compute
\[
\mathrm{scTM}(\widehat{\mathcal{B}})\;=\;\mathrm{TM\text{-}score}\!\big(\widetilde{\mathcal{B}},\ \widehat{\mathcal{B}}\big).
\]
We report the mean scTM over the $N$ samples.

\xhdr{(2) \textbf{Designability (\% with scTM $>0.5$)}}
A backbone is deemed \emph{designable} if its self-consistency exceeds the canonical threshold $0.5$:
\[
\mathrm{Designability}\;=\;\frac{1}{N}\sum_{n=1}^{N}\mathbf{1}\!\left\{\mathrm{scTM}(\widehat{\mathcal{B}}_n)>0.5\right\}\times 100\%.
\]
This is the fraction of samples for which a designed sequence refolds back to the generated backbone at the fold level. 
{We adopt the same workflow from \citet{trippe2022diffusion,wu2024protein}, where ProteinMPNN \citet{dauparas2022robust} proposes $8$ sequences per structure and OmegaFold \citet{wu2022high} is used to compute scTM.}

\xhdr{(3) \textbf{Diversity (mean pairwise TM)}}
To quantify sample-to-sample diversity, we compute the mean pairwise TM-score across the set (lower is more diverse):
\[
\mathrm{Diversity}\;=\;\frac{2}{N(N-1)}\sum_{1\le i<j\le N}
\mathrm{TM\text{-}score}\!\big(\widehat{\mathcal{B}}_i,\ \widehat{\mathcal{B}}_j\big).
\]
(When $N$ is large, we estimate this by uniform sub-sampling of pairs.)

\xhdr{(4) \textbf{Uniqueness (\% non-duplicates at TM $<0.5$)}}
We mark a sample as \emph{unique} if its nearest neighbor among the other generated backbones has TM-score $<0.5$:
\[
\mathrm{Uniqueness}\;=\;\frac{1}{N}\sum_{n=1}^{N}
\mathbf{1}\!\left\{\max_{m\ne n}\mathrm{TM\text{-}score}\!\big(\widehat{\mathcal{B}}_n,\widehat{\mathcal{B}}_m\big)\,<\,0.5\right\}\times 100\%.
\]
This measures the proportion of samples that are not near-duplicates under a fold-level threshold.

\xhdr{Reporting} For each model we report the four metrics above on the same set size $N$ (and the same sampling priors and temperature). Codebook size and token perplexity are \emph{not} used in these downstream comparisons.

\subsection{Generated Backbones}
\begin{figure}[h!]
  \centering
  \begin{subfigure}[t]{0.49\linewidth}
    \centering
    \includegraphics[width=\linewidth]{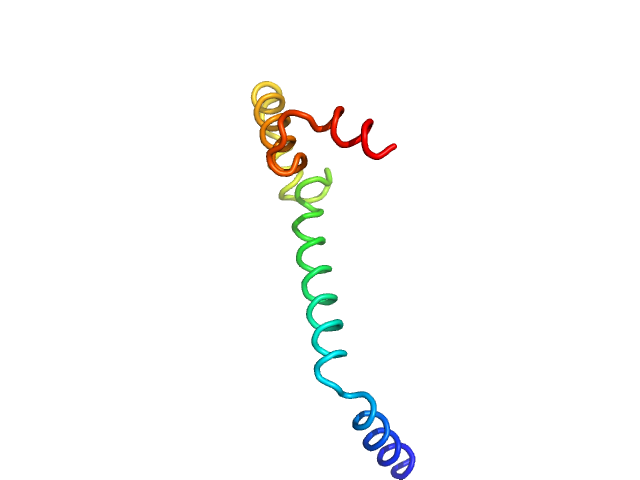}
    \caption{Generated backbone with scTM score 0.50}
    \label{fig:gen-1}
  \end{subfigure}
  \hfill
  \begin{subfigure}[t]{0.49\linewidth}
    \centering
    \includegraphics[width=\linewidth]{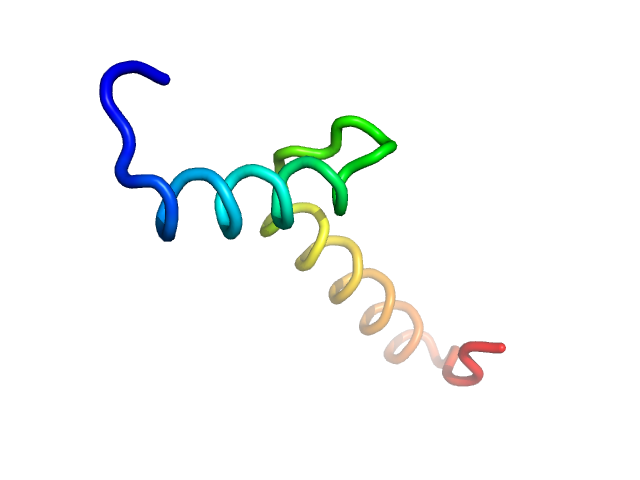}
    \caption{Generated backbone with scTM score 0.56}
    \label{fig:gen-2}
  \end{subfigure}
  \caption{We visualize two backbones generated by {\algo SSLM-Eval (with default settings in App. \ref{app:hyperparameters})}.}
  \label{fig:gen}
\end{figure}

In Fig. \ref{fig:gen-1}, we see a long, well-structured and assembled $\alpha$-helix, which is one of the most common and stable secondary structures in proteins. The curved helical cap at the top resembles a common N-terminal capping motif, which often stabilizes helices through hydrogen bonding networks or electrostatic interactions. Such elongated $\alpha$-helices are commonly found in transmembrane helices or coiled-coil domains which are involved in dimerization and DNA-binding. The overall curvature and spatial continuity also suggest potential compatibility with membrane proteins or structural scaffolds, especially behave as substance binding receptors as well as ion channels.

In Fig. \ref{fig:gen-2}, we see a structure that resembles DNA-binding motifs or cytokine folds, which are quite well-known for cellular signaling or regulation. The geometric density of this structure also suggests a pre-organized hydrophobic core, which is critical for proper folding and stability in the cytoplasmic environment. This structure exhibits a compact bundle of helices with apparent crossing angles which are similar to some small globular domains in common protein structures. The folding appears non-linear but in a quite controlled, manner which suggests potential tertiary structure forming interactions such as hydrophobic-hydrophobic interaction.

\subsection{Implementation Details}\label{app:vqvae-sampling}

\xhdr{SSLM-Eval \algo implementation and hardware details} We train a small autoregressive Transformer on discretized geometry tokens. We use a hidden size $d_{\text{model}}=256$, $L=8$ Transformer layers with \textsc{gelu} activations, $H=8$ attention heads, and feed-forward width $d_{\text{ff}}=1024$. Token and positional embeddings are summed, a LayerNorm is applied before the classifier, and the output projection is weight--tied to the token embedding. A causal attention mask enforces left-to-right prediction. Sequences are padded to a dataset-dependent maximum length (the 95th percentile of training lengths by default). We optimize cross-entropy loss with Adam (learning rate $1\times10^{-4}$), batch size $32$, for up to $100$ epochs with early stopping on validation perplexity. For unconditional generation, we sample $100$ sequences at temperature $1.0$, drawing target lengths from the empirical length prior (restricted to valid lengths by construction) and the first token from the empirical start-token prior; decoding proceeds token-by-token under the causal mask. On a single GPU, one epoch takes just under $10$ mins and converges in $\approx 60$ epochs (can vary across tokenizer settings). The data splits are the same as those for pretraining (see App. \ref{app:dataset}) -- $33992, 3810$ training/validation structures for \algo.

\xhdr{SSLM-Eval VQ-VAE implementation and hardware details} We extend the distributed Lightning setup of \citet{yuan2025protein} with a self-contained evaluation step at the end of each validation epoch. Using 4 ranks, each GPU accumulates the epoch’s quantized token sequences from training and validation; these are gathered and passed to a lightweight auxiliary trainer that uses the same SSLM-Eval script as \algo and hyperparameters. After convergence, we sample 100 new token sequences, decode them with the VQ-VAE decoder into backbone coordinates, and write PDBs to a directory named by the current epoch. We compute all non-SCTM metrics locally, then distribute a heavier SCTM evaluation across ranks on sharded PDB subsets. Each rank produces its shard’s results, and rank-0 merges them into a single summary that is logged to the trainer.

{
\section{Scaling \algo to Larger Datasets and Models}\label{app:scalability}
\xhdr{Goals \& Aims} The relatively minimalistic design of \algo SSLM in App. \ref{app:lm-eval} still begs the question whether \algo reliably scales to large datasets (e.g. ESM3 scale) as both a \emph{tokenizer} (\algo only) and the foundational component of a language model (\algo + SSLM). Thus, we attempt to bridge this gap by (i) tokenizing a $10$x larger dataset and benchmarking wall times, and (ii) training a $10$x larger SSLM model on the $10$x larger tokenized corpus.
}

{
\subsection{Tokenizing a $10$x Larger Dataset of Predicted Structures}

\xhdr{Setup} We downloaded the 550K Swiss-Prot structure predictions from AlphaFold DB \cite{uniprot, varadi2022alphafold}, a $>10$x increase from our PDB pretraining dataset. We adopt the pretrained $|V|=600$ tokenizer used in the paper (Fig. \ref{fig:pareto}, Tables \ref{tab:token-efficiency} \& \ref{tab:lm-eval}) as a baseline (i.e. tokenizers with larger $|V|$ will achieve lower distortion at the tradeoff of slower throughput).
}

{
\xhdr{Evaluation} We log both wall-time and thoroughput taken to tokenize all $550$k structures. We also report the distortion against AlphaFold DB predictions. We split into five 110K increments, and requested 5 jobs with 20 cores each. Each job writes each tokenized structure to a file, which allows us to log the running throughput from start to finish.
}

\begin{table}[h!]

\centering

\caption{{We report throughput and distortion tokenizing 550k Swiss-Prot predicted structures with 5 jobs} {of 20 cores each.}}
\label{tab:swissprot-tokenization}
{
\begin{tabular}{@{}lllllll@{}}
\toprule
                            & \multicolumn{6}{l}{Split (110K increments)}               \\
                            & 1     & 2     & 3     & 4     & 5     & Total (5 splits). \\ \midrule
Avg. Throughput (files/min) & 35.54 & 45.40 & 35.45 & 36.64 & 38.29 & 191.32            \\
RMSD                        & 1.52  & 1.54  & 1.54  & 1.53  & 1.53  & 1.53              \\
LDDT                        & 0.79  & 0.79  & 0.79  & 0.79  & 0.79  & 0.79              \\ \bottomrule
\end{tabular}
}
\end{table}

{
\xhdr{Results} 
The results are shown in Table. \ref{tab:swissprot-tokenization}. With pooled avg. throughput of 191.32, all $550$k structures were tokenized in $\sim 2$ days. Discrepancies in throughput between jobs likely explained by node traffic. We see all $5$ splits achieve the same LDDT ($0.79$) and within $0.01$ RMSD of the average RMSD over all 5 splits ($1.53$). These are comparable to the tokenizer's OOD test set distortions in Fig. \ref{fig:pareto} ($1.53$ RMSD, $0.72$ on CASP).
}

{
\xhdr{Conclusion}
Since \algo has shown strong OOD generalization from our findings in Sec. \ref{sec:experiments}, it is expected to not degrade in performance. Thus, the right approach is tokenization (Alg. \ref{alg:tokenize}) rather than retraining \algo from scratch. In contrast with \algo learning, tokenization is an \textit{embarrassingly parallel} procedure that easily scales with the number of cores. With $100$ cores, the entire process finished in $\sim 2$ days, or $200$ CPU days. Further scaling by another $10$x would only take $\sim 20$ days with $100$ CPUs, making \algo a scalable solution for tokenizing large databases of predicted structures.
}

{
\subsection{Structure Langauge Modeling at a $10$x Larger Scale}
Once we have the tokenized dataset of $550$k Swiss-Prot structures, we train a larger model and probe whether generation quality of \algo SSLM follows the expected improvements from scaling.
}

{
\xhdr{Setup} We increased our Transformer to $\sim 10$x parameters ($7.3M\rightarrow 65.9M$). We do so by widening the Transformer layer and deepening the model: $d_{\text{model}} \leftarrow 2d_{\text{model}}=512 $, $L\leftarrow 2.5L=20$ layers.  $H\leftarrow 2*H=16$ attention heads, $d_{\text{ff}}\leftarrow 2\cdot d_{\text{ff}}=2048$. The rest of SSLM remains the same (App. \ref{app:lm-eval}).
}

{
\xhdr{Evaluation} We used the same evaluation protocol (50-128 AAs, 10 each) of works such as ProtDiff \cite{trippe2022diffusion} and FoldingDiff \cite{wu2024protein}. Note that App. \ref{app:lm-eval} did not follow this protocol; we sampled from the size prior of our pretraining dataset (Alg. \ref{alg:length-start-priors}); the average generated protein was $\sim$ 214 AAs.
}

{
\xhdr{Results} GeoBPE+SSLM with $10$x more data and $10$x more parameters \textbf{achieves an average scTM of $0.4051$, with 20.8\% being Designable (scTM $>0.5$)}. This is notably higher than ProtDiff ($11.8\%$) and FoldingDiff ($14.2\%$). Scaling only by $10$x more data also delivers a respectable scTM of $0.376$, highlighting both scaling dimensions are throttles for generative performance. Uniqueness/diversity also remain high ($76.4\%$ and $0.73$).
}

{
\xhdr{Conclusion} This result confirms that \algo behaves according to scaling law expectations of language modeling. The significant increase in designable backbones $(4\%\rightarrow 21\%)$ from simply using more data and parameters justifies further scaling of data and training resources. We hope future works can adopt \algo as a foundational component in future large-scale models and explore the full potential of large-scale protein structure language model development.
}

\section{Expert Agreement Metrics}\label{app:expert-agreement-eval}
Our method segments a protein sequence into $M$ contiguous residue spans
$P_j=[p_j,q_j]$ with $q_j+1=p_{j+1}$ for $j=1,\dots,M-1$. We compare these
segments against $N$ ground-truth domain annotations $D_i=[s_i,e_i]$.
All sets below are sets of integer residue indices and $|\cdot|$ denotes
cardinality (length in residues). We report (i) \emph{domain-level} alignment
quality for each true domain using the best consecutive block of predicted
segments, and (ii) \emph{segment-level} detection statistics at an
Intersection-over-Union (IoU) threshold~$\tau$.
This combination captures both \emph{how well} each domain is covered and
\emph{how economically} the predicted segments explain the annotations, while
remaining robust to small boundary jitter.

\xhdr{Annotation source} Ground-truth domains come from \textbf{CATH FunFams} \cite{das2015cath}. They are functional families
defined by \emph{profile HMM} hits trained on primary-sequence data \cite{das2015functional, das2015cath}. Our evaluation
thus measures how well the predicted segmentation aligns with functionally
coherent families derived from sequence-based HMM models.

In our setting, individual predicted segments tend to be substantially shorter
than the curated domain annotations. A naive one-to-one comparison would
systematically penalize predictions that must be \emph{combined} to cover a
domain. To ensure a fair comparison, for each $D_i$ we first select the
single best \emph{consecutive} block of predicted segments $S_i$ that maximizes IoU
with $D_i$ (below), then compute per-domain scores and \emph{macro-average}
them so that each domain contributes equally, independent of its length.

\xhdr{Notation and best block per domain} For domains $D_i=[s_i,e_i]$ ($i=1{:}N$) and predicted segments $P_j=[p_j,q_j]$ ($j=1{:}M$, with $q_j{+}1=p_{j+1}$), define
\[
(a_i,b_i)\in\arg\max_{1\le m\le n\le M}\frac{\bigl|D_i\cap\bigcup_{k=m}^{n}P_k\bigr|}{\bigl|D_i\cup\bigcup_{k=m}^{n}P_k\bigr|},\qquad
S_i:=\bigcup_{k=a_i}^{b_i}P_k.
\]
(Ties may prefer the shortest $S_i$ or fewest segments.)

\xhdr{Domain-level scores (macro)}
Let $\mathrm{ov}_i=\lvert D_i\cap S_i\rvert$, $\lvert D_i\rvert=e_i-s_i+1$, $\lvert S_i\rvert=\sum_{k=a_i}^{b_i}(q_k-p_k+1)$. Then
\[
\mathrm{Recall}_i=\frac{\mathrm{ov}_i}{\lvert D_i\rvert},\quad
\mathrm{Precision}_i=\frac{\mathrm{ov}_i}{\lvert S_i\rvert},\quad
F_{1,i}=\frac{2\,\mathrm{Recall}_i\,\mathrm{Precision}_i}{\mathrm{Recall}_i+\mathrm{Precision}_i},\quad
\mathrm{IoU}_i=\frac{\mathrm{ov}_i}{\lvert D_i\rvert+\lvert S_i\rvert-\mathrm{ov}_i}.
\]
Macro-averages:
\[
\overline{\mathrm{Recall}}=\tfrac{1}{N}\sum_i\mathrm{Recall}_i,\quad
\overline{\mathrm{Precision}}=\tfrac{1}{N}\sum_i\mathrm{Precision}_i,\quad
\overline{F_1}=\tfrac{1}{N}\sum_i F_{1,i},\quad
\overline{\mathrm{IoU}}=\tfrac{1}{N}\sum_i\mathrm{IoU}_i.
\]
\emph{Interpretation:} recall rewards coverage; precision rewards compactness of $S_i$; $F_1$ balances both; IoU is thresholdable and scale-invariant.

\xhdr{Segment-level detection at IoU threshold $\tau$}
Let $\mathcal{U}=\bigcup_{i:\,\mathrm{IoU}_i\ge\tau}\{a_i,\dots,b_i\}$. Define
\[
\mathrm{SegPrec}=\frac{\lvert\mathcal{U}\rvert}{M},\qquad
\mathrm{SegRec}=\frac{\lvert\{\,i:\mathrm{IoU}_i\ge\tau\,\}\rvert}{N},\qquad
\mathrm{SegF_1}=\frac{2\,\mathrm{SegPrec}\,\mathrm{SegRec}}{\mathrm{SegPrec}+\mathrm{SegRec}}.
\]
\emph{Interpretation:} $\mathrm{SegPrec}$ penalizes unused segments; $\mathrm{SegRec}$ penalizes missed/poorly aligned domains. Sweeping $\tau$ yields a PR curve.

\xhdr{Randomization baseline and reporting}
Using $1000$ uniform random partitions into $M$ contiguous spans of the same sequence, recompute all metrics under the same best-block protocol and average over runs. We report using the format:
\[
\text{ours}\;(\text{random-avg}),
\]
e.g., $\overline{\mathrm{IoU}}=0.47\,(0.18)$.

\xhdr{Notes} In degenerate cases (e.g., $|S_i|=0$ or a zero denominator), we adopt the
standard convention of returning $0$ for the affected ratio or $F_1$ term.

\medskip
Together, the domain-level
(overlap-quality) metrics and the segment-level (parsimony and coverage) metrics
directly test the two desiderata of protein-domain segmentation:
(i) accurate coverage of each domain with minimal spillover, and
(ii) a parsimonious set of segments that explain as many domains as possible.
Macro-averaging after selecting the best block per domain ensures fairness when
predicted segments are shorter than annotated domains, and the permutation
baseline quantifies how far performance rises above chance given the same $M$.

\section{Expert Case Studies}\label{app:expert-case-study}

\begin{wrapfigure}{r}{0.3\textwidth}
    \includegraphics[width=\linewidth]{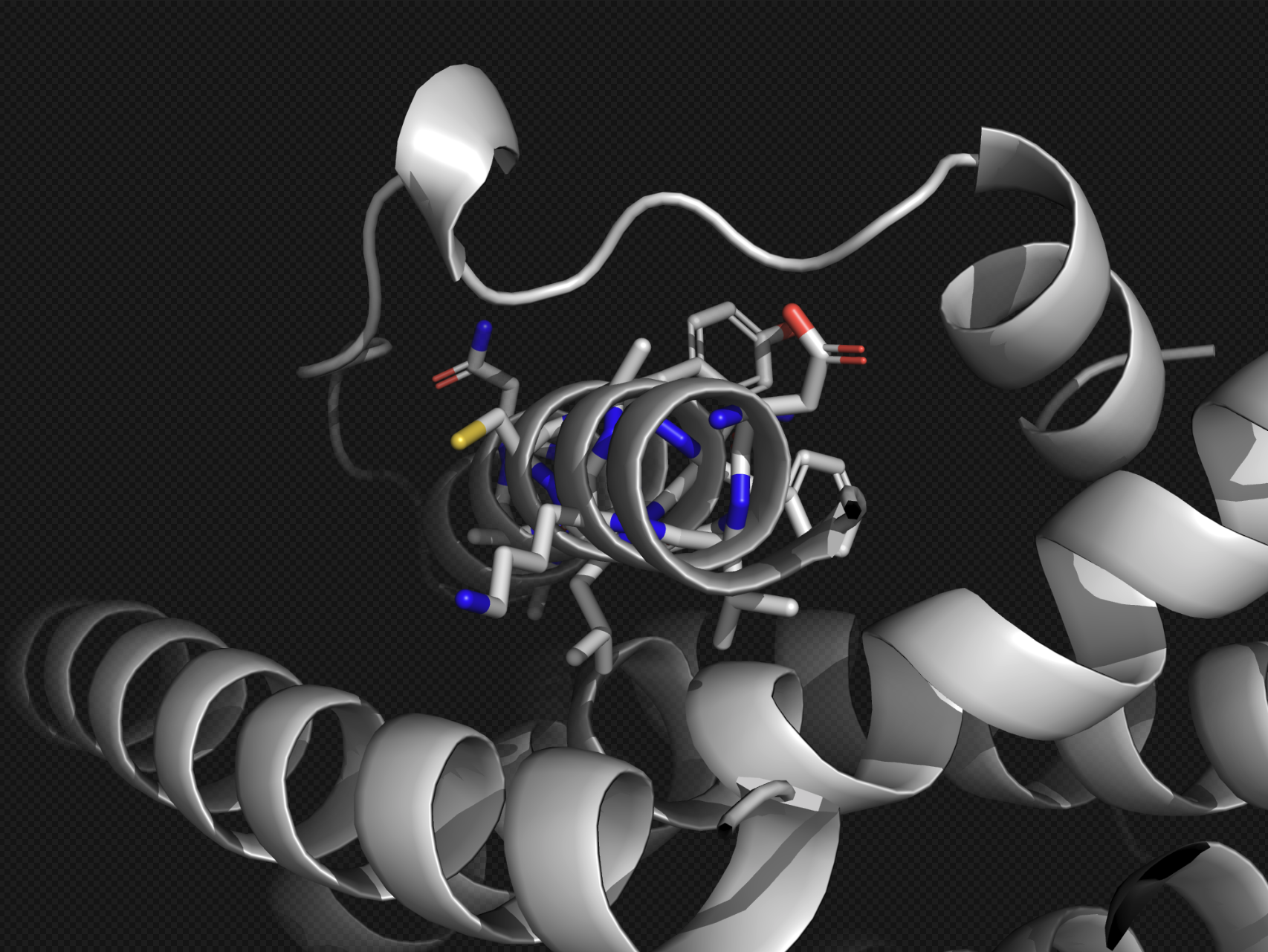}
    \caption{An exemplary \algo token spans backbone atoms of an alpha helix (colored).}
    \label{fig:lilmotif}
\end{wrapfigure}

\xhdr{Individual Tokens Correspond to Secondary Structures} Figure \ref{fig:lilmotif} is an example of a single token \algo discovers. It features an alpha helix that includes aromatic cage (formed by Tryptophan / Tyrosine) and hydrogen bonding residue. It can be a common structure in Nucleotide-recognition domains, especially the hydrogen bond donors/acceptors can serve for specific molecular recognition (e.g., methylated lysines, nucleotide bases or acetyl groups) as well as Neurotransmitter receptors. From interpretation, this motif is functionally specific. 
It can serve as ligand binding pocket, which is tightly packed and evolutionarily conserved. This could behave significantly in substance recognition. The tightly packed helical scaffold in this separated motif is likely stabilizing the motif’s geometry and ensuring specificity. Motif-scaffold synergy can also help to define a structure’s rigidity and flexibility.

\begin{figure}[h!]
\centering
\setlength{\fboxsep}{4pt}

    {%
      \begin{minipage}{0.98\linewidth}    
    \fcolorbox{black!20}{white}{\includegraphics[width=0.99\linewidth]{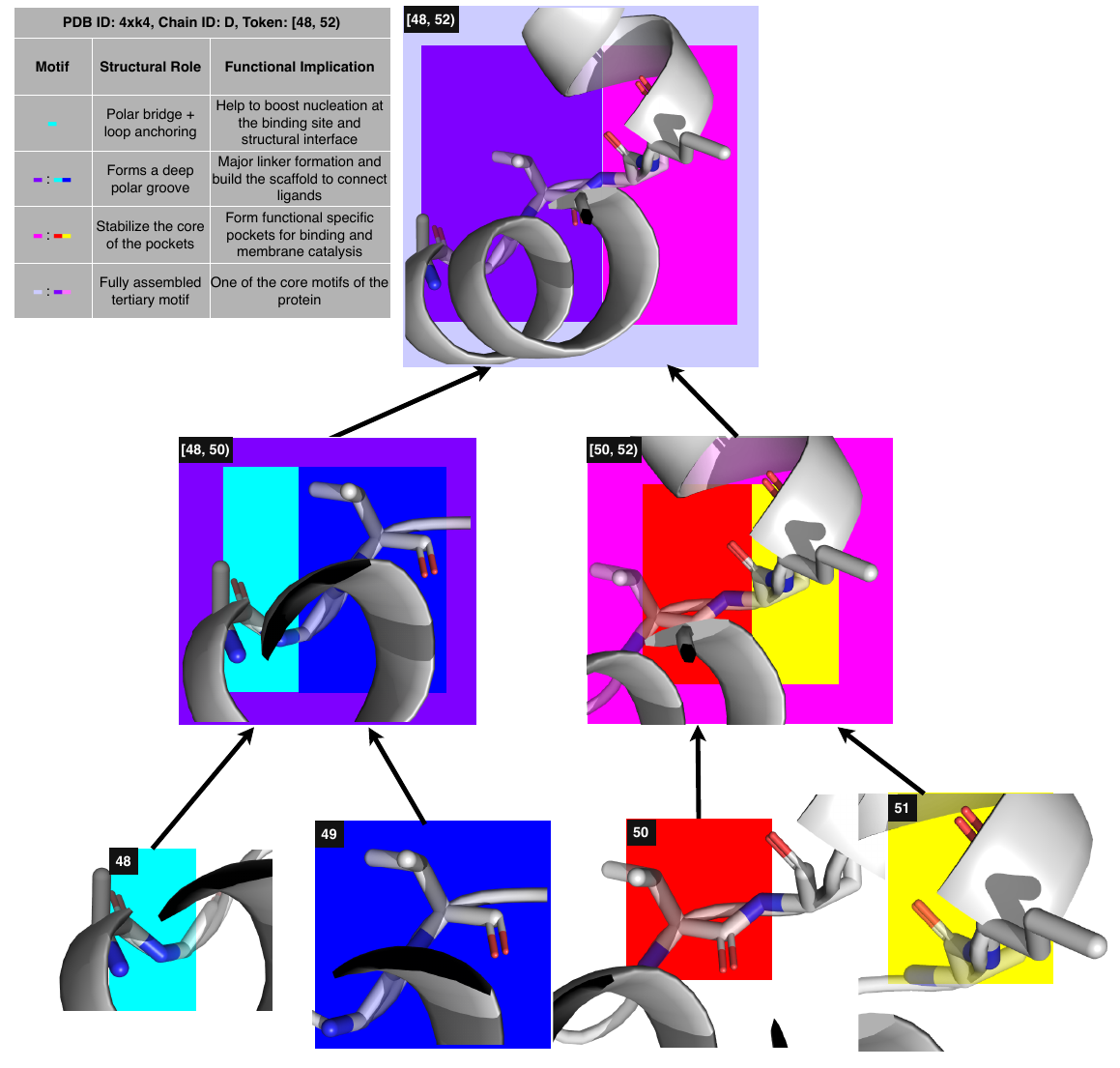}}
      \end{minipage}%
    }
    \caption{Chain D of PDB 4xk4. Hierarchical Merge Tree for Token [48, 52). \algo arrived at this token by merging [48, 49) with [49, 50), [50, 51) with [51, 52), and [48, 50) with [50, 52).}
    \label{fig:case-study-4xk4}
\end{figure}

\xhdr{Merge Hierarchy of \algo Reflects Combination of Secondary Structures for Driving Function}

\textbf{Figure \ref{fig:case-study-4xk4}.} 4xk4 is the human mitochondrial carrier protein SLC25A20 (carnitine/acylcarnitine translocase). It’s a transmembrane transport protein within the mitochondrial inner membrane, responsible for shuttling carnitine and acylcarnitine molecules across the membrane. This a process critical to fatty acid oxidation and energy metabolism. The core motif that the algorithm separated out contain three similar domains, each with two transmembrane helices and a loop. It appears to lie deep within the transmembrane domain, forming part of the central binding cavity. From this know-how information, the 48-4 motif is really significant in the following three aspects:
\begin{enumerate}[leftmargin=*,topsep=0pt,noitemsep]
    \item It will serve for substrate recognition where the internal polar residues bind to the acylcarnitine or carnitine head group via ionic and hydrogen bonds. It will also alter the transition state of the during the transport cycles. For example, this motif can play a role in shift conformation between open-to-cytoplasm and open-to-matrix states.
    \item We also observe similar motifs are found in other SLC25 family members (e.g., ADP/ATP carriers), indicating a shared mechanism of transport.
    \item While the broader transmembrane region is dominated by repetitive helices, this localized motif exhibits a unique composition of diverse side chains, polar residues, and tightly packed interactions, reinforcing its functional specificity.
\end{enumerate}

\begin{figure}[h!]
\centering
\setlength{\fboxsep}{4pt}

    {%
      \begin{minipage}{0.98\linewidth}        
    \fcolorbox{black!20}{white}{\includegraphics[width=0.99\linewidth]{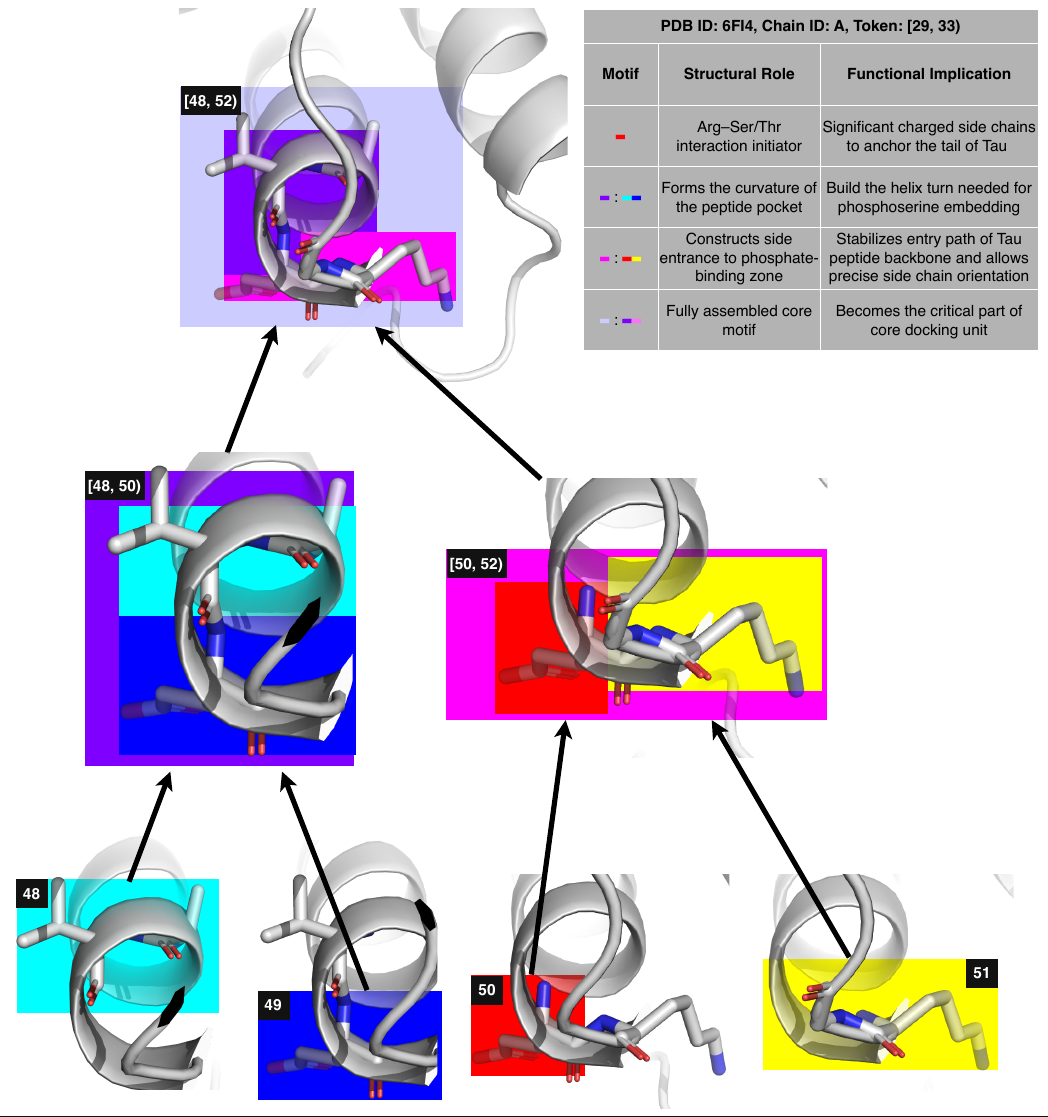}}
      \end{minipage}%
    }           
    \caption{Chain A of PDB 6FI4. Hierarchical Merge Tree for Token [29, 33). \algo arrived at this token by merging [29, 30) with [30, 31), [31, 32) with [32, 33), and [29, 31) with [31, 33).}
    \label{fig:case-study-6FI4}
\end{figure}

\textbf{Figure \ref{fig:case-study-6FI4}.} 6FI4 is the crystal structure of a hybrid peptide composed of a C-terminally modified Tau protein segment bound to the human 14-3-3$\sigma$ protein, solved at 2.0 Å resolution via X-ray crystallography. 14-3-3 proteins are a family of conserved regulatory molecules that bind phosphoserine/phosphothreonine-containing motifs on target proteins and are central to cell cycle control, apoptosis, transcriptional regulation, and signal transduction. The hybrid peptide mimics Tau phosphorylation, which is relevant to neurodegenerative disease pathology like Alzheimer's disease. 
From this know-how information, the 29-4 motif is significant in the following two aspects:

\begin{enumerate}[leftmargin=*,topsep=0pt,noitemsep]
    \item Phosphopeptide recognition and improve the binding Stability: This motif orchestrates recognition of the Tau-derived phosphoserine motif via a precise network of hydrogen bonds and electrostatic complementarity. The Lys/Arg residues (seen in blue) form salt bridges with phosphate groups, stabilizing the interaction. 
    \item The structure shared recognition fold across the 14-3-3 protein family: This motif, with its basic side chain tunnel and surrounding helices, represents a canonical recognition site. Similar structural motifs are observed in all 14-3-3 isozymes when binding phosphoproteins and will serve for post-translational modification signaling.
\end{enumerate}

\section{{Performance Across Protein Fold Types}}
{\xhdr{Setup} We evaluate robustness of \algo by computing distortion as defined in Sec. \ref{sec:pst} across fold types, unusualness, and size with per-chain metrics. Structural "unusualness" is computed from Foldseek’s TM-align mode as $100\times(1-\mathrm{TM})$ using the best hit against PDB. We also attach coarse labels: \emph{categories} by best-hit TM-score (\textbf{Near-identical} $\geq 0.90$, \textbf{Same fold} $0.50$–$0.89$, \textbf{Distant} $0.30$–$0.49$) and \emph{flags} indicating \textbf{very\_small} (chains with $<70$ residues) or \textbf{weak\_coverage} (low query coverage).}

\begin{figure*}[t]
  \centering
  \setlength{\fboxsep}{4pt}
            {%
      \begin{minipage}{0.98\linewidth}        
    \fcolorbox{black!20}{white}{
  \begin{subfigure}[t]{0.48\textwidth}
    \centering

    \includegraphics[width=\linewidth]{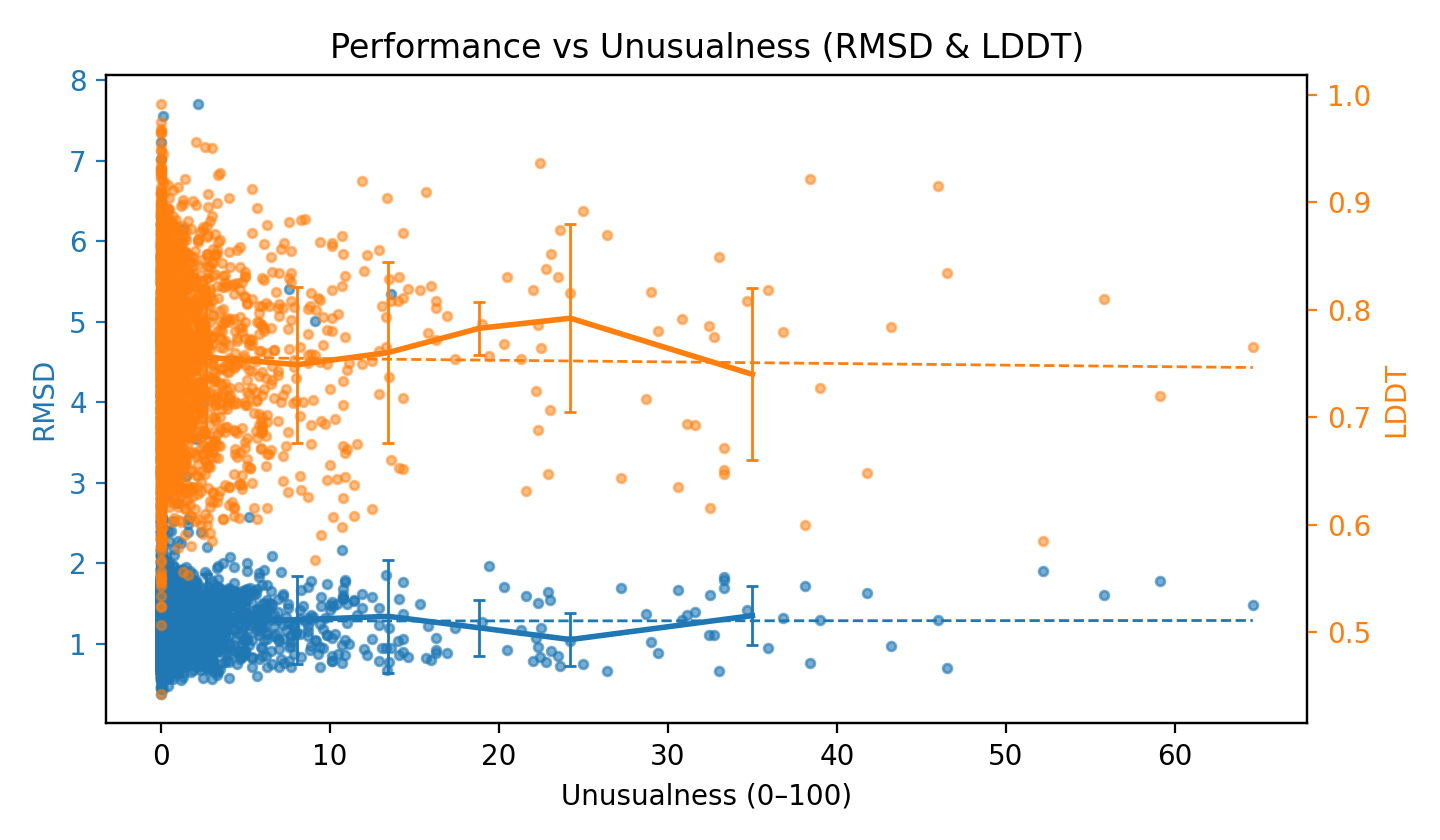}

    \caption{Dual-axis scatter vs.\ unusualness. RMSD (left, blue) and LDDT (right, orange), with bin means $\pm$ std and trend lines.}
    \label{fig:scatter-unusualness}
  \end{subfigure}\hfill
  \begin{subfigure}[t]{0.48\textwidth}
    \centering
    \includegraphics[width=\linewidth]{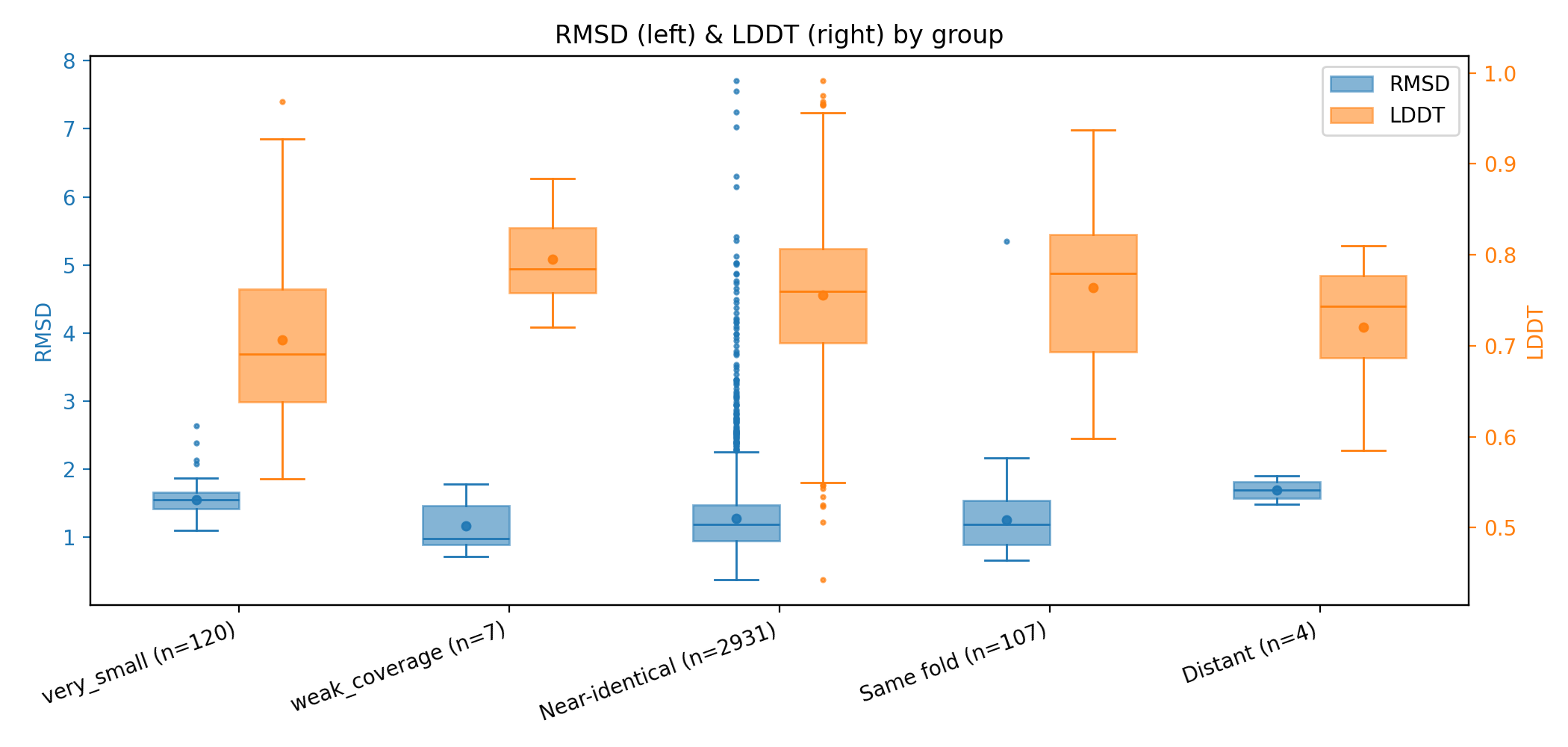}
    \caption{Group boxplots across \texttt{very\_small}, \texttt{weak\_coverage}, \texttt{Near-identical}, \texttt{Same fold}, \texttt{Distant}. Dual-$y$ boxes: RMSD (left, blue), LDDT (right, orange).}
    \label{fig:box-group}
  \end{subfigure}
}\end{minipage}}
  \vspace{0.6em}
            {%
      \begin{minipage}{0.98\linewidth}        
    \fcolorbox{black!20}{white}{
  \begin{subfigure}[t]{0.48\textwidth}
    \centering
    \includegraphics[width=\linewidth]{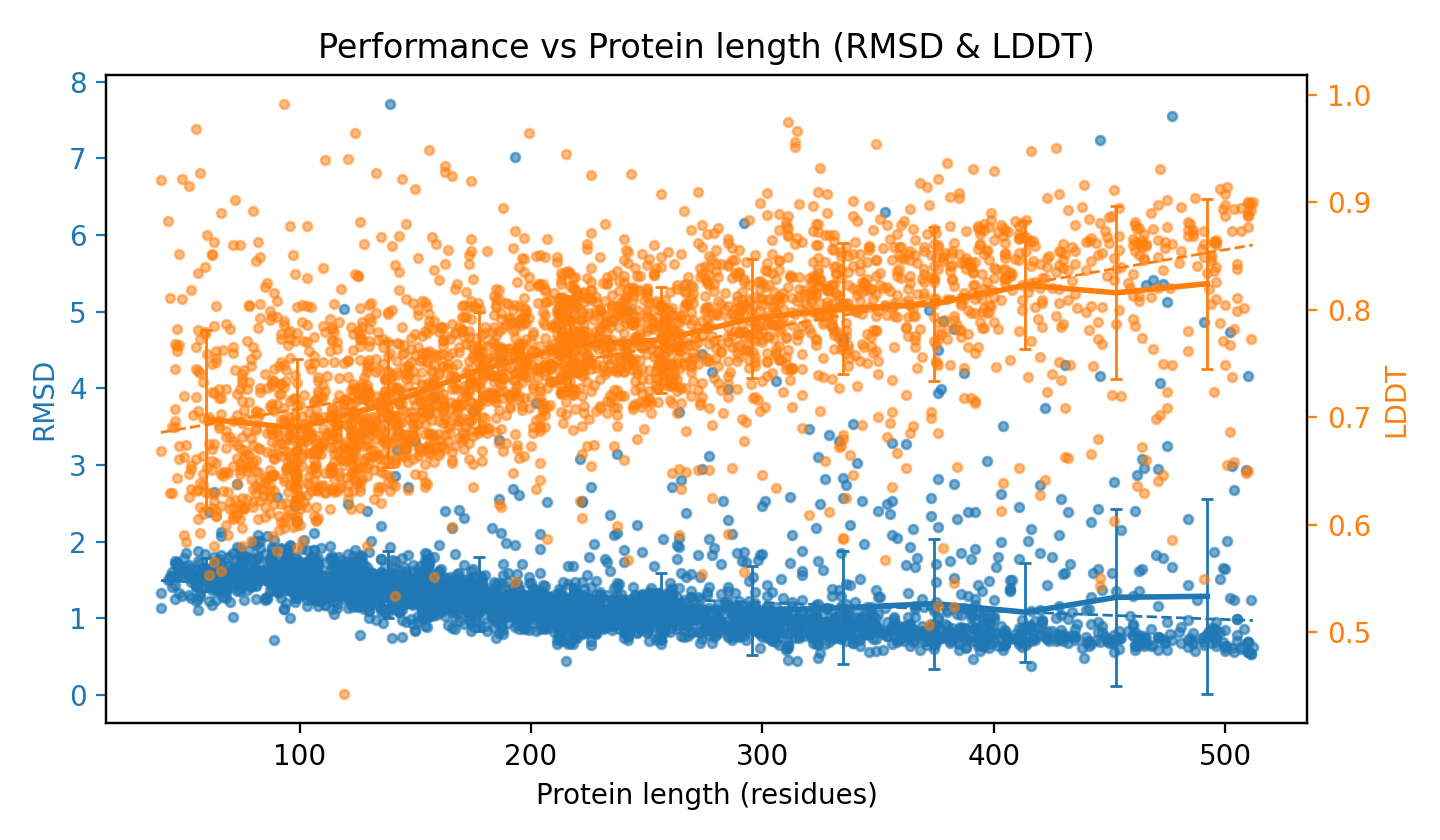}
    \caption{Dual-axis scatter vs.\ length (residues). RMSD (left, blue) and LDDT (right, orange), with bin means $\pm$ std and trend lines.}
    \label{fig:scatter-length}
  \end{subfigure}\hfill
  \begin{subfigure}[t]{0.48\textwidth}
    \centering
    \includegraphics[width=\linewidth]{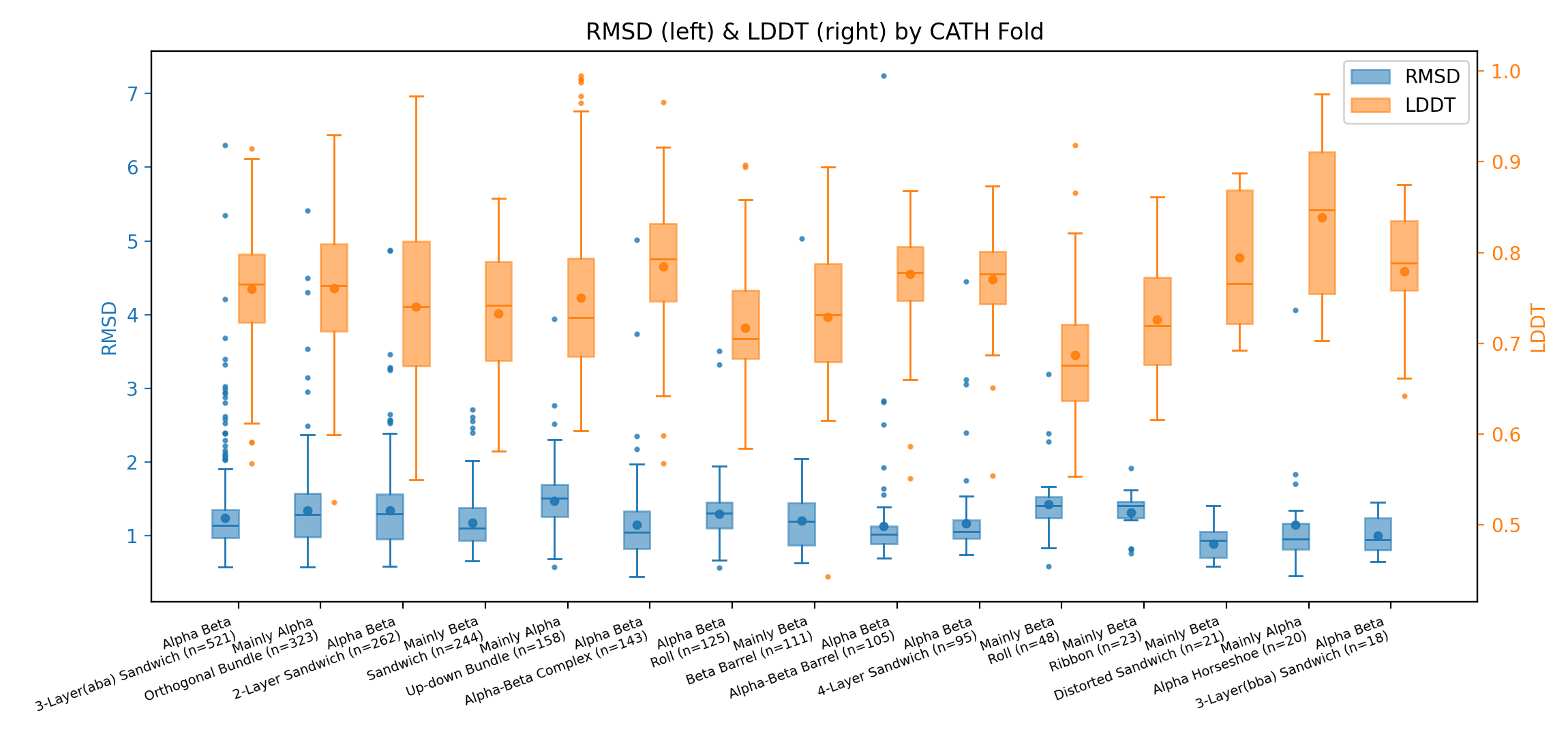}
    \caption{Fold boxplots (CATH). X-labels show Class\\Architecture. Dual-$y$ boxes: RMSD (left, blue), LDDT (right, orange).}
    \label{fig:box-fold}
  \end{subfigure}
    }\end{minipage}}
  \caption{Plots for \algo robustness evaluation. (A) vs.\ unusualness, (B) group distributions across flags and categories, (C) vs.\ length, (D) per-fold distributions. Numerical summaries (Pearson’s $r$ with 95\% CIs; group/fold means and medians) are in the accompanying CSVs.}
  \label{fig:eval-2x2}
\end{figure*}

{\xhdr{Fold type labels} For fold analyses, each chain is annotated with a CATH fold label (topology-level code) and rendered as a human-readable \emph{Class $\rightarrow$ Architecture} title (multi-line axis label). We report per-fold distributions and means for RMSD and LDDT.}

{\xhdr{Plots and statistics reported}
We use four concise views:
(i) \textbf{Dual-axis scatter vs.\ unusualness}: RMSD (left, blue) and LDDT (right, orange), with per-bin mean $\pm$ std and a fitted least-squares trend line.
(ii) \textbf{Group boxplots (shared $x$-axis)}: side-by-side, dual-$y$ box-and-whisker plots compare distributions across \{\texttt{very\_small}, \texttt{weak\_coverage}, \texttt{Near-identical}, \texttt{Same fold}, \texttt{Distant}\}; RMSD boxes map to the left axis (blue), LDDT boxes to the right axis (orange).
(iii) \textbf{Dual-axis scatter vs.\ length}: RMSD (left) and LDDT (right) versus protein length, again with per-bin mean $\pm$ std and a fitted trend line. 
(iv) \textbf{Fold boxplots}: side-by-side, dual-$y$ box-and-whisker plots per frequent CATH (Class, Architecture) (top-$N$ by support). Together, these views test how specific conditions (very small chains, weak alignment coverage, and decreasing fold similarity) modulate accuracy distributions, and whether protein size systematically correlates with errors. \\We report the following numbers and observations.}

{
\begin{itemize}[leftmargin=*,topsep=0pt,noitemsep]
\item \textbf{No degradation on unusual structures} In Fig. \ref{fig:scatter-unusualness}, we see \textit{no} correlation (Pearson’s $r$ of $0.0091$ for RMSD and $-0.0096$ for LDDT, $95\%$ intervals of $(-0.0346, 0.00365)$ and $(-0.0451, 0.0260)$) between Distortion and Unusualness. Consistent with our OOD results in Fig. \ref{fig:pareto}, we see \algo is robust to distributional shifts. As a geometry-grounded tokenizer, \algo captures energetically favorable motifs patterns, which are universal recurrences across all protein families and fold classifications.

\item \textbf{No degradation on less common folds} In Fig. \ref{fig:box-group}, we see distortion remain stable on near identical or same fold to those in FoldSeek-DB. In Fig. \ref{fig:box-fold}, there is no visual trend of degradation for less common fold types (left-to-right in Fig. \ref{fig:box-fold}) of the ones shown. Inter-fold discrepancy is also low: among folds with $n\ge 100$, the least faithfully preserved fold type suffers from $30.2\%$ higher RMSD than the most faithful.

\item \textbf{More faithful to larger folds than smaller folds} In Fig. \ref{fig:scatter-length}, we see \textit{weak} correlation (Pearson $r$ of $-0.2141$ for RMSD and $0.5627$ for LDDT, $95\%$ intervals of $(-0.2477, -0.1799)$ and $(0.5379, 0.5865)$) between Distortion and Length. In Fig. \ref{fig:box-group}, we see distortion slightly elevates for very small folds.  Among fold types with at least $n=100$ samples, \algo achieves lowest distortion ($1.130$ RMSD) on Alpha-Beta Barrels (cylindrically packed, stable folds) and highest distortion ($30\%$ higher RMSD) on Mainly Alpha Up-down Bundles (smaller folds primarily of alternating alpha helices). This suggests $\algo$ has a high propensity for packed but stable folds (sandwiches, barrels).
\end{itemize}
}

\section{Computational Complexity}\label{app:computational-complexity}
\xhdr{Notation} Let $\{t^{(\tau)}\}_{\tau=1}^T$ be $T$ backbones with lengths $N^{(\tau)}$, and let
$N\!\coloneqq\!\sum_{\tau=1}^T N^{(\tau)}$ be the total residues.
In each \textsc{Step} iteration, the most frequent geo-pair key has $M_t$ occurrences.
We use $K$ for the number of medoids produced when clustering a key’s occurrences (a small constant in practice).
For k-medoids we either: (i) cluster all $M_t$ items, or (ii) cap with $M_{\max}$ items.
Let $P$ be the \emph{period} at which \textsc{GlueOptAll} is invoked (see Alg. \ref{alg:step}), and let $C_{\mathrm{IK}}$ denote the cost of one global IK pass (see below).
The ordered map $\mathcal D$ stores key $\to$ occurrence-set with a priority
$(\rho, -|\mathcal O|, \kappa)$; each insert/erase in $\mathcal D$ costs $O(\log|\mathcal D|)=O(\log N)$.

\xhdr{Component building blocks}
\begin{itemize}[leftmargin=*,topsep=0pt,noitemsep]
\item \textbf{k-medoids on $m$ items:} $O(m^2)$ to build the pairwise RMSD matrix (constant fragment length),
plus a small constant number of assignment/update steps
\item \textbf{Priority map updates:} each merge touches $O(1)$ neighbor pairs; across the \emph{entire} run there are $O(N)$ merges $\Rightarrow O(N\log N)$ total map operations
Every merge eliminates one boundary and touches at most its two neighbors, so the total number of insert/erase operations in $\mathcal D$ across the full run is $O(N)$; with $O(\log N)$ per op, the total is $O(N\log N)$.
\item \textbf{Global IK (\textsc{GlueOptAll}) one pass:} forward kinematics is linear in links, so one pass costs
$C_{\mathrm{IK}} = O(N\cdot S_{\mathrm{FK}})$, where $S_{\mathrm{FK}}$ is the (small) number of optimizer steps $\times$ the constant forward/backward cost per link
 Periodic \textsc{GlueOptAll} adds $\tfrac{T}{P}\,O(N\log N)$ due to re-keying affected boundaries.
\end{itemize}

\xhdr{Worst-case complexity (no subsampling cap)}
\begin{itemize}[leftmargin=*,topsep=0pt,noitemsep]
  \item \textbf{ResInitTokens:}
    $O(N^2) + O(N\log N)$.
  \item \textbf{Step loop over all iterations:}
    $O\!\Big(\sum_t M_t^2\Big) + O(N\log N)$.
  \item \textbf{Periodic global glue opt:}
    $\dfrac{T}{P}\,\Big(C_{\mathrm{IK}} + O(N\log N)\Big)$.
  \item \textbf{Total (worst case):}
    $O(N^2) + O\!\Big(\sum_t M_t^2\Big) + O(N\log N)
    + \dfrac{T}{P}\,\Big(C_{\mathrm{IK}} + O(N\log N)\Big)$.
    \item \textbf{Total (with cap):}
    $O(M_{\max}^2) + O(T\,M_{\max}^2) + O(N\log N)
    + \dfrac{T}{P}\,\Big( C_{\mathrm{IK}} + O(N\log N) \Big)$.
\end{itemize}

In the worst case $M_t=\Theta(N)$ for many steps, $\sum_t M_t^2$ can reach $\Theta(N^2)$. Here $M_{\max}$ controls runtime. Putting it together, we can make the following statements about \algo's computational complexity:

\begin{itemize}[leftmargin=*,topsep=0pt,noitemsep]
\item \textbf{{(Alg. \ref{alg:GeoBPE})} Training (discovering the vocabulary):}
dominated by k-medoids calls and periodic IK:
\[
 O\big(T\,M_{\max}^2\big) + O(N\log N)
+ \tfrac{T}{P}\big(C_{\mathrm{IK}} + O(N\log N)\big).
\]
\item \textbf{{(Alg. \ref{alg:tokenize})} Tokenization (apply a learned vocabulary):} similar to training but without any k-medoids calls and in terms of $N^{(\tau)}$:
\[
 O(N^{(\tau)}\log N^{(\tau)})
+ \tfrac{T}{P}\big(C_{\mathrm{IK}} + O(N^{(\tau)}\log N^{(\tau)})\big).
\]
\item \textbf{{(Alg. \ref{alg:dequantize-assemble})} Detokenization (geometry reconstruction):}
forward kinematics per link is $O(1)$; reconstructing all atoms is $O(N)$.
\end{itemize}

\xhdr{Insights for efficient practice}
(i) Most structural variability concentrates in a small number of modes; a modest $M_{\max}$ suffices.  
(ii) Dictionary updates are \emph{incremental}; our implementation uses an ordered map.
(iii) In practice, we choose $P=10$; GlueOptAll calls are infrequent enough it does not become an issue. If this becomes the practical bottleneck, we recommend \textsc{GlueOpt} for local IK updates instead, which drops the $O(N\log N)$ term.

\xhdr{Distortion is insensitive to $M_{\max}$}
In App. \ref{app:GeoBPE-ablation}, we observe that increasing $M_{\max}$ yields no real gains beyond $5000$; any marginal gains are lost to the subsequent GlueOptAll call. This is
because medoids stabilize quickly on representative modes, capping clustering with $M_{\max}$ preserves reconstruction quality while bounding the dominant $O(M_{\max}^2)$ term. This is backed by observations made by \citet{de2002extension, mackenzie2016tertiary} and others that the structural universe of possible elements are captured by a exponentially smaller number of modes.

{
\xhdr{Increasing \textsc{glue\_opt\_every} does not significantly hurt performance} 
In Figs. \ref{fig:glue_opt_every=1} \& \ref{fig:glue_opt_every=10}, we see \algo behavior remains comparable between a run where the expensive glue optimization (all) is done every iteration, vs a run following our default recommendation of every ten iterations. The $\frac{T}{P} O(N\log N)$ term is often the key walltime bottleneck, as Table \ref{tab:walltimes} shows. Therefore, increasing $P$ would help amortize the expensive rigid body refinement routine across iterations.

}

\xhdr{{Wall times from our experiments}} In Table \ref{tab:walltimes}, we report empirical wall times from a sample run of \algo following default settings. The runtime can be accelerated with more CPUs.

\begin{table}[h!]
\centering

\caption{{Using $20$ CPUs, we report our job's wall-clock time. \underline{Underlined} steps perform periodic glue op-} {timization (period $P=10$). They are followed by $P-1$ \algo steps. We report wall times for steps $0, 10$,} {$20, 200$; omitted steps interpolate predictably.}}
\label{tab:walltimes}
{
\begin{tabular}{@{}clc@{}}
\toprule
Function            & \multicolumn{1}{c}{Paper Reference (Algo, Line)} & Time (HH:MM:SS) \\ \midrule
\_init\_thresholds  & Algo 1 L1(Empirical Quantizer Estimation)        & 00:01:33        \\
\_init\_res\_tokens & Algo 1 L2 (Per-residue Initialization)           & 02:16:02        \\
glue\_opt\_all      & Algo 1 L3 (Global glue refinement)               & 03:21:50        \\
{\ul Step 0}        & Algo 1 L7 (Step) w/ Algo 9 L13 (glue opt all)    & 02:36:21        \\
Steps 1-9   & Algo 1 L7 (Step)                                 & 01:32:50        \\
{\ul Step 10}       & Algo 1 L7 (Step) w/ Algo 9 L13 (glue opt all)    & 01:40:58        \\
Steps 11-19 & Algo 1 L7 (Step)                                 & 01:37:39        \\
{\ul Step 20}       & Algo 1 L7 (Step) w/ Algo 9 L13 (glue opt all)    & 01:27:39        \\
Steps 21-29         & Algo 1 L7 (Step)                                 & 01:36:41        \\
\multicolumn{3}{c}{...}                                                                  \\
{\ul Step 200}      & Algo 1 L7 (Step) w/ Algo 9 L13 (glue opt all)    & 00:56:56        \\
Steps 201-209       & Algo 1 L7 (Step)                                 & 00:50:06        \\
\multicolumn{3}{c}{...}                                                                  \\ \bottomrule
\end{tabular}
}
\end{table}

\section{Hyperparameter {Documentation and Guidelines}}\label{app:hyperparameters}
\subsection{{Main Hyperparameters and Reproducible Settings}}
We describe the key parameters that govern \algo's behaviors in Table \ref{tab:hyperparameters}. For each, we report the default setting used by GeoBPE across most key results of the paper: Fig. \ref{fig:pareto}, Tables \ref{tab:token-efficiency} \& \ref{tab:lm-eval} and App. \ref{app:GeoBPE-ablation}. We report any instances overriding the default settings here:
\begin{enumerate}[leftmargin=*,topsep=0pt,noitemsep]
    \item\label{hparams:1} Token efficiency / SSLM-Eval (Tables \ref{tab:token-efficiency}, \ref{tab:lm-eval}) set $\texttt{num\_p} \leftarrow \texttt{\{2:500,3:2000\}}, \texttt{bins} \leftarrow \texttt{\{1:1000\}}$ for codebook size $|\mathcal{V}|=2500$ and $\texttt{num\_p} \leftarrow \texttt{\{2:1000,3:5000\}}$, $\texttt{bins} \leftarrow \texttt{\{1:2000\}}$ for $|\mathcal{V}|=6000$.
    \item\label{hparams:2} Pareto-efficiency evaluation (Fig \ref{fig:pareto}) further add the setting for $|\mathcal{V}|=21000$ where $\texttt{num\_p} \leftarrow \texttt{\{2:1000,3:20000\}}$, $\texttt{bins} \leftarrow \texttt{\{1:2000\}}$. We vary \texttt{num\_p} elastically moves along the Pareto-efficiency plot, trading off BPR for distortion. {All runs use $w_t=1.0$, which we discover from ablation studies (see Tables \ref{tab:ik-weights-sensitivity} \& \ref{tab:ik-weights-sensitivity-full}) has better performance than $w_t=0.1$.}

    \item\label{hparams:3} Downstream transfer experiments (Tables \ref{tab:transfer}, \ref{tab:agreement-additional}) set $\texttt{num\_p} \leftarrow \texttt{\{2:2,3:5,5:1,6:2,8:1\}}$, {$\texttt{free\_bonds} \leftarrow$ False and $\texttt{bins}\leftarrow \{1:50\}$}, and $\texttt{bin\_strategy} \leftarrow \text{histogram-cover}$ to adaptively coarsen the resolution. \algo prioritizes learning fine-to-coarse hierarchical signals over low distortion for effective transfer.
\end{enumerate}

\begin{table}[h!]
\centering
\small
\caption{We report the main hyperparameters that affect \algo behavior. }
\label{tab:hyperparameters}
\resizebox{\textwidth}{!}{%
\begin{tabular}{@{}llll@{}}
\toprule
\textbf{Parameter} & \textbf{Value} & \textbf{Meaning} & \textbf{Default Behavior}\\
\midrule
\texttt{bin\_strategy} & histogram & Controls the strategy for empirical quantizer estimation (Alg. \ref{alg:GeoBPE}) & numpy.histogram with \texttt{bins}\\
\texttt{bins} & \texttt{\{1:500\}} & Controls the number of bins used by \texttt{bin\_strategy} & Uses 500 quantiles \\
\texttt{free\_bonds} & True & Whether to quantize bond lengths & Don't standardize  \\
& & Setting to False standardizes all bond lengths to precomputed values & Quantize with linear histograms \\
\texttt{glue\_opt} & True & Whether to do Glue Opt in Algs \ref{alg:res-init-tokens}, \ref{alg:step} & Do Glue Opt \\
\texttt{glue\_opt\_every} & 10 & How often to run global glue opt (final line of Alg. \ref{alg:step}) & Do every 10 iters \\
\texttt{glue\_opt\_method} & all & Whether to do batch glue opt (Alg. \ref{alg:glue-opt-all}) or single-boundary glue opt (Alg. \ref{alg:glue-opt}) & Do batch glue opt \\
\texttt{glue\_opt\_prior} & 1.0 & Prior weight encouraging optimized glues to match empirical distribution & 1.0 \\
{\texttt{w\_R, w\_t}} & {1.0, 0.1} & {Rotation, translation loss term weights to IK loss (Alg. \ref{alg:glue-opt})} & {Weigh rotation error 10x translation error}\\
\texttt{max\_num\_strucs} & 5000 & Max number of occurrences for clustering ($M_{\text{max}}$ in Alg. \ref{alg:rmsd-partition}) & 5000 \\
\texttt{num\_p} & \texttt{\{2:100,3:500}, & $K$ determined by span length $L$ in Alg. \ref{alg:rmsd-partition}, $L$ not in $\texttt{num\_p}$ round down to nearest key & Use $K=100$ when $L=2$ \\ 
& \texttt{ 5:20,6:100\}}& & Use $K=500$ when $L\ge 3$ \\
\texttt{rmsd\_super\_res} & True & Whether to use occurrences from \textit{original} backbone $t_{\tau}$ or current backbone in Alg. \ref{alg:step} & Use original states\\
\bottomrule
\end{tabular}
}
\end{table}

\subsection{{Hyperparameter Selection Guidelines}}\label{sec:guidelines}
{
\xhdr{Which ones to prioritize} Only a few parameters in Table \ref{tab:hyperparameters} dictate overall behavior, performance and runtime. Essential knobs are:
\begin{itemize}[leftmargin=*,topsep=0pt,noitemsep]
    \item \textbf{Vocabulary Growth}
    \begin{itemize}
        \item \texttt{num\_p} (number of medoids)
        \item \texttt{bins} (quantizer strength)
        \item \texttt{max\_iter}(or \# iterations to run)          
    \end{itemize}
    \item \textbf{Compression/Runtime Tradeoff}
    \begin{itemize}
        \item \texttt{glue\_opt}/\texttt{glue\_opt\_method}/\texttt{glue\_opt\_every} (glue optimization)
    \end{itemize}
\end{itemize} 

Aside from these, we suggest leaving the rest to default values.
}

{
\xhdr{Choosing \texttt{num\_p} (medoids per step) and \texttt{bins} (angle/length quantization strength)}
\label{sec:guidelines-num-p-bins}

We define \texttt{num\_p} via a step-wise schedule over motif sizes. For example,
\[
\{2:2,\; 3:5,\; 5:10\}
\]
(passed as \texttt{\textbf{\texttt{--num-p} 2-2:3-5:5-10}}) means:
\begin{itemize}
  \item introduce $2$ tokens for geometric keys with $2$ bonds (C-terminal residue orientations),
  \item $5$ tokens for keys with $3$ bonds (all non-terminal residues),
  \item $10$ tokens for all merged geometric keys with $5$ or more bonds (every GeoBPE step after residue initialization).
\end{itemize}

The \texttt{bins} parameter uses the same syntax as \texttt{num\_p}. For example,
\[
\{1:100,\; 3:10\}
\]
(passed as \texttt{\textbf{\texttt{--bins} 1-100:3-10}}) introduces $100$ bins to discretize the angular histogram at initialization, with $10$ bins for keys of size $\ge 3$. For brevity, $\textsc{bins}=n$ is shorthand for $\textsc{bins}= \{1:n\}$.

The binning strategy is controlled by \texttt{bin-strategy}. If glue optimization produces angles outside the supported range, we snap them to the closest bin. In practice, we recommend increasing or decreasing \texttt{bins} in tandem with \texttt{num\_p}.

Below we give practical recommendations by downstream use case.
}

{
\xhdr{GeoBPE for compression / reconstruction}

\emph{Intuition.} Larger \texttt{num\_p} values $\rightarrow$ more medoids per step, which improves reconstruction quality (RMSD/LDDT) at the cost of a larger vocabulary and noisier merges. Empirically, we observe diminishing returns in reconstruction beyond settings such as
\[
\texttt{num\_p} = \{2:200,\; 3:1000\},
\]
consistent with there being only a limited number of modes in the conformational variability of energetically favored backbone regions (Ramachandran landscape).

\emph{Recommendation.} Use relatively large \texttt{num\_p} to maximize reconstruction fidelity, but pair it with a high (yet not extreme) \texttt{bins[size]} to avoid a combinatorial explosion in the space of geometric keys. A good default for reconstruction-oriented use is
\[
\texttt{bins} = \{1:500\},
\]
combined with moderately large \texttt{num\_p}.
}

{
\xhdr{GeoBPE for representation learning}

\emph{Intuition.} GeoBPE emits both a sequence of tokens and a merge hierarchy, with the hierarchy providing the main inductive bias for downstream representation learning from residue to protein level. A useful hierarchy should:
\begin{itemize}
  \item capture higher-level patterns (from basic secondary structure elements to functional sites),
  \item avoid overfitting to high-frequency local vibrations.
\end{itemize}
Here the goal shifts from pure compression to \emph{coarsening}: we want motifs that aggregate meaningful local structure without being overly fine-grained.

\emph{Recommendation.} Use relatively small \texttt{num\_p} values and correspondingly small \texttt{bins}. For example, the configuration used in our paper for representation learning was
\[
\texttt{num\_p} = \{2:2,\; 3:5,\; 5:1,\; 6:2,\; 8:1\},
\]
paired with
\[
\texttt{bins} = \{1:50\}
\]
and the \texttt{histogram-cover} strategy. This yields coarser motifs and hierarchies that are better suited to downstream predictive tasks.
}

{
\xhdr{Choosing the number of merge iterations}
\label{sec:guidelines-num-iters}

At iteration $t$, GeoBPE increases the vocabulary size by looking up \texttt{num\_p}:
\[
|\mathcal{V}_{\text{final}}|
\;\approx\;
|\mathcal{V}_{\text{init}}| + \sum_{t=1}^{T} \texttt{num\_p}[|\texttt{key}^{(t)}|],
\]
where $T$ is the number of merge iterations. More iterations yield a larger and more varied vocabulary, but each ``word'' (motif) is then used less frequently.

The optimal stopping point depends on the downstream application.
}

{
\xhdr{GeoBPE for representation learning}

For representation learning, the merge hierarchy serves as an inductive bias: merged token pairs tend to correspond to secondary structure segments and align with domain or homology hits. Here GeoBPE should \emph{coarsen} high-resolution details into higher-level motifs instead of growing an extremely large vocabulary.

\emph{Recommendation.} Use a moderate number of iterations, stopping once downstream validation metrics (e.g., AUROC, Spearman $\rho$, Macro-F1) plateau. In practice, this typically occurs well before exhausting all possible merges; beyond that point, additional iterations mainly create very specific, low-usage motifs that add complexity without improving downstream performance.
}

{
\xhdr{GeoBPE for compression / reconstruction}

GeoBPE is closest to its BPE origins when used as a compression algorithm: the goal is to reduce sequence length (increase compression), preserving geometry (minimize distortion), while monitoring the amortized bits to store the growing vocabulary in the background.

\emph{Recommendation.} Allow fewer iterations and monitor the trade-off between bit-rate (i.e., BPR) and reconstruction error (RMSD/LDDT). A general rule-of-thumb is to continue merging as long as additional iterations lowers \textit{either} BPR or distortion. Later, one can choose the right iteration checkpoint to navigate the tradeoff. Thus, it is wise to stop immediately when both metrics begin degrading simultaneously. On very high resolutions and a moderate dataset (e.g. our pretraining dataset), this happens early on. The amortized bits used to grow the vocabulary generally outpace the bits saved from decreasing the number of tokens per structure; this relationship reverses on lower resolutions and larger datasets.}

{
\xhdr{GeoBPE for language modeling}

When using GeoBPE as a tokenizer for protein language models, a common heuristic adapted from NLP is to set the final vocabulary size such that
\[
\frac{|\mathcal{V}|}{L} \approx \frac{N}{1000},
\]
where $L$ is the average number of motifs per structure and $N$ is the number of structures. Equivalently, the total number of tokens
\[
T \approx L \times N
\]
suggests a target vocabulary size $|\mathcal{V}| \approx T / 1000$.

Table \ref{tab:lm_heuristic_vocab} shows concrete numbers for different LM scales.}

\begin{table}[h!]
  \centering
  
  \caption{{Heuristic vocabulary sizes $|\mathcal{V}|$ for GeoBPE when used as a tokenizer for protein language models} {at different data and model scales.}}
  \label{tab:lm_heuristic_vocab}
\label{tab:sec-agreement-avg}
\resizebox{\textwidth}{!}{
{
\begin{tabular}{@{}lcccc@{}}
\toprule
LM scale           & \# structures $N$ & $L$ (motifs / structure) & $|\mathcal{V}| \approx (L \times N)/1000$ & Model params   \\ \midrule
Toy / demo         & $10^3$            & $100$                    & $\sim 10^2$                               & $\sim 10^6$    \\
Small / usable     & $10^4$            & $100$                    & $\sim 10^3$                               & $10$--$50$M    \\
Base ``GPT-small'' & $10^5$            & $100$                    & $\sim 10^4$                               & $\sim 10^8$    \\
Mid-scale          & $10^6$            & $100$                    & $\sim 10^5$                               & $\gtrsim 10^9$ \\ \bottomrule
\end{tabular}

  }}
\end{table}

\begin{figure}[h!]
  \setlength{\fboxsep}{4pt}
    {%
      \begin{minipage}{0.98\linewidth}        
  \centering
  \includegraphics[width=0.6\linewidth]{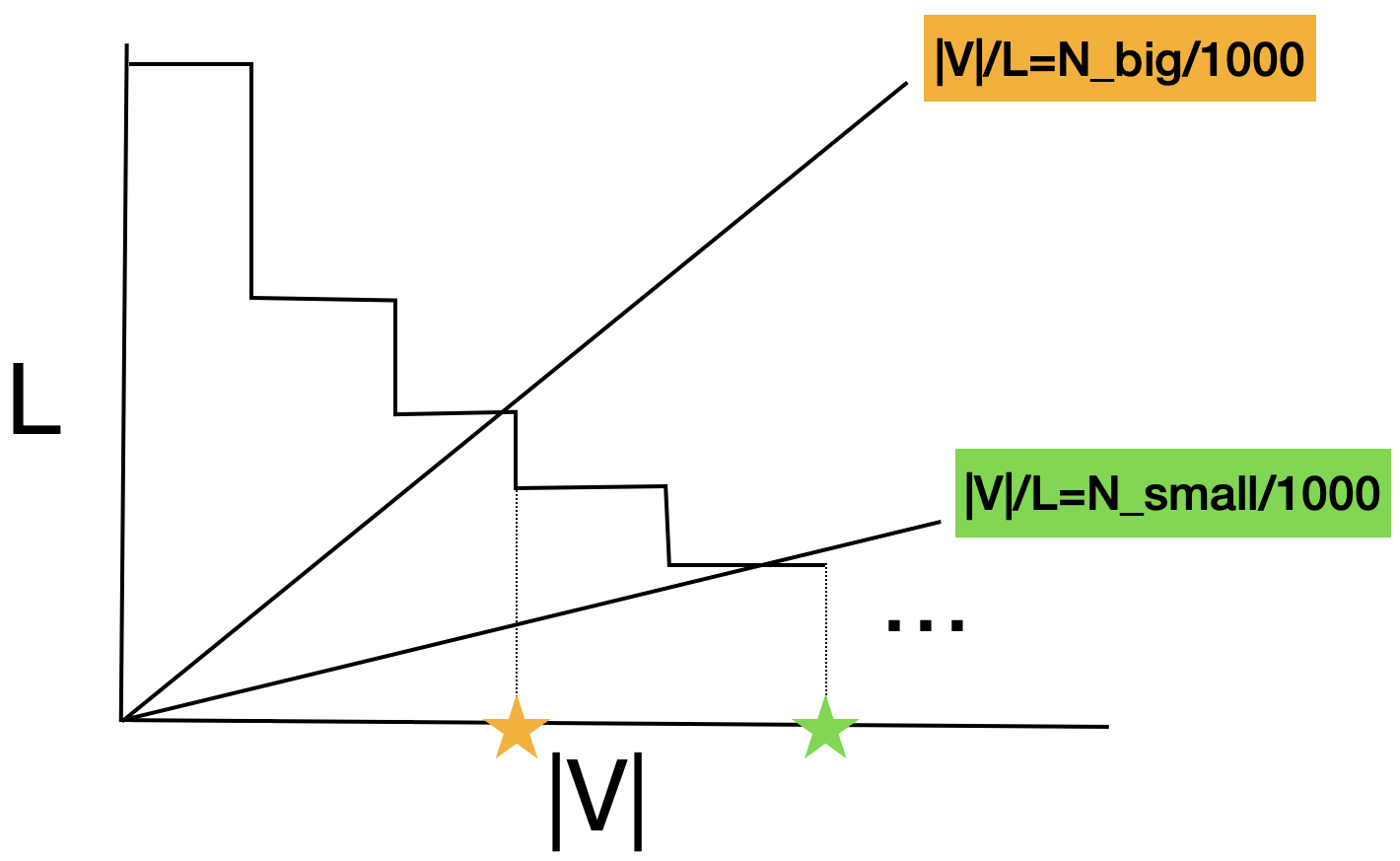}

  \caption{Illustration of the heuristic for choosing the number of merge iterations based on the target vocabulary size $|\mathcal{V}|$ for language modeling. The marked point indicates the recommended stopping iteration for a dataset with $N$ structures.}
  \label{fig:lm_heuristic}
    \end{minipage}
  }  
\end{figure}

{
We implement a stopping criterion based on this heuristic: during training we track $|\mathcal{V}|$ as merges accumulate ($T$ decreases) and mark the iteration where the target $|\mathcal{V}|$ meets $T/1000$. In Fig. \ref{fig:lm_heuristic}, this iteration is highlighted (e.g., with a star) and concretely in the \texttt{run\_\{iter\}.png} plots produced in each run directory by our code.}
{

\xhdr{Practical tip} In practice, you can set a relatively large \texttt{max\_iter} and let GeoBPE proceed for many iterations while logging checkpoints. After training, select the checkpoint whose vocabulary size and downstream metrics best match your target (compression, representation quality, or LM tokenizer size), rather than trying to tune the exact stopping iteration a priori.

}
\subsection{{Sensitivity Studies}}

\begin{figure*}[t]
  \centering
  \setlength{\fboxsep}{4pt}
    {
      \begin{minipage}{0.98\linewidth}        
    \fcolorbox{black!20}{white}{
  \begin{subfigure}[t]{0.48\textwidth}
    \centering
    \includegraphics[width=\linewidth]{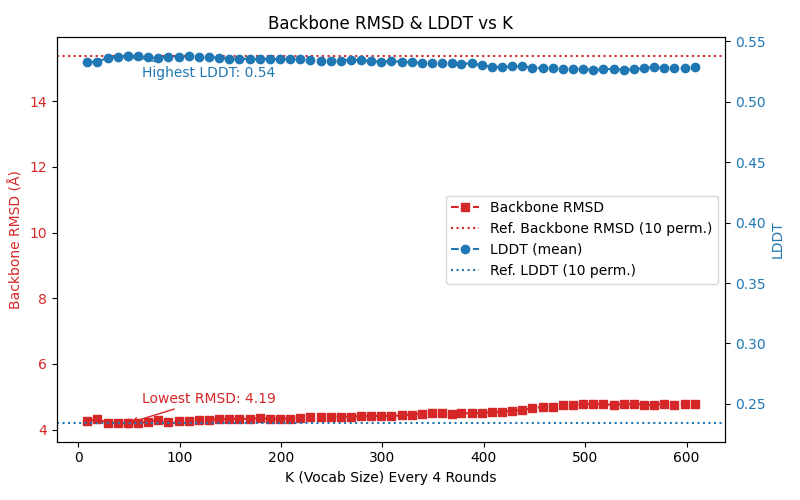}
    \caption{$\textsc{bins}=100, \textsc{free\_bonds}=\text{True}$}
    \label{fig:glue_opt_every=10}
  \end{subfigure}\hfill
  \begin{subfigure}[t]{0.48\textwidth}
    \centering
    \includegraphics[width=\linewidth]{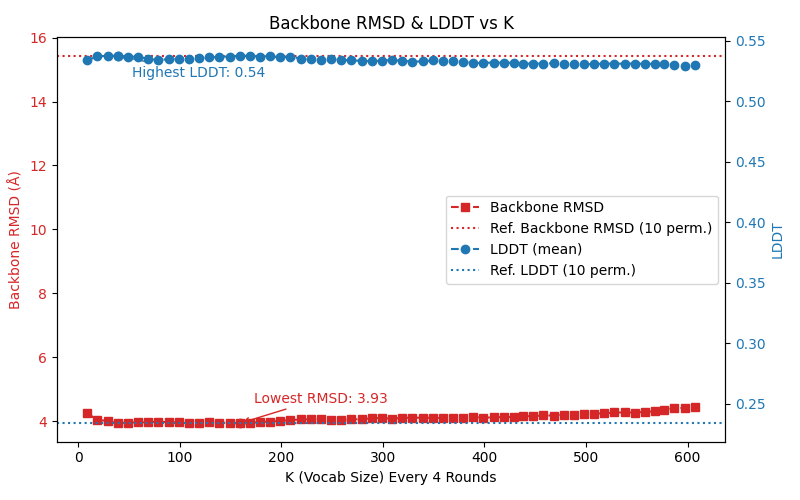}
    \caption{$\textsc{bins}=100, \textsc{free\_bonds}=\text{True}, \textsc{glue\_opt\_every}=1$}
    \label{fig:glue_opt_every=1}
  \end{subfigure}
}
\end{minipage}
}

    {
      \begin{minipage}{0.98\linewidth}        
    \fcolorbox{black!20}{white}{
  \begin{subfigure}[t]{0.97\textwidth}
    \centering
    \includegraphics[width=\linewidth]{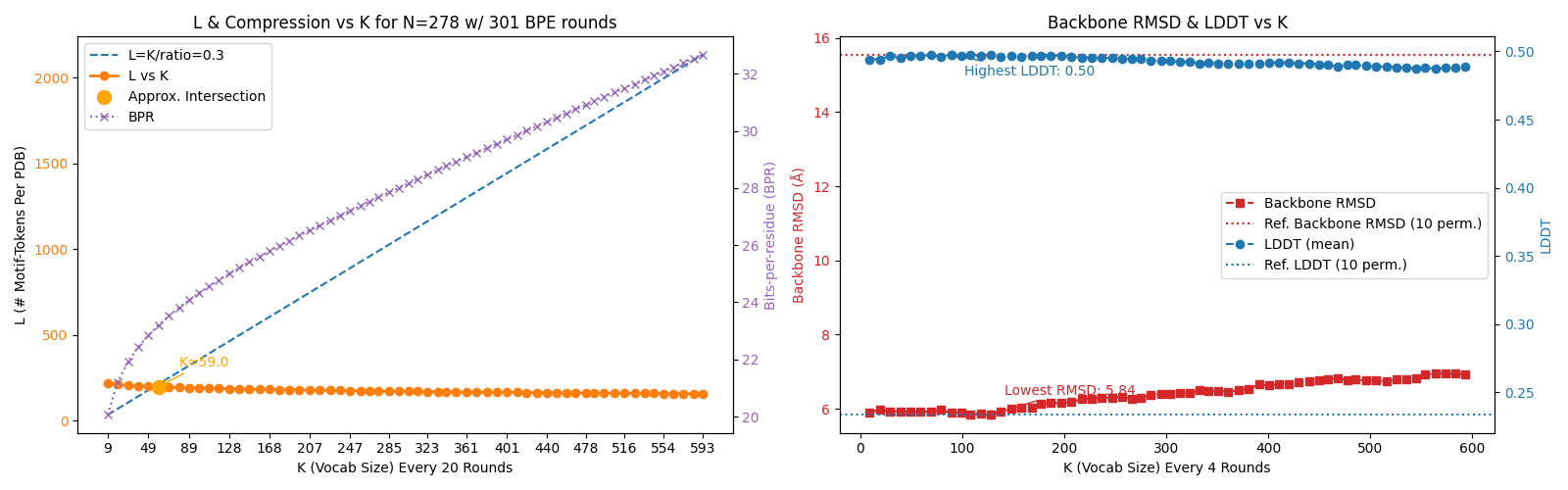}
    \caption{Baseline for Sensitivity Ablations (default settings)}
    \label{fig:fixed_bonds}
  \end{subfigure}\hfill
  }\end{minipage}
  }

    {
      \begin{minipage}{0.98\linewidth}        
    \fcolorbox{black!20}{white}{
  \begin{subfigure}[t]{0.97\textwidth}
    \centering
    \includegraphics[width=\linewidth]{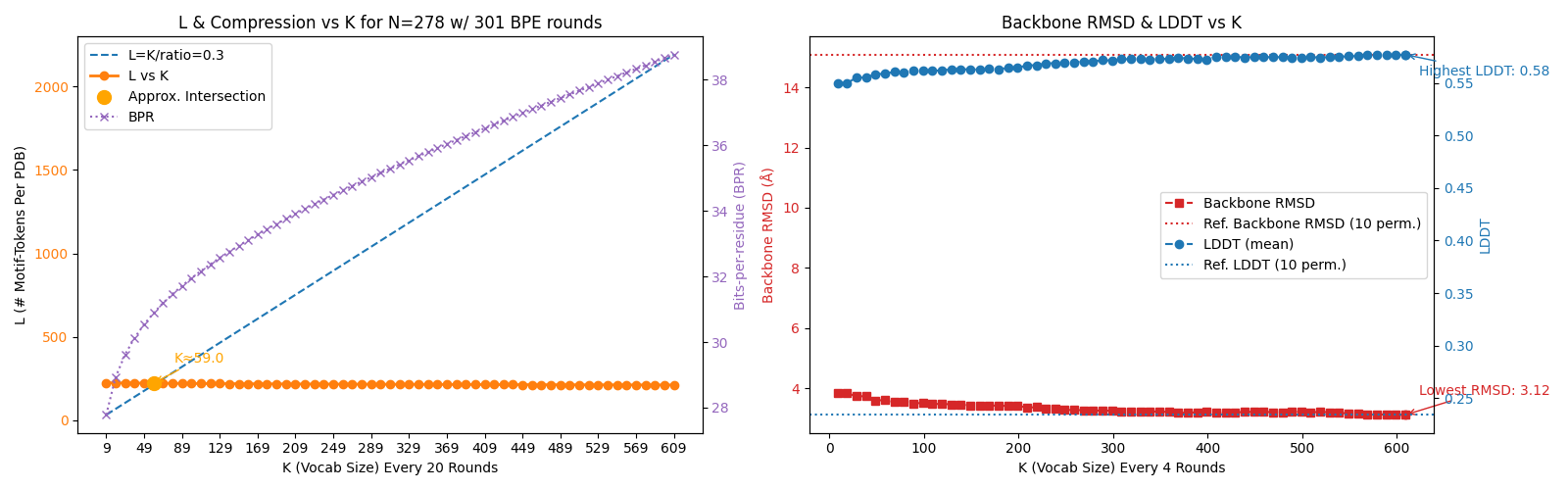}
    \caption{$\textsc{bins}=300, \textsc{free\_bonds}=\text{True}$}
    \label{fig:bins=300}
  \end{subfigure}\hfill
    }\end{minipage}}

    {
      \begin{minipage}{0.98\linewidth}        
    \fcolorbox{black!20}{white}{
  \begin{subfigure}[t]{0.97\textwidth}
    \centering
    \includegraphics[width=\linewidth]{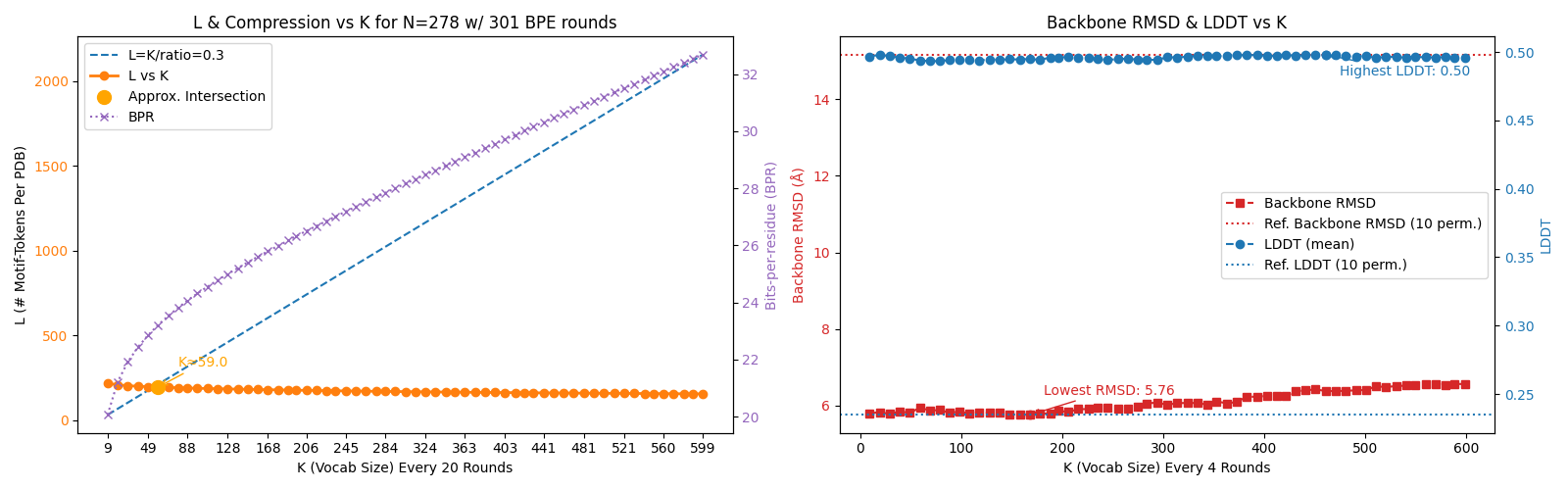}
    \caption{$\textsc{free\_bonds}=\text{True}$}
    \label{fig:bins=50}
  \end{subfigure}\hfill
    }\end{minipage}
}

{\caption{Sensitivity analysis on $\textsc{bins} \leftarrow \{50~(\text{default}), 100, 300\}$), $\textsc{glue\_opt\_every} \leftarrow \{1, 10~(\text{default})\}$ and $\textsc{free\_bonds}\leftarrow \{\text{True}, \text{False}~(\text{default})\}$. Rest of hyperparameters match defaults (Settings \ref{hparams:3}). We show \algo ($1\%$) progress plots at $\textsc{iter}=300$ for all ablation settings. $L$ is \# avg. tokens per structure; $K$ is $|V|$; see \ref{sec:guidelines} for details.}}
  \label{fig:sensitivity}
  
\end{figure*}
{
We show how sensitive \algo behavior is to key hyperparameters by running ablation experiments for selected hyperparameters, one at a time.
}
\begin{table}[h!]
\centering

\caption{{Cluster of runs that vary $\textsc{num\_p}$; each row is a run; lower resolution runs include periodic check-} {points to see how RMSD/LDDT/BPR changes over iterations.}}
\label{tab:vocab-sensitivity}
{
\begin{tabular}{@{}l|llll@{}}
\toprule
\textsc{num\_p}                                              & $|V|$   & RMSD        & LDDT         & BPR         \\ \midrule
\{2: 100, 3: 500, 5: 20, 6: 100, ...\}              & 600   & 1.66 & 0.73  & 36.02 \\ \midrule
\{2: 500, 3: 2000, 5: 100, 6: 500, ...\}            & 2500  & 1.41 & 0.75 & 41.11 \\ \midrule
\{2: 1000, 3: 5000, 5: 200, 6: 1000, ...\}          & 6000  & 1.37  & 0.76 & 45.44  \\ \midrule
\{2: 1000, 3: 20000\}                               & 21000 & 1.21 & 0.76 & 47.62 \\ \midrule
\{2: 50, 5: 20, 6: 100, ...\}                       & 5200  & 2.11 & 0.68 & 37.24 \\ \midrule
\multirow{7}{*}{\{2: 10, 3: 50, 5: 1, 6: 5, 8: 1\}} & 65    & 1.78  & 0.71   & 30.81 \\
                                                    & 237   & 1.77  & 0.70   & 34.00 \\
                                                    & 388   & 1.72  & 0.70   & 35.88 \\
                                                    & 521   & 1.71   & 0.70    & 37.33 \\
                                                    & 631   & 1.69  & 0.70   & 38.47  \\
                                                    & 739   & 1.71  & 0.70   & 39.54 \\
                                                    & 845   & 1.73 & 0.69   & 40.57 \\ \midrule
\multirow{4}{*}{\{2: 2, 3: 5, 5: 1, 6: 2, 8: 1\}}   & 109   & 3.96  & 0.53   & 27.26 \\
                                                    & 309   & 4.07  & 0.53   & 30.81 \\
                                                    & 508   & 4.23  & 0.53   & 33.46 \\
                                                    & 707   & 4.50 & 0.53 & 35.93 \\ \bottomrule
\end{tabular}
}
\end{table}

{
\xhdr{$|V|$ (\textsc{num\_p}, \# \text{iterations}) varies}

In Table \ref{tab:vocab-sensitivity} is an ablation study varying $\textsc{num\_p}$ across runs and \# \text{iterations} per run; $|V|$ depends on both. We also combine both into the throttle $|V|$. We make the following observations:
\begin{enumerate}[leftmargin=*,topsep=0pt,noitemsep]
    \item As \textsc{num\_p} values become high, the unique GeoPair keys increase exponentially. Since each iteration only looks at one key, the number of merges done falls off. Empirically, the top rows show only marginal changes to distortion as iteration increases. We omitted them for brevity.
    \item As \textsc{num\_p} values drop too low, \algo becomes more of a coarsening algorithm (lots of merges, repetitive patterns are preserved but higher frequencies are lost). When merges happen more often, more drift is introduced, so error quickly accumulates. We can see distortion monotonically increase for the last run.
    \item There exists a tension between \textsc{num\_p} and merge frequency, but glue opt is still potent enough to manage drift accumulation. We see error decrease before increasing again, when eventually merges overwhelm.
\end{enumerate}
}

{
\xhdr{$\textsc{bins} \in \{50,100,300\}$}
\textsc{bins} controls the quantization of bond lengths, bond angles and torison angles connecting motifs; it trades off structural fidelity for better coarsening. Geo-Pair keys are of the form ($\mathcal{M}_{p:q},\Gamma_q,\mathcal{M}_{q:r}$), and the space of $\gamma_q$ has size $\sim (\textsc{bins})^3$ (since there are $3$ glue angles). Importantly, it is orthogonal to $\textsc{num\_p}$, which control the id's of $\mathcal{M}_{p:q}$ and $\mathcal{M}_{q:r}$, so it can be tuned independently. Fixing Hyperparameter Setting \ref{hparams:3}, we \textit{increase} the number of bins used to discretize $\theta^{C\!N\!CA}, \omega, \phi$ angles angles by $2$x, $6$x. Fig. \ref{fig:bins=50} uses the default value $\textsc{bins}=50$; Figs. \ref{fig:glue_opt_every=10} \& \ref{fig:bins=300} use $\textsc{bins}=100, 300$. Increasing $\textsc{bins}$ decreases frequency of merges by around the same factor, so we observe $L \text{ vs } K$ is flatter for higher \textsc{bins} settings $L$ decreases slower. Since $\textsc{bins}$ is an important control of resolution, decreasing it increases distortion (e.g. $3.12\rightarrow 4.19 \rightarrow 5.76$ RMSD). Distortion is not a priority consideration for transfer experiments. Since the goal is to compress local noise into meaningful global hierarchies, introducing distortion is \textit{necessary} to cluster common motifs. Setting $\textsc{bins}$ too low can \textit{misrepresent} the overall structure, so we recommend $\textsc{bins}=50$ as a good starting value.
}

{
\xhdr{$\textsc{glue\_opt\_every} \in \{1, 10\}$}
Fig. \ref{fig:glue_opt_every=1} ($\textsc{glue\_opt\_every}=1$) only shows a $6.2\%$ decrease in RMSD and comparable LDDT vs Fig. \ref{fig:glue_opt_every=10} ($\textsc{glue\_opt\_every}=10$). As the wall times in \ref{tab:walltimes} shows, decreasing the frequency of glue\_opt significantly accelerates \algo, regardless of how many cores are available. App. \ref{app:computational-complexity} reveals glue\_opt period $P$ to directly dictate a rate-limiting term in $\algo$'s complexity. Thus, we adopt $\textsc{glue\_opt\_every}=10$ as the default setting. We also suggest \algo users to try setting $\textsc{glue\_opt\_every}>10$ to balance the tradeoffs.
}
\begin{table}[h!]
\centering

\caption{{We performed the following sweep over $(w_R, w_t)$ (order of magnitude changes to $w_T/w_R$); remain-} {ing settings match defaults (App. \ref{app:hyperparameters}).}}
\label{tab:ik-weights-sensitivity}
{
\begin{tabular}{@{}l|llllll@{}}
\toprule
$(w_R, w_t)$ & \multicolumn{2}{l}{Train} & \multicolumn{2}{l}{CAMEO} & \multicolumn{2}{l}{CASP14} \\ \midrule
GeoBPE (1\%)       & RMSD        & LDDT        & RMSD        & LDDT        & RMSD         & LDDT        \\ \midrule
$(10, 0.1)$          & 2.846       & 0.615       & 2.767       & 0.601       & 2.608        & 0.587       \\
$(1, 0.1)$-default   & 1.718       & 0.739       & 1.656       & 0.734       & 1.526        & 0.721       \\
$(1.0, 1.0)$         & 1.552       & 0.764       & 1.546       & 0.755       & 1.412        & 0.743       \\
$(0.1, 1.0)$         & 1.537       & 0.767       & 1.532       & 0.758       & 1.396        & 0.745       \\
$(0.1, 10)$          & 1.533       & 0.768       & 1.533       & 0.758       & 1.407        & 0.745       \\ \bottomrule
\end{tabular}
}
\end{table}

\begin{table}[h!]
\centering

\caption{{We compare the default $(w_R, w_t)$ setting with $(1.0,1.0)$, which in Table \ref{tab:ik-weights-sensitivity} resulted in lower distor-} {tion for \algo (1\%).}}
\label{tab:ik-weights-sensitivity-full}
{
\resizebox{\textwidth}{!}{%
\begin{tabular}{@{}llllllllll@{}}
\toprule
\multicolumn{2}{l}{GeoBPE} & Train &      & Valid &      & CAMEO &      & CASP14 &      \\ \midrule
$|V|$   & $(w_R, w_t)$          & RMSD  & LDDT & RMSD  & LDDT & RMSD  & LDDT & RMSD   & LDDT \\ \midrule
600    & ${(1, 0.1)}$-default  & 1.66  & 0.73 & 1.71  & 0.72 & 1.77  & 0.72 & 1.53   & 0.72 \\
       & ${(1.0, 1.0)}$        & 1.55  & 0.75 & 1.58  & 0.74 & 1.65  & 0.74 & 1.39   & 0.74 \\
2500   & ${(1, 0.1)}$-default  & 1.41  & 0.75 & 1.50  & 0.74 & 1.57  & 0.74 & 1.51   & 0.73 \\
       & ${(1.0, 1.0)}$        & 1.29  & 0.78 & 1.36  & 0.77 & 1.41  & 0.77 & 1.33   & 0.76 \\
6000   & ${(1, 0.1)}$-default  & 1.37  & 0.76 & 1.46  & 0.75 & 1.52  & 0.74 & 1.54   & 0.72 \\
       & ${(1.0, 1.0)}$        & 1.23  & 0.78 & 1.30  & 0.78 & 1.37  & 0.77 & 1.35   & 0.75 \\
21000  & ${(1, 0.1)}$-default  & 1.21  & 0.77 & 1.28  & 0.76 & 1.40  & 0.75 & 1.55   & 0.72 \\
       & ${(1.0, 1.0)}$        & 1.05  & 0.80 & 1.12  & 0.79 & 1.25  & 0.78 & 1.35   & 0.76 \\ \bottomrule
\end{tabular}
}
}
\end{table}

{
\xhdr{$\textsc{w\_R}/\textsc{w\_t} \in \{10^{-2},\ldots, 10^2\}$}
Table \ref{tab:ik-weights-sensitivity} shows $w_t$ is relatively more important than $w_R$ for reconstruction. The interpretation is correct positions are more critical than correct orientations. We observe diminishing returns once $w_t\ge w_R$ ($|\Delta_{LDDT}|\approx 10^{-3}, |\Delta_{RMSD}|\approx 10^{-2}$).
}

{
\xhdr{$\textsc{free\_bonds} \in \{\text{False}, \text{True}\}$}
\textsc{free\_bonds} decides whether bond lengths are free variables, or standardized to fixed values. Generally, the backbone bond lengths are very close to fixed and most workflows (e.g. X-ray diffraction, NMR, Cryo-EM) that solve structures make such assumptions. \algo is designed to be fully general, allowing variable bond lengths. In lieu of the known fact that they have relatively narrow ranges, we ran a sanity check to see if \algo is sensitive to \textsc{free\_bonds}. Comparing Figs. \ref{fig:fixed_bonds} \& \ref{fig:bins=50}, we see the run with free bonds achieves only $1.69\%$ lower RMSD, which is negligible.
}

\section{Large Language Model Usage}
We used LLMs mainly for polishing the writing, including prompts to check for grammar mistakes, improving clarity of mathematical notation, and formatting the text to save space.

\section{Algorithmic Details}\label{app:algo-details}

{
\xhdr{{Additional notation for algorithms}}

We reuse all geometric and GeoBPE notation from Secs.~\ref{sec:prelims}--\ref{sec:algorithm}.
For convenience we collect the additional symbols that appear only inside the
algorithmic pseudocode.

\begin{description}[leftmargin=0em]
\item[$\mathcal S,\mathcal A$]
Set of motif (or motif--pair) occurrences and its sampled subset used by
\textsc{RMSD\_Partition} (Alg. \ref{alg:rmsd-partition}); each $u\in\mathcal S$ indexes a motif $\mathcal M^{(t_u)}_{i_u{:}k_u}$.

\item[$\mathcal S_3,\mathcal S_2$]
Collections of interior and terminal bond--residue occurrences used to
build residue-level codebooks $\mathcal A_3,\mathcal A_2$ (Algo.~\ref{alg:res-init-tokens}).

\item[$\widehat{\mathcal M}$, $c(\cdot)$]
Medoid set and assignment map returned by \textsc{RMSD\_Partition}, used as inputs
to glue-optimization routines (Algos.~\ref{alg:glue-opt},\ref{alg:glue-opt-all}).

\item[$\mathcal D^{(\star)},\ \mathcal O^{(\star)}(\kappa)$]
Single-backbone geo-pair map and occurrence sets for a new backbone $t^\star$
during tokenization (Algo.~\ref{alg:tokenize}).

\item[$\Sigma,\Sigma_{\mathrm{med}}$]
Token dictionary used for geometric language modeling (Algos.~\ref{alg:build-joint-vocab},
\ref{alg:backbone-to-seq}); $\Sigma_{\mathrm{med}}$ contains only motif tokens; $\Sigma$ also includes glue angle tokens.

\item[$\mathrm{id}_{\mathrm{med}},\mathrm{id}_{\mathrm{bin}}$]
Integer maps assigning token IDs to motif medoids $(\kappa,j)$ and to
glue-angle bins $(\mathrm{type},b)$, respectively.

\item[$x^{(\tau)}$]
Token sequence encoding backbone $t_\tau$ obtained by alternating motif and
glue-bin tokens (Algo.~\ref{alg:backbone-to-seq}).

\item[$h_i^\uparrow,c_i^\uparrow,\bar h_i,\bar c_i$]
Upward and downward TreeLSTM states at node $i$ in the up--down encoder
(Algos.~\ref{alg:treelstm-cell}--\ref{alg:updown-encoder}).

\item[$z^{\mathrm{res}}_{\tau,i},\ z^{\mathrm{prot}}_\tau$]
Final residue-level and protein-level embeddings produced by the up--down encoder
on the merge hierarchy $\mathcal F^{(\tau)}$ (Algo.~\ref{alg:updown-encoder}).
\end{description}
}

\begin{algorithm}[h!]
\caption{\textsc{RMSD\_Partition} on motif--pair occurrences}
\label{alg:rmsd-partition}
\begin{algorithmic}[1]
\REQUIRE Motif--pair occurrences $\mathcal{S}=\{u=1,\dots,M\}$ with $\mathcal{M}^{(t_u)}_{i_u{:}k_u}$, common span length $L=k_u-i_u+1$, and either $\forall u,\,k_u=N^{(t_u)}$ or $\forall u,\,k_u<N^{(t_u)}$; target $K\ge1$; optional $M_{\max},T,\varepsilon$.
\ENSURE Medoids $\widehat{\mathcal{M}}=\{\widehat m_1,\dots,\widehat m_K\}\subseteq\mathcal{S}$ and assignments $c:\{1,\dots,M\}\to\{1,\dots,K\}$.

\STATE For each $u\in\mathcal{S}$, compute $\mathbf{X}_u\in\mathbb{R}^{3L\times3}$ via \textsc{COMPUTE\_COORDS}$(i_u{:}k_u)$.
\STATE Let $\mathcal{A}\subseteq\mathcal{S}$ be a uniform sample without replacement of size $\min(M,M_{\max})$ (or $\mathcal{A}=\mathcal{S}$).
\STATE Build $D\in\mathbb{R}^{|\mathcal{A}|\times|\mathcal{A}|}$ with $D_{uv}=\textsc{KABSCH\_RMSD}(\mathbf{X}_u,\mathbf{X}_v)$ for $u,v\in\mathcal{A}$.
\STATE Initialize $\mathcal{M}\leftarrow\{m_1,\dots,m_K\}$ as $K$ distinct uniform indices from $\{1,\dots,|\mathcal{A}|\}$.
\FOR{$t=1$ \TO $T$}
  \STATE Assign: $c(u)\leftarrow\arg\min_{j\in\{1,\dots,K\}} D_{u,m_j}$ for all $u\in\mathcal{A}$.
  \STATE Update each $j$: $\mathcal{C}_j=\{u\in\mathcal{A}:c(u)=j\}$. If $\mathcal{C}_j=\emptyset$, reseed $m_j$ uniformly from $\mathcal{A}$; else
  \[
  m_j' \leftarrow \arg\min_{u\in\mathcal{C}_j} \sum_{v\in\mathcal{C}_j} D_{uv}.
  \]
  \STATE If $\sum_{j=1}^K D_{m_j,m_j'}<\varepsilon$ \textbf{break}; else set $m_j\leftarrow m_j'$ for all $j$.
\ENDFOR
\STATE Map $\mathcal{M}=\{m_1,\dots,m_K\}$ (indices in $\mathcal{A}$) to $\widehat{\mathcal{M}}=\{\widehat m_1,\dots,\widehat m_K\}$ (indices in $\mathcal{S}$).
\STATE For each $u\in\mathcal{S}$, set $c(u)\leftarrow\arg\min_{j\in\{1,\dots,K\}}\textsc{KABSCH\_RMSD}(\mathbf{X}_u,\mathbf{X}_{\widehat m_j})$.
\STATE \textbf{return} $\widehat{\mathcal{M}}$ and $c(\cdot)$.
\end{algorithmic}
\end{algorithm}
\begin{algorithm}[h!]
\caption{\textsc{KABSCH\_RMSD}$(\mathbf{P},\mathbf{Q})$}
\label{alg:kabsch-rmsd}
\begin{algorithmic}[1]
\REQUIRE $\mathbf{P},\mathbf{Q}\in\mathbb{R}^{n\times 3}$ with $n=3L$.
\STATE $\bar{\mathbf{p}}=\frac{1}{n}\sum_i \mathbf{P}_i$, $\bar{\mathbf{q}}=\frac{1}{n}\sum_i \mathbf{Q}_i$
\STATE $\tilde{\mathbf{P}}=\mathbf{P}-\bar{\mathbf{p}}$, \ \ $\tilde{\mathbf{Q}}=\mathbf{Q}-\bar{\mathbf{q}}$
\STATE $\mathbf{H}=\tilde{\mathbf{P}}^\top \tilde{\mathbf{Q}}$, \ \ $\mathbf{U}\mathbf{\Sigma}\mathbf{V}^\top=\mathrm{SVD}(\mathbf{H})$
\STATE $\mathbf{R}=\mathbf{U}\mathbf{V}^\top$; if $\det(\mathbf{R})<0$, set $\mathbf{V}_{:,3}\leftarrow -\mathbf{V}_{:,3}$ and recompute $\mathbf{R}=\mathbf{U}\mathbf{V}^\top$
\STATE $\mathbf{Q}_{\text{aligned}}=(\mathbf{Q}-\bar{\mathbf{q}})\mathbf{R}^\top+\bar{\mathbf{p}}$
\STATE \textbf{return} $\sqrt{\frac{1}{n}\sum_{i=1}^{n}\|\mathbf{P}_i-\mathbf{Q}_{\text{aligned},i}\|^2}$
\end{algorithmic}
\end{algorithm}
\begin{algorithm}[h!]
\caption{\textsc{K\_MEDOIDS} on a precomputed distance matrix}
\label{alg:k-medoids}
\begin{algorithmic}[1]
\REQUIRE Symmetric $D\in\mathbb{R}^{N\times N}$, number of clusters $K$, iterations $T$, tolerance $\varepsilon$.
\ENSURE Medoid set $\{m_1,\dots,m_K\}$ and assignments $c(\cdot)$ on $\{1,\dots,N\}$.
\STATE Initialize medoids $\{m_j\}$ as $K$ distinct random indices.
\FOR{$t=1$ \TO $T$}
  \STATE $c(u)\leftarrow \arg\min_{j} D_{u,m_j}$ for all $u$
  \FOR{$j=1$ \TO $K$}
    \STATE $\mathcal{C}_j=\{u:c(u)=j\}$; if $\mathcal{C}_j=\emptyset$, re-seed $m_j$ at random
    \STATE $m_j'\leftarrow \arg\min_{u\in\mathcal{C}_j}\sum_{v\in\mathcal{C}_j} D_{uv}$
  \ENDFOR
  \STATE If $\sum_{j=1}^K D_{m_j,m_j'}<\varepsilon$, \textbf{break}; else $m_j\leftarrow m_j'$ for all $j$
\ENDFOR
\STATE \textbf{return} $\{m_j\}$ and $c(\cdot)$
\end{algorithmic}
\end{algorithm}

\begin{algorithm}[h!]
\caption{\textsc{Step} — one \algo merge iteration}
\label{alg:step}
\begin{algorithmic}[1]
\REQUIRE 
Current segmentations $\{\mathcal P^{(\tau)}\}_{\tau=1}^T$ and merge hierarchies $\{\mathcal F^{(\tau)}\}_{\tau=1}^T$ (frontier leaves of $\mathcal F^{(\tau)}$ equal $\mathcal P^{(\tau)}$);
priority-ordered map $\mathcal D$ with keys $\pi(\kappa)=(\rho(\kappa),-|\mathcal O(\kappa)|,\kappa)$ and values $\mathcal O(\kappa)$;
current vocabulary $\mathcal V$ (map: key $\to$ prototype set);
boundary-glue quantizers $Q_{\theta^{C\!N\!CA}},Q_{\omega},Q_{\phi}$; optional glue mode $\in\{\textit{none},\textit{each},\textit{all}\}$ and, if \textit{all}, a period.
\ENSURE Updated $(\{\mathcal P^{(\tau)}\},\{\mathcal F^{(\tau)}\},\mathcal D,\mathcal V)$.

\STATE \textbf{Select the merge key.} 
\[
\big((\rho^\star,-c^\star,\kappa^\star),\,\mathcal O(\kappa^\star)\big) \leftarrow \textsc{Front}(\mathcal D).
\]
Write each occurrence as $(\mathcal L,\mathcal R)\in\mathcal O(\kappa^\star)$ with
$\mathcal L=\mathcal M^{(t_\tau)}_{p{:}q}$ and $\mathcal R=\mathcal M^{(t_\tau)}_{q+1{:}r}$.

\STATE \textbf{Prototype assignment (create-or-assign).}
\IF{$\rho^\star=1$ \ (\emph{no prototypes yet})}
  \STATE Gather concatenated spans $\{\mathcal M^{(t_\tau)}_{p{:}r}\}$ from \textit{original} $t_\tau$ for all $(\mathcal L,\mathcal R)\in\mathcal O(\kappa^\star)$ (identical length).
  \STATE Run \textsc{RMSD\_Partition} (Alg.~\ref{alg:rmsd-partition}) to obtain medoids and $c:\mathcal O(\kappa^\star)\!\to\!\{1,\dots,K_{|\kappa^\star|}\}$.
  \STATE Define $\mathcal A_{\kappa^\star}=\{\Pi^{(\kappa^\star)}_j\}_{j=1}^{K_{|\kappa^\star|}}$ (medoid spans’ internal-parameter tuples).
  \STATE Update vocabulary: $\mathcal V[\kappa^\star]\leftarrow \mathcal A_{\kappa^\star}$ and set $\rho^\star\leftarrow 0$.
\ELSE
  \STATE For each occurrence, set $c(\mathcal L,\mathcal R)=\arg\min_j \mathrm{RMSD}\big(\mathcal M^{(t_\tau)}_{p{:}r},\Pi^{(\kappa^\star)}_j\big)$ using $\mathcal V[\kappa^\star]$.
\ENDIF

\STATE \textbf{Greedy, non-overlapping merges (and hierarchy updates).}
For each backbone $t_\tau$, sort occurrences by $p$ and choose a maximal disjoint subset $S^{(\tau)}$ left-to-right.
For every $(\mathcal L,\mathcal R)\in S^{(\tau)}$ with label $j=c(\mathcal L,\mathcal R)$:
\begin{enumerate}[leftmargin=*,topsep=0pt,noitemsep]
  \item \textbf{Form merged motif} $\widetilde{\mathcal M}=\mathcal M^{(t_\tau)}_{p{:}r}$ and \emph{overwrite} its internals by the prototype:
  \[
    \big(\ell,\theta,\psi,\omega,\phi,\{\Gamma_i\}\big)\Big|_{\widetilde{\mathcal M}}\ \leftarrow\ \Pi^{(\kappa^\star)}_j.
  \]
  \item \textbf{Update segmentation} $\mathcal P^{(\tau)}$: replace $(\mathcal L,\mathcal R)$ by $\widetilde{\mathcal M}$.
  \item \textbf{Update hierarchy} $\mathcal F^{(\tau)}$: add a \emph{parent} node for span $[p{:}r]$ with left child the node of $\mathcal L$ and right child the node of $\mathcal R$; update the \emph{frontier} (replace the two leaves by their parent so the frontier again equals $\mathcal P^{(\tau)}$).
  \item \textbf{(Optional) single-boundary glue opt} at link $p{-}1\!\to p$ if mode=\textit{each}; re-snap the three boundary angles.
\end{enumerate}

\STATE \textbf{Update counts and priorities in $\mathcal D$.}
For each merged $(\mathcal L,\mathcal R)$:
\begin{enumerate}[leftmargin=*,topsep=0pt,noitemsep]
  \item \textbf{Merged pair decrement:} remove this occurrence from $\mathcal O(\kappa^\star)$; let the new count be $c_{\text{new}}$.
  Erase $\pi_{\text{old}}=(0,-c^\star,\kappa^\star)$ and, if $c_{\text{new}}>0$, insert $(0,-c_{\text{new}},\kappa^\star)\mapsto \mathcal O(\kappa^\star)$.
  \item \textbf{Neighbor decrements:} with neighbors $\mathcal L^{-}$ and $\mathcal R^{+}$ (when defined), compute
  $k_L=\textsc{ComputeGeoKey}(\mathcal L^{-},\mathcal L)$ and
  $k_R=\textsc{ComputeGeoKey}(\mathcal R,\mathcal R^{+})$.
  For each $k\in\{k_L,k_R\}$ whose count decreases to $c_{\text{new}}$, erase $(\rho(k),-c_{\text{old}},k)$ and, if $c_{\text{new}}>0$, insert $(\rho(k),-c_{\text{new}},k)$.
  \item \textbf{Neighbor increments:} compute
  $k_L'=\textsc{ComputeGeoKey}(\mathcal L^{-},\widetilde{\mathcal M})$ and
  $k_R'=\textsc{ComputeGeoKey}(\widetilde{\mathcal M},\mathcal R^{+})$ (when defined);
  increment their counts and (re)insert with priorities $(\rho(k),-c_{\text{new}},k)$, where $\rho(k)=\mathbf{1}[k\notin\mathrm{dom}(\mathcal V)]$.
\end{enumerate}

\STATE \textbf{(Optional periodic global glue opt).}
If mode=\textit{all} and the schedule triggers, apply \textsc{GlueOptAll} (Alg.~\ref{alg:glue-opt-all}) to all modified backbones; 
recompute keys for their adjacent pairs, and for every affected key $k$, perform the same erase/insert priority update with $\rho(k)=\mathbf{1}[k\notin\mathrm{dom}(\mathcal V)]$.
If $\textsc{Front}(\mathcal D)$ then exposes a recurring key $(\rho=0)$ promoted by glue refinement, immediately re-invoke \textsc{Step} (no new clustering).

\STATE \textbf{return} $\{\mathcal P^{(\tau)}\}$, $\{\mathcal F^{(\tau)}\}$, $\mathcal D$, and $\mathcal V$.
\end{algorithmic}
\end{algorithm}

\begin{algorithm}[t]
{
\begin{minipage}{0.96\linewidth}

\caption{{\textsc{Tokenize} — use learned \algo vocabulary to tokenize new backbone}}
\label{alg:tokenize}
\begin{algorithmic}[1]
\REQUIRE New backbone $t^\star$ (length $N$); learned residue codebooks $\mathcal A_3,\mathcal A_2$; learned vocabulary $\mathcal V$ whose \emph{pair-keys} are ordered by training insertion, written
\[
\mathrm{Order}(\mathcal V)=\big\langle \kappa_1,\kappa_2,\ldots,\kappa_{|\mathcal{V}|}\big\rangle,
\]
with each $\kappa_\ell$ mapped to its fixed prototype set $\mathcal A_{\kappa_\ell}=\{\Pi^{(\kappa_\ell)}_j\}_{j=1}^{K_{|\kappa_\ell}|}$; quantizers $Q_{\theta^{C\!N\!CA}},Q_{\omega},Q_{\phi}$; optional glue mode $\in\{\textit{none},\textit{each},\textit{all}\}$ and (if \textit{all}) a period $P$.
\ENSURE Tokenized segmentation $\mathcal P^{(\star)}$ and merge hierarchy $\mathcal F^{(\star)}$ for $t^\star$.
\STATE \textbf{Per-residue init (no new clustering).}
Set $\mathcal P^{(\star)}\!\leftarrow\!(\mathcal M^{(t^\star)}_{1{:}1},\ldots,\mathcal M^{(t^\star)}_{N{:}N})$.
Assign each residue motif to the nearest element of $\mathcal A_3$ (interior) or $\mathcal A_2$ (terminal) by Kabsch-aligned RMSD, and overwrite its internal parameters accordingly.
\STATE \textbf{Initialize hierarchy.}
Let $\mathcal F^{(\star)}$ be a binary forest whose leaves (in order) are $\{\mathcal M^{(t^\star)}_{i{:}i}\}_{i=1}^N$; its frontier equals $\mathcal P^{(\star)}$.
\STATE \textbf{Optional one-time global glue.}
If mode$=$\textit{all}, run \textsc{GlueOptAll} (Alg.~\ref{alg:glue-opt-all}) on $t^\star$ once; snap $(\theta^{C\!N\!CA},\omega,\phi)$ to $Q$.
\STATE \textbf{Build single-structure geo-pair map.}
Compute $\mathcal D^{(\star)}\!\leftarrow\!\textsc{BinHelper}\big(t^\star,\mathcal P^{(\star)},Q\big)$ (Alg.~\ref{alg:bin-helper}), which maps any geo-pair key $\kappa$ to its occurrence set $\mathcal O^{(\star)}(\kappa)$ on $t^\star$.  \hfill{(Uses \textsc{ComputeGeoKey} with raw medoid internals and quantized boundary glue.)}
\STATE \textbf{Apply learned merges in training order (no new keys).}
\FOR{$s=1$ \TO ${|\mathcal{V}|}$}
  \STATE $\kappa\leftarrow \kappa_s$ \quad (the $s$-th key in $\mathrm{Order}(\mathcal V)$).
  \IF{$\kappa\notin \mathcal D^{(\star)}$} \STATE \textbf{continue} \ENDIF
  \STATE \textbf{Assign prototypes (no clustering).} For each $(\mathcal L,\mathcal R)\in \mathcal O^{(\star)}(\kappa)$ with $\mathcal L=\mathcal M^{(t^\star)}_{p{:}q}$, $\mathcal R=\mathcal M^{(t^\star)}_{q+1{:}r}$, set
  \[
  c(\mathcal L,\mathcal R)\;=\;\arg\min_{j\in\{1,\ldots,K_{|\kappa|}\}}
  \mathrm{RMSD}\big(\mathcal M^{(t^\star)}_{p{:}r},\ \Pi^{(\kappa)}_j\big).
  \]
  \STATE \textbf{Greedy disjoint merges \& hierarchy updates (left$\to$right).}
  Order $\mathcal O^{(\star)}(\kappa)$ by increasing $p$; select the maximal disjoint subset $S^{(\star)}$.
  For each $(\mathcal L,\mathcal R)\in S^{(\star)}$ with label $j=c(\mathcal L,\mathcal R)$:
  \begin{enumerate}
    \item Form $\widetilde{\mathcal M}=\mathcal M^{(t^\star)}_{p{:}r}$ and overwrite its internals by $\Pi^{(\kappa)}_j$.
    \item Update segmentation $\mathcal P^{(\star)}$: replace $(\mathcal L,\mathcal R)$ by $\widetilde{\mathcal M}$.
    \item \emph{Update hierarchy} $\mathcal F^{(\star)}$: add a parent for span $[p{:}r]$ with left child the node of $\mathcal L$ and right child the node of $\mathcal R$; update the frontier so it equals the new $\mathcal P^{(\star)}$.
    \item If mode$=$\textit{each} and the boundary $p{-}1\!\to p$ exists, apply \textsc{GlueOpt} (Alg.~\ref{alg:glue-opt}); snap its three angles to $Q$.
  \end{enumerate}
  \STATE \textbf{Maintain the single-structure map.}
  For each merged $(\mathcal L,\mathcal R)\in S^{(\star)}$, update $\mathcal D^{(\star)}$ locally:
  remove the occurrence of $\kappa$; decrement keys of neighbors $(\mathcal L^{-},\mathcal L)$ and $(\mathcal R,\mathcal R^{+})$ (when defined); insert the new neighbor keys $(\mathcal L^{-},\widetilde{\mathcal M})$ and $(\widetilde{\mathcal M},\mathcal R^{+})$ using \textsc{ComputeGeoKey}.
  \STATE \textbf{Optional periodic global glue.}
  If mode$=$\textit{all} and $s\bmod P=0$, run \textsc{GlueOptAll} on $t^\star$; then recompute keys adjacent to changed boundaries via \textsc{ComputeGeoKey} and refresh their occurrences in $\mathcal D^{(\star)}$.
\ENDFOR
\STATE \textbf{return} $\mathcal P^{(\star)}$ and $\mathcal F^{(\star)}$.
\end{algorithmic}
\end{minipage}
}
\end{algorithm}

\begin{algorithm}[t]
{
\begin{minipage}{0.96\linewidth}

\caption{{\textsc{BinHelper} — build geo-pair map for one backbone}}
\label{alg:bin-helper}
\begin{algorithmic}[1]
\REQUIRE Backbone $t^\star$ with current segmentation $\mathcal P^{(\star)}$;
quantizers $Q_{\theta^{C\!N\!CA}},Q_{\omega},Q_{\phi}$.
\ENSURE $\mathcal D^{(\star)}:\kappa\mapsto \mathcal O^{(\star)}(\kappa)$.

\STATE Initialize $\mathcal D^{(\star)}\leftarrow \emptyset$.
\FOR{each adjacent pair $(\mathcal L,\mathcal R)$ in $\mathcal P^{(\star)}$}
  \STATE $\kappa\leftarrow \textsc{ComputeGeoKey}(\mathcal L,\mathcal R)$ using:
  \begin{itemize}
    \item \emph{Raw} medoid internals for $\mathcal L$ and $\mathcal R$ (as assigned in initialization or prior merges);
    \item boundary glue $(\theta^{C\!N\!CA},\omega,\phi)$ snapped by $Q$.
  \end{itemize}
  \STATE Insert the occurrence $(\mathcal L,\mathcal R)$ into $\mathcal O^{(\star)}(\kappa)$.
\ENDFOR
\STATE \textbf{return} $\mathcal D^{(\star)}$.
\end{algorithmic}
\end{minipage}
}
\end{algorithm}

\begin{algorithm}[h!]
\caption{\textsc{GlueOptAll} — global differentiable inverse kinematics over glue angles}
\label{alg:glue-opt-all}
\begin{algorithmic}[1]
\REQUIRE Medoids $\widehat{\mathcal M}$ and assignments $c(\cdot)$ from \textsc{RMSD\_Partition}; occurrences $\mathcal S=\{u\}$ with spans $\mathcal M^{(t_u)}_{i_u{:}k_u}$; target frames $F^{\star,(t)}_i=(R^{\star,(t)}_i,t^{\star,(t)}_i)$ (with $F^{\star,(t)}_1$ from \textsc{SeedTriad}); weights $(w_R,w_t)$; optimizer steps $T$ and step size $\eta$
\ENSURE Updated glues $\{\Gamma^{(t)}_i\}$ and frames $\{\widehat F^{(t)}_i\}$

\STATE \textbf{Snap internals:} for $u\in\mathcal S$, set internals of $\mathcal M^{(t_u)}_{i_u{:}k_u}\leftarrow$ those of its medoid $m(u)=\widehat m_{c(u)}$
\STATE \textbf{Init glues:} copy original $\Gamma^{(t)}_i$ for all backbones $t$ and links $i=1{:}N^{(t)}{-}1$ (these are the optimization variables)
\STATE \textbf{Loss:}
\[
\mathcal L(\Gamma)=\sum_t\sum_{i=2}^{N^{(t)}}\!\Big(
w_R\big\|\log((\widehat R^{(t)}_i)^\top R^{\star,(t)}_i)\big\|_2^2
+w_t\big\|\widehat t^{(t)}_i-t^{\star,(t)}_i\big\|_2^2\Big)
\]
\STATE \textbf{Forward kinematics (FK):} with $\widehat F^{(t)}_1=F^{\star,(t)}_1$,
\[
\widehat F^{(t)}_{i+1}=\widehat F^{(t)}_i\,\widehat G^{(t)}_{i}\big(\Gamma^{(t)}_i;\ \text{current internals}\big),\quad
\widehat G^{(t)}_{i}\ \text{from internals and}\ \Gamma^{(t)}_i=\{\theta^{CNC\!A}_i,\psi_i,\phi_{i+1}\}\footnotemark
\]
\STATE \textbf{Optimize glues (autodiff):} for $s=1{:}T$:
\STATE \quad run FK, evaluate $\mathcal L$; backprop $\nabla_{\Gamma}\mathcal L$; update all $\Gamma^{(t)}_i$
\STATE \quad wrap $\psi,\phi\!\in\!(-\pi,\pi]$; project $\theta^{CNC\!A}\!\in\!(0,\pi)$
\STATE \textbf{return} $\{\Gamma^{(t)}_i\}$, $\{\widehat F^{(t)}_i\}$
\end{algorithmic}
\end{algorithm}

\begin{algorithm}[h!]
\caption{\textsc{Binary Tree–LSTM Cell} (Tai et al., 2015)}
\label{alg:treelstm-cell}
\begin{algorithmic}[1]
\REQUIRE Left $(h_\ell,c_\ell)\in\mathbb{R}^d\times\mathbb{R}^d$, right $(h_r,c_r)$; $W\in\mathbb{R}^{5d\times2d}$, $b\in\mathbb{R}^{5d}$
\ENSURE $(h_p,c_p)\in\mathbb{R}^d\times\mathbb{R}^d$
\STATE $u\!\leftarrow\!\begin{bmatrix}h_\ell\\h_r\end{bmatrix}$;\quad $\begin{bmatrix}i\\f_\ell\\f_r\\o\\g\end{bmatrix}\!\leftarrow\!Wu+b$
\STATE $i,f_\ell,f_r,o\!\leftarrow\!\sigma(\cdot)$;\quad $g\!\leftarrow\!\tanh(g)$
\STATE $c_p\!\leftarrow\!f_\ell\odot c_\ell+f_r\odot c_r+i\odot g$
\STATE $h_p\!\leftarrow\!o\odot\tanh(c_p)$
\STATE \textbf{return} $(h_p,c_p)$
\end{algorithmic}
\end{algorithm}
\begin{algorithm}[h!]
\caption{\textsc{Downward Binary Tree–LSTM Cell}}
\label{alg:treelstm-down-cell}
\begin{algorithmic}[1]
\REQUIRE Parent downward $(\bar h_p,\bar c_p)\in\mathbb{R}^d\times\mathbb{R}^d$, sibling upward $(h_s,c_s)$; $\widetilde W\in\mathbb{R}^{5d\times2d}$, $\widetilde b\in\mathbb{R}^{5d}$
\ENSURE $(\bar h_c,\bar c_c)\in\mathbb{R}^d\times\mathbb{R}^d$
\STATE $u\!\leftarrow\!\begin{bmatrix}\bar h_p\\ h_s\end{bmatrix}$;\quad $\begin{bmatrix}\bar i\\ \bar f_p\\ \bar f_s\\ \bar o\\ \bar g\end{bmatrix}\!\leftarrow\!\widetilde W u+\widetilde b$
\STATE $\bar i,\bar f_p,\bar f_s,\bar o\!\leftarrow\!\sigma(\cdot)$;\quad $\bar g\!\leftarrow\!\tanh(\bar g)$
\STATE $\bar c_c\!\leftarrow\!\bar f_p\odot\bar c_p+\bar f_s\odot c_s+\bar i\odot\bar g$
\STATE $\bar h_c\!\leftarrow\!\bar o\odot\tanh(\bar c_c)$
\STATE \textbf{return} $(\bar h_c,\bar c_c)$
\end{algorithmic}
\end{algorithm}
\begin{algorithm}[h!]
\caption{\textsc{Up–Down Tree Encoder on a Forest} (one protein)}
\label{alg:updown-encoder}
\begin{algorithmic}[1]
\REQUIRE Protein $t_\tau$ with $N^{(\tau)}$ residues; binary forest $\mathcal F^{(\tau)}=(V^{(\tau)},E^{(\tau)})$ whose frontier (in order) is $\mathcal P^{(\tau)}$; leaf embeddings $\{e^{(\tau)}_i\in\mathbb{R}^d\}_{i=1}^{N^{(\tau)}}$ (e.g., ESM3)\footnotemark; internal-edge topological order $E^{(\tau)}=\{(p,\ell,r)\}$; roots $R^{(\tau)}\subset V^{(\tau)}$; parameters $\Theta=\{W,b,\widetilde W,\widetilde b\}$; combiner $\oplus\in\{\text{concat},\text{sum}\}$.
\ENSURE $z^{\mathrm{prot}}_\tau\in\mathbb{R}^{d_z}$; $\{z^{\mathrm{res}}_{\tau,i}\}_{i=1}^{N^{(\tau)}}$.

\STATE \textbf{Upward.} For leaves $i\le N^{(\tau)}$: $h_i^\uparrow\!\leftarrow\!e^{(\tau)}_i$, $c_i^\uparrow\!\leftarrow\!0$. For $(p,\ell,r)\in E^{(\tau)}$ in order:
\[
(h_p^\uparrow,c_p^\uparrow)\leftarrow \textsc{TreeLSTMCell}(h_\ell^\uparrow,c_\ell^\uparrow,\ h_r^\uparrow,c_r^\uparrow;\ W,b)\ \text{(Alg.~\ref{alg:treelstm-cell})}.
\]

\STATE \textbf{Super-root.} $h_{\mathrm{SR}}^\uparrow \leftarrow |R^{(\tau)}|^{-1}\!\sum_{r\in R^{(\tau)}}\! h_r^\uparrow$; set node $\mathrm{SR}$ with $(h_{\mathrm{SR}}^\uparrow,c_{\mathrm{SR}}^\uparrow{=}0)$.

\STATE \textbf{Downward.} $(\bar h_{\mathrm{SR}},\bar c_{\mathrm{SR}})\leftarrow(0,0)$. For each tree rooted at $r\in R^{(\tau)}$, recurse: for internal $p$ with children $(\ell,r)$ and given $(\bar h_p,\bar c_p)$,
\[
\begin{aligned}
(\bar h_\ell,\bar c_\ell)&\leftarrow \textsc{DownTreeLSTM}\!\big((\bar h_p,\bar c_p),\ (h_r^\uparrow,c_r^\uparrow);\ \widetilde W,\widetilde b\big),\\
(\bar h_r,\bar c_r)&\leftarrow \textsc{DownTreeLSTM}\!\big((\bar h_p,\bar c_p),\ (h_\ell^\uparrow,c_\ell^\uparrow);\ \widetilde W,\widetilde b\big)\ \text{(Alg.~\ref{alg:treelstm-down-cell})}.
\end{aligned}
\]

\STATE \textbf{Representations.} For any node $v$: $u_v^\downarrow\leftarrow \bar h_v$.
\IF{$\oplus=\text{concat}$}
  \STATE $z^{\mathrm{res}}_{\tau,i}\leftarrow [\,h_i^\uparrow;\,u_i^\downarrow\,]\in\mathbb{R}^{2d}\ (i{=}1{:}N^{(\tau)})$;\quad $z^{\mathrm{prot}}_\tau\leftarrow [\,h_{\mathrm{SR}}^\uparrow;\,u_{\mathrm{SR}}^\downarrow\,]\in\mathbb{R}^{2d}$
\ELSE
  \STATE $z^{\mathrm{res}}_{\tau,i}\leftarrow h_i^\uparrow+u_i^\downarrow\in\mathbb{R}^{d}\ (i{=}1{:}N^{(\tau)})$;\quad $z^{\mathrm{prot}}_\tau\leftarrow h_{\mathrm{SR}}^\uparrow+u_{\mathrm{SR}}^\downarrow\in\mathbb{R}^{d}$
\ENDIF
\STATE \textbf{return} $z^{\mathrm{prot}}_\tau$, $\{z^{\mathrm{res}}_{\tau,i}\}_{i=1}^{N^{(\tau)}}$.
\end{algorithmic}
\end{algorithm}
\begin{algorithm}[h!]
\caption{\textsc{BuildJointVocab} — medoids then glue–angle bins}
\label{alg:build-joint-vocab}
\begin{algorithmic}[1]
\REQUIRE \algo vocab $\mathcal V=\{\kappa\mapsto\mathcal A_\kappa\}$ with key introduction order $(\kappa^{(1)},\dots,\kappa^{(S)})$; medoids $\mathcal A_\kappa=\{\Pi^{(\kappa)}_{j}\}_{j=1}^{K_{|\kappa|}}$; glue quantizers $Q_{\theta^{C\!N\!CA}},Q_{\omega},Q_{\phi}$ with bin centers $\{\mu_b^{\theta}\}_{b=1}^{B_\theta}$, $\{\mu_b^{\omega}\}_{b=1}^{B_\omega}$, $\{\mu_b^{\phi}\}_{b=1}^{B_\phi}$
\ENSURE Dictionary $\Sigma$; maps $\mathrm{id}_{\mathrm{med}}:(\kappa,j)\!\mapsto\!\{1,\dots,|\Sigma_{\mathrm{med}}|\}$ and $\mathrm{id}_{\mathrm{bin}}:(\mathrm{type}\!\in\!\{\theta,\omega,\phi\},b)\!\mapsto\!\{|\Sigma_{\mathrm{med}}|{+}1,\dots,|\Sigma|\}$

\STATE \textbf{Medoids (in introduction order):} $\Sigma_{\mathrm{med}}\leftarrow[\,]$.
\FOR{$s=1$ \TO $S$}\FOR{$j=1$ \TO $K_{|\kappa^{(s)}|}$}
  \STATE Append $\langle\kappa^{(s)},j\rangle$ to $\Sigma_{\mathrm{med}}$; set $\mathrm{id}_{\mathrm{med}}(\kappa^{(s)},j)$ to its index.
\ENDFOR\ENDFOR

\STATE \textbf{Glue bins (appended after medoids):} let $M=|\Sigma_{\mathrm{med}}|$.
\STATE $\mathrm{id}_{\mathrm{bin}}(\theta,b)=M+b$;\quad
       $\mathrm{id}_{\mathrm{bin}}(\omega,b)=M+B_\theta+b$;\quad
       $\mathrm{id}_{\mathrm{bin}}(\phi,b)=M+B_\theta+B_\omega+b$.

\STATE $\Sigma\leftarrow \Sigma_{\mathrm{med}}\ \cup\ \{\text{all glue-bin tokens}\}$ \ (\emph{optional}: add BOS/EOS)
\STATE \textbf{return} $\Sigma$, $\mathrm{id}_{\mathrm{med}}$, $\mathrm{id}_{\mathrm{bin}}$
\end{algorithmic}

\end{algorithm}
\begin{algorithm}[h!]
\caption{\textsc{BackboneToSequence} — tokenize a segmented backbone}
\label{alg:backbone-to-seq}
\begin{algorithmic}[1]
\REQUIRE Protein $t_\tau$ with segmentation $\mathcal P^{(\tau)}=(\mathcal M^{(t_\tau)}_{p_1{:}q_1},\ldots,\mathcal M^{(t_\tau)}_{p_M{:}q_M})$; for each $\mathcal M^{(t_\tau)}_{p_m{:}q_m}$ its key $\kappa_m$ and medoid $j_m$ (prototype $\Pi^{(\kappa_m)}_{j_m}$); boundary glue $\Gamma_{q_m}=\{\theta^{C\!N\!CA}_{q_m},\omega_{q_m},\phi_{q_{m}+1}\}$ for $m=1{:}M{-}1$; quantizers $Q_{\theta},Q_{\omega},Q_{\phi}$; token id maps from Alg.~\ref{alg:build-joint-vocab}.
\ENSURE Token sequence $x^{(\tau)}=(x_1,\ldots,x_L)\in\Sigma^L$.

\STATE $x^{(\tau)}\leftarrow[\,]$ \ (optionally prepend BOS/append EOS)
\FOR{$m=1$ \TO $M$}
  \STATE \textbf{Motif:} $x^{(\tau)}.\mathrm{append}\big(\mathrm{id}_{\mathrm{med}}(\kappa_m,j_m)\big)$
  \IF{$m<M$}
    \STATE \textbf{Glue quantize:} $b_\theta\!\leftarrow\!Q_\theta(\theta^{C\!N\!CA}_{q_m}),\ \ b_\omega\!\leftarrow\!Q_\omega(\omega_{q_m}),\ \ b_\phi\!\leftarrow\!Q_\phi(\phi_{q_{m}+1})$
    \STATE \textbf{Emit (fixed order):} $x^{(\tau)}.\mathrm{append}\big(\mathrm{id}_{\mathrm{bin}}(\theta,b_\theta)\big)$; $x^{(\tau)}.\mathrm{append}\big(\mathrm{id}_{\mathrm{bin}}(\omega,b_\omega)\big)$; $x^{(\tau)}.\mathrm{append}\big(\mathrm{id}_{\mathrm{bin}}(\phi,b_\phi)\big)$
  \ENDIF
\ENDFOR
\STATE \textbf{return} $x^{(\tau)}$
\end{algorithmic}

\end{algorithm}

\begin{algorithm}[h!]
\caption{\textsc{ResInitTokens} — initialize bond–residue codebook and quantize all residues}
\label{alg:res-init-tokens}
\begin{algorithmic}[1]
\REQUIRE Backbones $\{t^{(1)},\dots,t^{(T)}\}$ with lengths $N^{(\tau)}$; targets $K_3$ (interior bond–residues), $K_2$ (terminal bond–residues)
\ENSURE Codebooks $\mathcal{A}_3=\{\Pi^{(3)}_j\}_{j=1}^{K_3}$, $\mathcal{A}_2=\{\Pi^{(2)}_j\}_{j=1}^{K_2}$; labels $c^{(3)},c^{(2)}$; backbones with per-residue internals set to their prototypes

\STATE \textbf{Collect occurrences:}
\[
\mathcal{S}_3=\{u\!\equiv\!(\tau,i):1\!\le\! i\!<\!N^{(\tau)},\ \mathcal{M}^{(t_\tau)}_{i:i}\ \text{interior}\},\quad
\mathcal{S}_2=\{u\!\equiv\!(\tau,i):i\!=\!N^{(\tau)},\ \mathcal{M}^{(t_\tau)}_{i:i}\ \text{terminal}\}.
\]

\STATE \textbf{Cluster interiors:} \textsc{RMSD\_Partition}$(\mathcal{S}_3,K_3)$ $\to$ medoids $\widehat{\mathcal{M}}_3=\{\widehat m^{(3)}_j\}_{j=1}^{K_3}$ and labels $c^{(3)}:\mathcal{S}_3\to\{1,\dots,K_3\}$

\STATE \textbf{Define interior prototypes:} for $j=1{:}K_3$, let $u^\star=\widehat m^{(3)}_j$ and
\[
\Pi^{(3)}_j=\big(\ell^{N\!-\!CA}_{i_{u^\star}},\ \ell^{CA\!-\!C}_{i_{u^\star}},\ \ell^{C\!-\!N}_{i_{u^\star}},\
\theta^{N\!CA\!C}_{i_{u^\star}},\ \theta^{CA\!C\!N}_{i_{u^\star}},\ \psi_{i_{u^\star}},\ \omega_{i_{u^\star}},\ \phi_{i_{u^\star}}\big)
\]
(\emph{omit undefined terms for a single residue if using a minimal parameterization}).

\STATE \textbf{Quantize interiors:} for $u=(\tau,i)\in\mathcal{S}_3$ with $j=c^{(3)}(u)$,
\[
(\ell,\theta,\psi,\omega,\phi)\big|_{\mathcal{M}^{(t_\tau)}_{i:i}}\leftarrow \Pi^{(3)}_j.
\]

\STATE \textbf{Cluster terminals:} \textsc{RMSD\_Partition}$(\mathcal{S}_2,K_2)$ $\to$ medoids $\widehat{\mathcal{M}}_2=\{\widehat m^{(2)}_j\}_{j=1}^{K_2}$ and labels $c^{(2)}:\mathcal{S}_2\to\{1,\dots,K_2\}$

\STATE \textbf{Define terminal prototypes \& quantize:} for $j=1{:}K_2$, let $u^\star=\widehat m^{(2)}_j$ and set the appropriate terminal tuple (e.g., $\ell^{N\!-\!CA}_i,\ \ell^{CA\!-\!C}_i,\ \theta^{N\!CA\!C}_i$) as $\Pi^{(2)}_j$; for $u=(\tau,N^{(\tau)})$ with $j=c^{(2)}(u)$,
\[
(\ell,\theta)\big|_{\mathcal{M}^{(t_\tau)}_{N^{(\tau)}:N^{(\tau)}}}\leftarrow \Pi^{(2)}_j.
\]

\STATE \textbf{return} $\mathcal{A}_3,\mathcal{A}_2,\ c^{(3)},c^{(2)}$, and updated backbones
\end{algorithmic}

\end{algorithm}
\begin{algorithm}[h!]
\caption{\textsc{GlueOpt} — single-boundary IK to absorb one rounding drift}
\label{alg:glue-opt}
\begin{algorithmic}[1]
\REQUIRE Occurrence $u$ with motif $\mathcal M^{(t_u)}_{i_u{:}k_u}$ and medoid $\widehat m_{c(u)}$ from \textsc{RMSD\_Partition}; frames $\{F^{\star,(t_u)}_i\}$ with $F^{\star,(t_u)}_1$ from \textsc{SeedTriad}; weights $(w_R,w_t)$; steps $T$, step size $\eta$
\ENSURE $\Gamma^{(t_u)}_{i_u-1}$, $\widehat F^{(t_u)}_{k_u}$

\STATE \textbf{Snap internals:} replace $\mathcal M^{(t_u)}_{i_u{:}k_u}$ by its medoid $\mathcal M^{(t_{\widehat m_{c(u)}})}_{i_{\widehat m_{c(u)}}{:}k_{\widehat m_{c(u)}}}$; set $T^{\text{med}}_u\!\leftarrow\!T^{\mathrm{int}}_{i_{\widehat m_{c(u)}}{:}k_{\widehat m_{c(u)}}}$
\STATE \textbf{Drift:} $T^{\text{occ}}_u\!\leftarrow\!T^{\mathrm{int}}_{i_u{:}k_u}$; $\Delta T_u\!\leftarrow\!T^{\text{occ}}_u (T^{\text{med}}_u)^{-1}$
\STATE \textbf{Vars:} $\Gamma^{(t_u)}_{i_u-1}=\{\theta^{C\!N\!CA}_{i_u-1},\omega_{i_u-1},\varphi_{i_u}\}$ are the \emph{only} optimization variables\footnotemark; init to originals
\STATE \textbf{FK:} keep $F^{\star,(t_u)}_{i_u-1}$ fixed; for any $\Gamma^{(t_u)}_{i_u-1}$,
\[
\widehat F^{(t_u)}_{k_u}=F^{\star,(t_u)}_{i_u-1}\,\widehat G^{(t_u)}_{i_u-1}\!\big(\Gamma^{(t_u)}_{i_u-1}\big)\,T^{\text{med}}_u
\]
\STATE \textbf{Loss:}
\[
\mathcal L_u(\Gamma^{(t_u)}_{i_u-1})=
w_R\big\|\log((\widehat R^{(t_u)}_{k_u})^\top R^{\star,(t_u)}_{k_u})\big\|_2^2+
w_t\big\|\widehat t^{(t_u)}_{k_u}-t^{\star,(t_u)}_{k_u}\big\|_2^2
\]
\STATE \textbf{Optimize (autodiff):} for $s=1{:}T$: run FK \& evaluate $\mathcal L_u$; compute $\nabla_{\Gamma^{(t_u)}_{i_u-1}}\mathcal L_u$; update $\Gamma^{(t_u)}_{i_u-1}$ (e.g., Adam, lr $\eta$); wrap $\psi,\varphi\!\in\!(-\pi,\pi]$, project $\theta^{C\!N\!CA}\!\in\!(0,\pi)$
\STATE \textbf{return} $\Gamma^{(t_u)}_{i_u-1}$, $\widehat F^{(t_u)}_{k_u}$
\end{algorithmic}

\end{algorithm}
\begin{algorithm}[h!]
\caption{\textsc{GlueOptAll (wrapper)} — apply rounding, then call the core global IK over glues}
\label{alg:glue-opt-all-wrapper}
\begin{algorithmic}[1]
\REQUIRE Medoids $\widehat{\mathcal M}$ and assignments $c(\cdot)$ from \textsc{RMSD\_Partition};
         a set of occurrences $\mathcal S=\{u\}$ to round, each $\mathcal M^{(t_u)}_{i_u{:}k_u}$;
         cached original exit frames for each backbone $t$ (seeded by \textsc{SeedTriad});
         histogram bin centers \& thresholds for $(\omega,\theta^{C\!N\!CA},\varphi)$; prior weights; loss weights $(w_R,w_t)$.
\ENSURE Updated glue angles (snapped to bins) for all boundaries in all affected backbones; recomputed frames $\{\widehat F^{(t)}_i\}$.

\STATE \textbf{(Quantize internals)} For each $u\in\mathcal S$, replace
\[
\mathcal M^{(t_u)}_{i_u{:}k_u}\ \longleftarrow\
\mathcal M^{(t_{\widehat m_{c(u)}})}_{\,i_{\widehat m_{c(u)}}{:}k_{\widehat m_{c(u)}}}
\]
by copying the medoid’s internal coordinates (hard assignment).
\STATE \textbf{(Ensure targets are cached)} For each backbone $t$, ensure original exit frames $\{(R^{\star,(t)}_{\text{occ}}[j],t^{\star,(t)}_{\text{occ}}[j])\}_{j=1}^{N^{(t)}-1}$ are available (compute once if missing).
\STATE \textbf{(Global glue optimization)} \emph{Call the core routine} \textsc{GlueOptAll} on all backbones:
\[
\textsc{GlueOptAll}\Big(\{t\},\ \text{BinCenters},\ \text{Thresholds},\ \text{GluePrior},\ w_R,\ w_t\Big),
\]
which jointly optimizes every boundary’s glue triplet $\Gamma_i=\{\theta^{C\!N\!CA}_i,\ \omega_i,\ \varphi_{i+1}\}$ via differentiable FK and snaps each angle to the nearest histogram bin.
\STATE \textbf{return} updated glues and frames $\{\widehat F^{(t)}_i\}$.
\end{algorithmic}
\end{algorithm}
\begin{algorithm}[h!]
\caption{\textsc{BinGeoPairs} — build the dictionary of geo-pair occurrences (with hierarchy)}
\label{alg:bin}
\begin{algorithmic}[1]
\REQUIRE For each backbone $t^{(\tau)}$: its current segmentation
$\mathcal P^{(\tau)}=(\mathcal M^{(t_\tau)}_{p_1{:}q_1},\ldots,\mathcal M^{(t_\tau)}_{p_{M_\tau}{:}q_{M_\tau}})$
\emph{and} its merge hierarchy $\mathcal F^{(\tau)}$ whose frontier leaves, in order, equal $\mathcal P^{(\tau)}$;
precomputed boundary-glue quantizers $Q_{\theta^{C\!N\!CA}},Q_{\omega},Q_{\phi}$ (used only at pair boundaries).
\ENSURE A \emph{priority-ordered} map $\mathcal D$ from keys to occurrence sets, with ordered keys
$\pi(\kappa)=(\rho(\kappa),-|\mathcal O(\kappa)|,\kappa)$ where
$\rho(\kappa)=\mathbf{1}[\kappa\notin\mathrm{dom}(\mathcal V)]$ indicates if the key already has prototypes.

\STATE Initialize an empty ordered map $\mathcal D$.
\FOR{$\tau=1$ \TO $T$}
  \STATE Let $(\mathcal M^{(t_\tau)}_{p_1{:}q_1},\ldots,\mathcal M^{(t_\tau)}_{p_{M_\tau}{:}q_{M_\tau}})$ be the frontier leaves of $\mathcal F^{(\tau)}$ (these equal $\mathcal P^{(\tau)}$).
  \FOR{$j=1$ \TO $M_\tau-1$}
     \STATE $(\mathcal L,\mathcal R)\gets\big(\mathcal M^{(t_\tau)}_{p_j{:}q_j},\ \mathcal M^{(t_\tau)}_{p_{j+1}{:}q_{j+1}}\big)$
     \STATE $\kappa \leftarrow \textsc{ComputeGeoKey}(\mathcal L,\mathcal R)$ \hfill (raw medoid internals inside $\mathcal L,\mathcal R$; boundary glue quantized by $Q$)
     \STATE Insert the occurrence $(\mathcal L,\mathcal R)$ into $\mathcal O(\kappa)$.
  \ENDFOR
\ENDFOR
\FOR{each key $\kappa$ with nonempty $\mathcal O(\kappa)$}
  \STATE Set $\rho(\kappa)\leftarrow \mathbf{1}[\kappa\notin \mathrm{dom}(\mathcal V)]$.
  \STATE Insert $\big((\rho(\kappa),-|\mathcal O(\kappa)|,\kappa)\mapsto \mathcal O(\kappa)\big)$ into $\mathcal D$.
\ENDFOR
\STATE \textbf{return} $\mathcal D$.
\end{algorithmic}
\end{algorithm}
\begin{algorithm}[h!]
\caption{\textsc{ComputeGeoKey} — discrete key for an adjacent motif pair}
\label{alg:compute-geo-key}
\begin{algorithmic}[1]
\REQUIRE Adjacent motifs $\mathcal L=\mathcal M^{(t)}_{p{:}q}$, $\mathcal R=\mathcal M^{(t)}_{q+1{:}r}$ on backbone $t$
\ENSURE Canonical, hashable key $\kappa$ for the geo-pair

\STATE \textbf{Interiors (as stored post-quantization):}
\[
\mathrm{Int}(\mathcal L)=\Big(\{\ell,\theta,\psi,\omega,\phi\}\big|_{i=p}^{q},\ \{\Gamma_i\}_{i=p}^{q-1}\Big),\quad
\mathrm{Int}(\mathcal R)=\Big(\{\ell,\theta,\psi,\omega,\phi\}\big|_{i=q+1}^{r},\ \{\Gamma_i\}_{i=q+1}^{r-1}\Big)
\]
(kept unchanged in the key).

\STATE \textbf{Boundary glue (quantized):}
\[
\Gamma_q=(\theta^{C\!N\!CA}_q,\omega_q,\phi_{q+1}),\qquad
\widetilde{\Gamma}_q=\big(Q_\theta(\theta^{C\!N\!CA}_q),\ Q_\omega(\omega_q),\ Q_\phi(\phi_{q+1})\big)
\]
($Q_\bullet$ wrap to a fixed $2\pi$ interval before snapping).

\STATE \textbf{Canonical record \& hash:}
\[
\mathrm{rec}=\big(\mathrm{Int}(\mathcal L),\ \widetilde{\Gamma}_q,\ \mathrm{Int}(\mathcal R)\big),\qquad
\kappa\leftarrow \textsc{Hash}(\mathrm{rec})
\]
\STATE \textbf{return} $\kappa$
\end{algorithmic}

\end{algorithm}

\begin{algorithm}[h!]
\caption{\textsc{Compute\_Coords} — Internal $\to$ Cartesian for a bond--residue motif $i{:}j$}
\label{alg:motif-to-cart}
\begin{algorithmic}[1]
\REQUIRE Internal geometry for $r=i,\dots,j$ (bond lengths/angles/dihedrals).
\ENSURE $\mathbf{X}\in\mathbb{R}^{3(j-i+1)\times 3}$ for $(N_i,\mathrm{CA}_i,C_i,\dots,N_j,\mathrm{CA}_j,C_j)$

\STATE \textbf{Seed residue $i$:} $(N_i,\mathrm{CA}_i,C_i)\gets \textsc{SeedTriad}(i)$
\STATE $\mathcal{C}\gets [N_i,\mathrm{CA}_i,C_i]$
\FOR{$r=i+1$ \TO $j$}
\STATE $N_r \gets \textsc{PlaceDihedral}\bigl(\mathcal{C}[-3],\mathcal{C}[-2],\mathcal{C}[-1];\ \ell^{C\!-\!N}_{r-1},\ \theta^{CA\!C\!N}_{r-1},\ \psi_{r-1}\bigr)$; $\mathcal{C}.\mathrm{append}(N_r)$;\\
\STATE $\mathrm{CA}_r \gets \textsc{PlaceDihedral}\bigl(\mathcal{C}[-3],\mathcal{C}[-2],\mathcal{C}[-1];\ \ell^{N\!-\!CA}_r,\ \theta^{C\!N\!CA}_{r-1},\ \omega_{r-1}\bigr)$; $\mathcal{C}.\mathrm{append}(\mathrm{CA}_r)$\\
\STATE $C_r \gets \textsc{PlaceDihedral}\bigl(\mathcal{C}[-3],\mathcal{C}[-2],\mathcal{C}[-1];\ \ell^{CA\!-\!C}_r,\ \theta^{N\!CA\!C}_r,\ \varphi_r\bigr)$; $\mathcal{C}.\mathrm{append}(C_r)$
\ENDFOR

\STATE $\mathbf{X} \gets [\, N_i,\ \mathrm{CA}_i,\ C_i,\ \dots,\ N_j,\ \mathrm{CA}_j,\ C_j \,]$
\RETURN $\mathbf{X}$
\end{algorithmic}
\end{algorithm}
\begin{algorithm}[h!]
\caption{\textsc{PlaceDihedral}$(a,b,c;\ L,\ \beta,\ \tau)$}
\label{alg:place-dihedral}
\begin{algorithmic}[1]
\STATE Right-handed local frame at $c$:
\[
\hat{\mathbf{b}}=\frac{c-b}{\|c-b\|},\quad
\mathbf{n}=\frac{(b-a)\times \hat{\mathbf{b}}}{\|(b-a)\times \hat{\mathbf{b}}\|},\quad
\tilde{\mathbf{n}} = \mathbf{n}\times \hat{\mathbf{b}}.
\]
\STATE Local offset:
\[
\mathbf{d}
= \bigl[-L\cos\beta\bigr]\hat{\mathbf{b}}
 + \bigl[L\cos\tau\sin\beta\bigr]\tilde{\mathbf{n}}
 + \bigl[L\sin\tau\sin\beta\bigr]\mathbf{n}.
\]
\STATE Return $d = c + \mathbf{d}$.
\end{algorithmic}
\end{algorithm}
\begin{algorithm}[h!]
\caption{\textsc{SeedTriad}$(r)$ — seed triad for residue $r$}
\label{alg:seedtriad}
\begin{algorithmic}[1]
\REQUIRE Canonical seed $(N_\star,\mathrm{CA}_\star,C_\star)$; target
$L_{CA\!-\!C}=\ell^{CA\!-\!C}_r$, $L_{N\!-\!CA}=\ell^{N\!-\!CA}_r$, and $\theta_{N\!CA\!C}=\theta^{N\!CA\!C}_r$.
\STATE Place $\widetilde{\mathrm{CA}}_r$ on the ray from $C_\star$ toward $\mathrm{CA}_\star$ at distance $L_{CA\!-\!C}$.
\STATE Let $\mathbf{u}=N_\star-\widetilde{\mathrm{CA}}_r$ and $\mathbf{v}=C_\star-\widetilde{\mathrm{CA}}_r$.
Rotate $\mathbf{u}$ about axis $\mathbf{u}\times\mathbf{v}$ to achieve angle $\theta_{N\!CA\!C}$, then rescale to length $L_{N\!-\!CA}$; translate by $\widetilde{\mathrm{CA}}_r$ to get $N_r$.
\STATE Set $C_r\gets C_\star$ and return $(N_r,\widetilde{\mathrm{CA}}_r,C_r)$.
\end{algorithmic}
\end{algorithm}
\newpage

\end{document}